\documentclass{aa}  

\usepackage{amsmath}	
\usepackage{relsize}
\usepackage{booktabs}
\usepackage{graphicx}
\usepackage{hyperref}
\usepackage{tablefootnote}

\hypersetup{
    colorlinks = true, 
    urlcolor = cyan, 
    linkcolor = blue, 
    citecolor = blue 
}
\usepackage{txfonts}

\usepackage{xcolor}
\usepackage{lscape}
\begin{document} 

   \title{Searching for exotic object companions \\ in the dense core of NGC 362}
    \titlerunning{Exotic object companions in NGC 362}
   \subtitle{A multi-wavelength and multi-epoch photometric analysis}
   \titlerunning{Searching for exotic object companions in NGC 362}

   \author{Greta Ettorre,
          \inst{1,2}
          Emanuele Dalessandro,
          \inst{2} 
          Cristina Pallanca,
          \inst{1,2}
          Mario Cadelano, 
          \inst{1,2}
          Gourav Kumawat, 
          \inst{3}\\
          Craig O. Heinke,
          \inst{3}
          Sebastian Kamann,
          \inst{4}
          Mattia Libralato,
          \inst{5}
          Phyllis M. Lugger,
          \inst{6}
          Haldan N. Cohn,
          \inst{6}
          Stefan Dreizler
          \inst{7}
          }

   \authorrunning{Ettorre et al.}         
   \institute{Department of Physics and Astronomy “Augusto Righi”, University of Bologna, Via Gobetti 93/2, 40129 Bologna, Italy\\
              \email{greta.ettorre@inaf.it}
         \and
             INAF – Astrophysics and Space Science Observatory of Bologna, Via Gobetti 93/3, 40129 Bologna, Italy
        \and
            Department of Physics, University of Alberta, Edmonton, AB T6G 2G7, Canada
        \and
            Astrophysics Research Institute, Liverpool John Moores University, 146 Brownlow Hill, Liverpool L3 5RF, UK
        \and
            INAF-Osservatorio Astronomico di Padova, Via dell’Osservatorio 5, 35122 Padova, Italy
        \and
            Department of Astronomy, Indiana University, 727 E. Third St., Bloomington, IN 47405, USA
        \and
            Institut f\"ur Astrophysik, Georg-August-Universit\"at G\"ottingen, Friedrich-Hund-Platz 1, 37077 G\"ottingen, Germany
            }
   \date{Received XXX; accepted YYY}

  \abstract
  {The dense cores of globular clusters (GCs) are efficient environments for the production of exotic stellar populations, including millisecond pulsars (MSPs), low-mass X-ray binaries (LMXBs) and cataclysmic variables (CVs). Most of these objects likely form through two- and three-body interactions and are useful tracers of the cluster's dynamical evolution. In this work we explore the exotic object population in the galactic GC NGC 362, searching for the optical counterpart of 33 X-ray sources identified within $1 \arcmin$ from the cluster center. To this aim, we exploit a large \textit{Hubble Space Telescope} (\textit{HST}) dataset, obtained in eight different epochs and covering a wavelength range from near UV to the optical I band. To identify the most promising counterparts we follow a multi-step analysis, which is based on four main ingredients, namely positional coincidence, position in the colour-magnitude-diagrams (CMDs), H$\alpha$ excess and photometric variability.
  In addition, we complement the photometric analysis with spectroscopic information coming from the analysis of MUSE radial velocity (RV) curves. 
  Thanks to this multi-diagnostic approach, we are able to identify 28 high-confidence optical counterparts, including several candidate MSPs, active binaries (ABs) and CVs.  The most intriguing counterparts include a candidate black widow (BW) system, an eclipsing binary blue straggler, and a system in outburst, potentially representing either a LMXB or a nova eruption from a CV.
  The candidate MSPs proposed in this work will contribute to ongoing radio analyses with MeerKAT for the identification and detailed study of MSPs in NGC 362.
  
  }


   \maketitle

\section{Introduction}\label{intro}
It is well established that frequent stellar dynamical interactions in the high-density cores of globular clusters (GCs) make them efficient factories for forming exotic objects, including millisecond pulsars (MSPs), low-mass X-ray binaries (LMXBs) and cataclysmic variables (CVs) \citep{Paresce1992,Bailyn1995,Bellanzini1995,Ferraro2001,Pooley2006,Freire2008}. Most of these objects are believed to result from the evolution of binary systems \citep{Ivanova2006}. Their existence is therefore closely tied to the dynamics of the parent cluster, thus making them useful indicators for tracing its dynamical evolution \citep{Ferraro2009,Ferraro2012,dalessandro2008a,dalessandro2008b,Lanzoni2016,Kiziltan2017,Dickson2024}. Moreover, these objects are responsible for most of the X-ray emission detected in GCs \citep{Clark1975,Hertz1983}.

LMXBs are binary systems in which a neutron star (NS) or a black hole (BH) accretes mass via Roche lobe overflow from a low-mass companion star through the formation of an accretion disc. When accretion onto the NS is low or non-existent, these systems are classified as quiescent LMXBs (qLMXBs). The latter are usually the brightest X-ray sources in the $0.5-2.5$ keV band, with $\mathrm{L_{X}} \gtrsim 10^{32}$ erg/s, and show soft blackbody-like spectra \citep{Heinke2003,Verbunt2008}.  
Interestingly, LMXBs have long been considered the progenitors of MSPs, as mass accretion from the evolving companion eventually spins up the NS to millisecond periods \citep{Ferraro2015b}. The number of MSPs per unit mass in Galactic GCs is about $10^3$ times larger than in the Galactic field \citep{Hui2010,Turk2013,ZhaoHeinke2022}. This is due to the fact that dynamical interactions in the dense cluster cores can promote the formation of binaries suitable for recycling NSs into MSPs. This recycling phase can last up to approximately 1 Gyr, ultimately producing a rapidly spinning MSP with spin periods typically shorter than 10 ms. The companion star becomes a low-mass white dwarf (WD) with a He core or, more rarely, with a CO core \citep[e.g.][]{TaurisSavoije1999} ; or with non-degenerate companions, the so-called “spiders” \citep{Roberts2013}.

These “spiders” are divided into two main classes, Black Widows (BWs) and Redbacks (RBs). The distinction between the two types of spiders lies in the mass of their companion stars. RBs have companions with masses ranging from $0.1-0.5$ $\mathrm{M_{\odot}}$, whereas BWs have companions with masses smaller than $0.1$ $\mathrm{M_{\odot}}$. 
The evolutionary scenario of such systems is still under debate. According to the simulations performed by \cite{Chen2013}, the strong MSP wind may be the discriminant factor in the evolution of the PSR companion into a RB or a BW, suggesting that they are the result of two different evolutionary paths. A strong evaporation scenario is also adopted by \cite{Smedley2015}. On the other side, \cite{Benvenuto2014} suggests that the RBs may either evolve into BWs or into canonical MSPs with a He WD companion, adopting an irradiation feedback scenario \citep{Benvenuto2015,Devito2020}. 
The connection between LMXBs and MSPs was confirmed by the discovery of a new class of binary systems known as "transitional MSPs" (tMSPs), which alternate between a purely accretion-powered state and a purely rotation-powered state, where they appear as RBs \citep{Papitto2013,Pallanca2013,DeMartino2015, Archibald2009}.

In addition to their radio emission, MSPs also show X-ray emission \citep{Saito1997,Verbunt1996,Zavlin2002}. The majority of MSPs appear as soft thermal X-ray sources, where the emission is mostly dominated by thermal blackbody from the heated polar caps of the NS \citep{Bogdanov2006}. However, some MSPs may show sharp X-ray pulsations, which are indicative of non-thermal processes occurring in the system, as magnetospheric emission \citep{Chatterjee2007} or non-thermal shock radiation resulting from the collision between the pulsar wind and the stellar wind \citep{Wadiasingh2017}.
MSPs with degenerate counterparts typically do not show strong H$\alpha$ emission, because accretion is inhibited by their strong magnetic fields.
In contrast, those with non-degenerate companions (spiders) may show strong H$\alpha$ emission \citep[e.g. the RB system PSR J1740-5340A in NGC 6397;][]{Ferraro2001}. This is particularly true for tMSPs, which can have accretion disks, as found for the two tMSPs PSR J1824-2452I in M28 \citep{Pallanca2013} and PSR J1023+0038 found in the Galactic field \citep{Bond2002,Archibald2009}.

CVs are binary systems where a WD accretes material from a secondary companion star, either a main sequence (MS) star or a subgiant, through Roche lobe overflow. These systems are important for searching for progenitors of type Ia supernovae, as all type Ia supernova progenitor populations are believed to be closely related to CVs, and some CVs might themselves be progenitors of type Ia supernovae \citep{Knigge2012}. These objects exhibit hard X-ray spectra generated by thermal emission from hot plasma \citep{Heinke2005}. 
Identifying CVs in GCs by using X-ray positions to search for \textit{Hubble Space Telescope} (\textit{HST}) counterparts has been highly successful \citep{Cool1995,Pooley2002,Edmonds2003,Cohn2010}. Photometric identification of a star in an X-ray error circle as a CV has typically relied on using some combination of blue colours, variability, and/or H$\alpha$ emission. All three of these signatures are produced by hot, variable, often optically thin accretion disks produced by mass transfer.
In particular, in the case of a magnetically-weak CV, the line emission may originate from the optically thin regions of the accretion disc, whereas for a magnetic CV, it may come from the accretion stream \citep{Witham2006}.

Finally, active binaries (ABs) are also known to contribute to the X-ray emission observed from GCs. In these systems, the emission is due to the coronal activity of MS, giant, or subgiant stars, where the magnetic corona contains plasma at temperatures exceeding $10^6$ K \citep{Gudel2004}. Examples of these systems include sub-subgiants (SSGs), red straggler stars (RSSs), BY Dra binaries, and RS CVn binaries.  SSGs are observed below and redwards of the sub-giant branch in optical colour-magnitude diagrams (CMDs), while RSSs are observed to the red of the normal Red Giant Branch \citep[RGB;][]{Geller2017}. RS CVn binaries involve at least one evolved star showing enhanced coronal activity, while BY Dra binaries include two main-sequence stars, where the enhanced activity is due to the tidal locking of a close binary producing fast rotation, that increases coronal activity \citep{Dempsey1997}.
These systems represent the faintest X-ray sources in GCs, often characterized by soft X-ray spectra \citep{Verbunt2008}. In the sample studied by \cite{Geller2017}, $58\%$ of the SSGs are characterized by X-ray emission with $0.5-2.5$ keV luminosities of the order $10^{30}-10^{31}$ erg/s and $33\%$ of them are H$\alpha$ emitters. In fact, chromospheric activity in ABs also leads to the presence of H$\alpha$ emission. However, the intensity of this emission is significantly lower in ABs compared to CVs and LMXBs, due to the different physical processes giving rise to it \citep[see for example the results found in 47 Tucanae by][]{Beccari2013}.

Identifying optical counterparts to exotic objects is essential for understanding their formation and characterizing their properties in terms of both their degenerate and non-degenerate components. In particular, it has been shown that the adoption of a multi-wavelength approach, including X-ray and optical observations, is extremely effective in characterizing the properties and nature of exotic objects \citep{Cadelano2019,Cadelano2020,Zhang2023}. In fact, each of these observational methods offers unique and complementary insights into the same system, allowing for a more comprehensive understanding of its nature. On one side, X-ray observations provide insights into the possible presence of a compact object, as well as any ongoing accretion processes or coronal activity, while the optical emission is dominated by the companion non-degenerate star.

In this framework, we explore the exotic object population in the Galactic massive GC NGC 362. This cluster is located at 8.8 kpc distance, it has an estimated age of $11.5$ Gyrs \citep{Dotter2010} and metallicity [Fe/H]=$-1.3$ \citep{Carretta2009}. Its kinematic properties \citep{Bianchini2018,Libralato2018}, as well as the presence of a double blue straggler stars sequence \citep{Dalessandro2013}, suggest that NGC 362 is in a post-core-collapse state. Moreover, 12 pulsars have been identified (but not yet published) in NGC 362, mostly via the TRAPUM project using the MeerKAT radio array, with nine of them residing in binary systems. \footnote{\url{https://www3.mpifr-bonn.mpg.de/staff/pfreire/GCpsr.html}}

The starting point of our analysis is the list of 33 X-ray sources identified by \cite{Kumawat2024} within $1 \arcmin$ from the cluster center, using 78.8 kiloseconds of \textit{Chandra} data taken in 2004.  In their work, \cite{Kumawat2024} used the \textit{HST} UV Globular Cluster Survey (HUGS) catalogue \citep{Piotto2015,Nardiello2018} to search for optical counterparts to these X-ray sources. They found 15 candidate counterparts, including two background active galactic nuclei (AGN), three sub-subgiants, two RSS counterparts, five candidate ABs and two objects shifting between the bluer and redder parts of the RGB. According to their results, the majority of these bright optical counterparts are likely to be powered by coronal activity.

In this work, we expand on the search for optical counterparts to the X-ray sources in NGC 362. 

While \citet{Kumawat2024} identified optical counterparts by performing astrometric matching between \textit{Chandra} observations and the HUGS catalog,  here we make use of a larger multi-wavelength, multi-instrument and multi-epoch \textit{HST} dataset and we adopt a data-analysis approach specifically tailored to identify possibly faint objects in strongly crowded fields. As a result, we obtain deeper CMDs—particularly in the bluer filters where many stars were not detected in HUGS—we incorporate H-alpha photometric analysis and investigate photometric variability of the counterparts across individual exposures. Additionally, we use the spectroscopic information secured by the integral field spectrograph Multi Unit Spectroscopic Explorer (MUSE) at the Very Large Telescope \citep[VLT;][]{Bacon2010}, including radial velocities (RVs).  
The combination of the adopted approach and extended dataset enables a deep and comprehensive search for optical counterparts, strengthening the identification of the most likely ones and providing useful information for the characterization of their properties.

The paper is organized as follows: in Sect.~\ref{analysis}, we describe the \textit{HST} dataset and detail the photometric analysis, including the calibration and astrometric procedures used during the reduction process. In Sect.~\ref{methods}, we describe our approach for identifying the optical counterparts of the X-ray sources, including their identification in the optical CMD, and the analysis of their variability and H$\alpha$ emission. The results of our analysis are summarised in Sect.~\ref{results}, highlighting the most interesting counterparts we discovered. In Sect.~\ref{spectroscopy}, we present the insights obtained from spectroscopy, including a comparison with a recent spectroscopic study and our analysis of RV curves for a subset of the high-confidence counterparts. Finally, in Sect.~\ref{conclusions} we compare our results with the reference work by \cite{Kumawat2024} and draw our conclusions.

\section{\textit{HST} observations and data analysis}\label{analysis}
\begin{figure}
    \centering
    \includegraphics[width=\hsize]{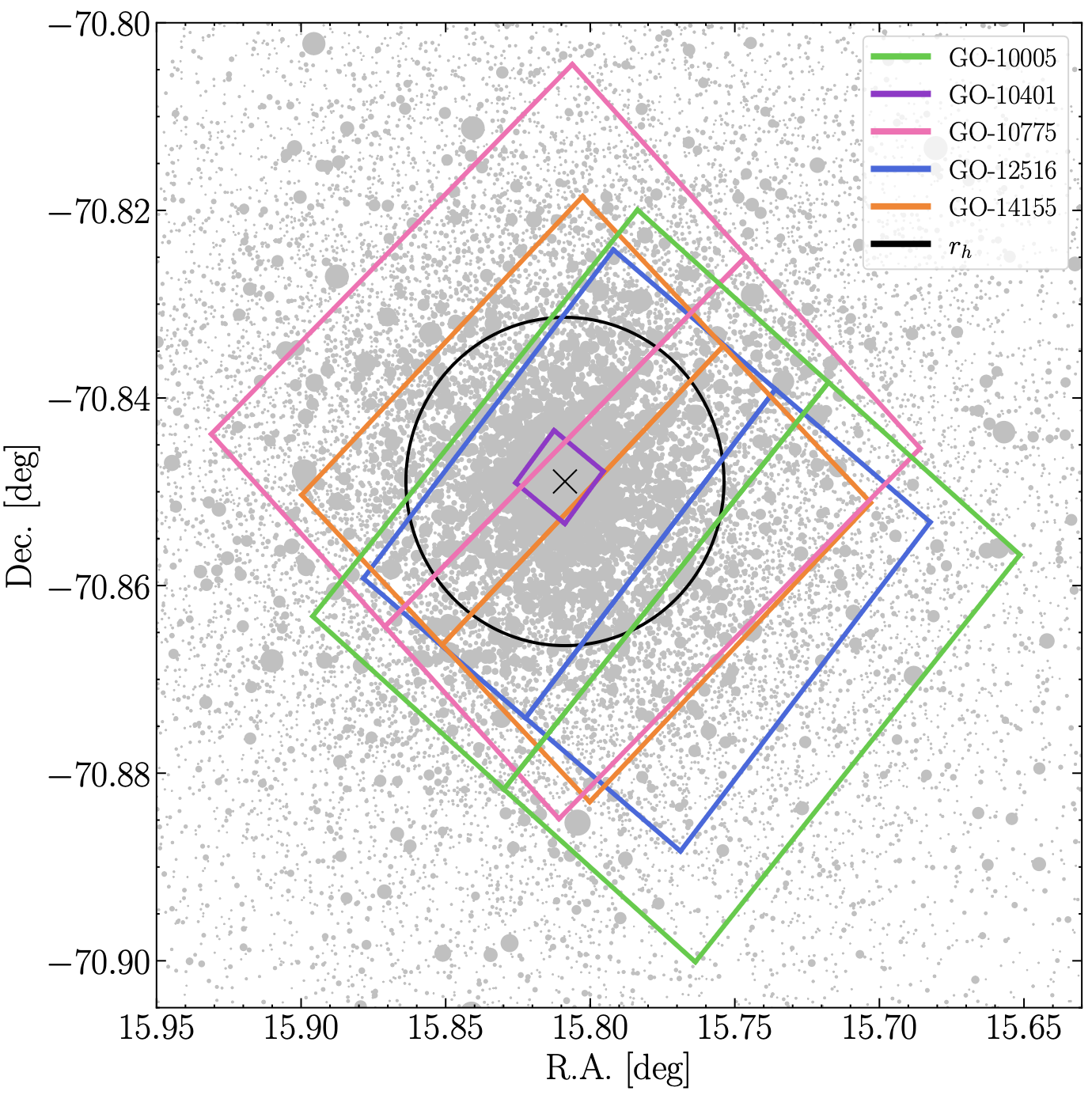}
    \caption{Footprints of the FoVs covered by the \textit{HST} observations listed in Table~\ref{dataset}. The underlying stars come from the Gaia DR3 dataset \citep{Gaia2023} centered on NGC 362, obtained with a search radius of 1 degree and using the Gaia \textit{G}-band magnitude to scale the size of the data points. Each FoV is colour-coded according to the GO proposal number, listed in Table~\ref{dataset}. The black cross marks the centre\citep[R.A. = 15.8087453, decl. = -70.8489012,][]{Dalessandro2013} while the black circle represents the half-mass radius, $\mathrm{r_h}$ \citep[$63\arcsec\!.5$,][]{Dalessandro2013}.}
    \label{map}
\end{figure}

The photometric dataset used in this work consists of high-resolution \textit{HST} images ranging from the near-UV (F225W) to the optical I band (F814W), obtained using the ultraviolet-visible (UVIS) channel of the Wide Field Camera 3 (WFC3) and the Advanced Camera for Surveys (ACS), both with the High Resolution Channel (HRC) and Wide Field Channel (WFC). Fig.~\ref{map} shows the Field of View (FoV) covered by all the \textit{HST} observations used in this work, while Table~\ref{dataset} provides detailed information about them in terms of the adopted filters, number of images and exposure time.

When searching for optical counterparts of exotic objects, two major challenges have to be faced: the faintness of the companions, such as in the case of MSPs or CVs, and the significant crowding of the dense cores of GCs where these objects are typically located. As a result, achieving high photometric accuracy and completeness at faint magnitudes in crowded regions is crucial for this kind of analysis. Thus, our approach in the photometric reduction described in the following sections is guided by these requirements.

\begin{table*}[h]
\caption{\label{dataset}\textit{HST} observations of NGC 362 used in this work.}
\centering
\begin{tabular}{lcccccc}
\hline\hline
Instrument & Date & Filter & N of images & Proposal ID & PI & $\mathrm{N} \times \mathrm{Exp.\, time \, (s)}$\\
\hline
ACS/WFC &   2003 Dec 04  &F435W & 4   & 10005   & Lewin & $4\times 340$\\
ACS/WFC &   2003 Dec 04  &F625W  & 4   & 10005  & Lewin & $2\times 110$, $2\times 120$ \\
ACS/WFC &   2003 Dec 04  &F658N & 4   & 10005  & Lewin & $2\times 440$, $2\times 500$ \\
ACS/HRC     &   2004 Dec 06  &F435W & 16   & 10401 & Chandar & $16 \times 85$\\
ACS/WFC &   2006 Jun 02  &F606W & 5  & 10775 $*$   & Sarajedini & $4\times 150$, $1\times 10$\\
ACS/WFC &   2006 Jun 02  &F814W & 5   & 10775 $*$  & Sarajedini &  $4\times 170$, $1\times 10$\\
WFC3/UVIS   &   2012 Apr 13  &F390W & 14   & 12516 & Ferraro & $14 \times 348$\\
WFC3/UVIS    &   2012 Apr 13  &F555W & 10   & 12516  & Ferraro & $ 1 \times 160$, $ 1 \times 200$, $ 1 \times 145$, $ 1 \times 144$, $6 \times 150$\\
WFC3/UVIS    &   2012 Apr 13  &F814W  & 15   & 12516  & Ferraro & $3 \times 390$, $12 \times 348$\\
WFC3/UVIS    &   2016 Sept 18  &F225W  & 18   & 14155  & Kalirai & $3 \times 30$, $3 \times 680$, $8 \times 700$, $3 \times 650$, $1 \times 580$\\
WFC3/UVIS    &   2016 Sept 18  &F275W  & 7   & 14155  & Kalirai & $2 \times 30$, $ 680$, $2 \times 700$, $2 \times 650$\\
\hline
\end{tabular}
\tablefoot{From left to right: \textit{HST} instrument name, date of observation, filter of the observation, number of images, proposal ID, principal investigator and exposure time in seconds for each image. The asterisk (*) in the fifth column denotes the \textit{HST} observations that were also used by \citet{Kumawat2024}. Conversely, the observations used in their work but not used in our analysis are the WFC3/UVIS images from Proposal GO-12605, specifically in the F275W, F336W, and F438W filters.\\
All WFC3/UVIS and ACS/HRC observations listed in the table were processed following the photometric reduction described in Sects.~\ref{wfc3data} and \ref{hrcdata}, and the resulting catalogues were subsequently used in the analysis. The only external catalogue used in this study is the one from \citet{Libralato2018,Libralato2022}, constructed from the ACS/WFC observations.}
\end{table*}

\subsection{WFC3 data}\label{wfc3data}
\begin{figure}
    \centering    \includegraphics[width=\hsize]{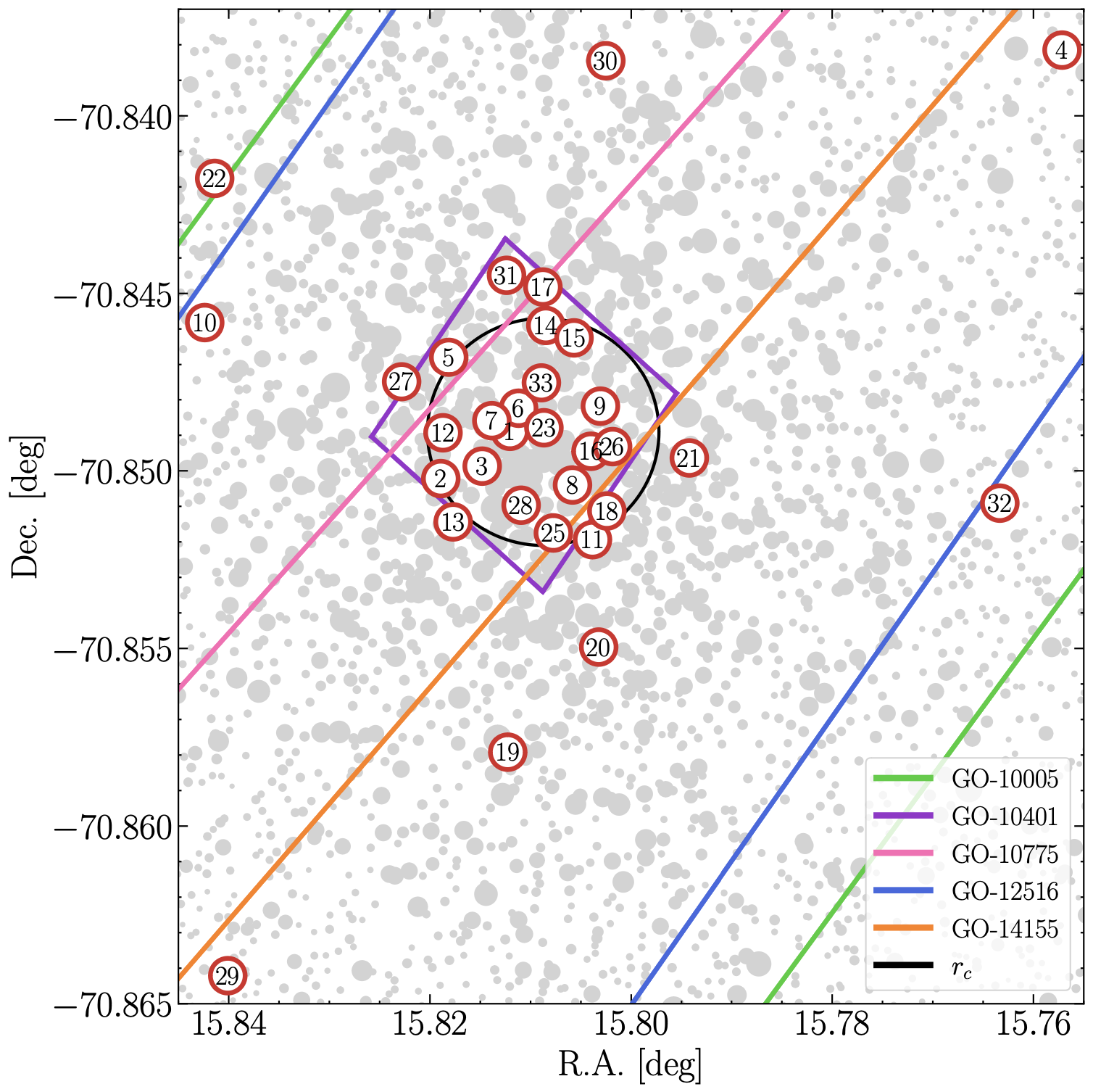}
    \caption{Zoomed-in version of Fig.~\ref{map}.
    Red circles represent the 33 X-ray sources identified by \cite{Kumawat2024}. 
    Each source is labelled with a number, following the nomenclature established by \cite{Kumawat2024}. The black circle marks the core radius, $\mathrm{r_c}$ \citep[$13\arcsec\!.0$,][]{Dalessandro2013}.}
    \label{map_zoom}
\end{figure}

The WFC3 dataset consists of images obtained in two different epochs. The first contains 39 images obtained on 2012 April 13 as part of the proposal 12516 (PI: Ferraro) with the F390W, F555W, and F814W filters. The cluster core falls entirely within chip $\#$1, and 32 out of 33 X-ray sources are located within the FoV (blue solid line in Figs.~\ref{map} and \ref{map_zoom}).
The second sample contains 25 images taken on 2016 September 18, in the filters F225W and F275W as part of the proposal 14155 (PI: Kalirai). In this case the cluster core is almost entirely contained within chip $\#$2 and all the 33 X-ray sources fall within the FoV (orange solid line in Figs.~\ref{map} and \ref{map_zoom}).
In both cases, the reduction was performed on UVIS calibrated exposures that include the charge transfer efficiency (CTE) correction (\texttt{$\_$flc} images). Additionally, for the F225W and F275W images, cosmic rays were removed using the Python package \texttt{lacosmic} \citep{VanDokkum2001}.
After correcting the frames for the  ``Pixel-Area-Map'' (PAM), we conducted the photometric analysis using the \texttt{DAOPHOT} package \citep{Stetson1987} separately for the two datasets but following the same steps, which are described below.

As a first step, we created a list of stars using the \texttt{FIND} package, including only sources having peak counts larger than a given threshold above the local background level. This limit was set at $5\sigma$ for the first dataset and $3\sigma$ for the F225W and F275W images, where $\sigma$ is the standard deviation of the local background noise. Using \texttt{PHOTO} we obtained the concentric aperture photometry of each source. For each image we selected a large number ($\sim 200$) of bright, unsaturated and isolated stars. 
These stars were then used to derive the best-fit point-spread function (PSF) through the package \texttt{PSF}. The PSF model that best reproduces the observed PSF is a Penny function \citep{Penny1976} for the F814W images and a Moffat function \citep{Moffat1969} with $\beta=1.5$ and $\beta=2.5$ for the F555W and F390W, F225W, F275W frames, respectively. The best-fit PSF models were then applied to all the sources using \texttt{ALLSTAR}. Finally, for each of the two datasets, we took full advantage of the \texttt{ALLFRAME} \citep{Stetson1994} routine to simultaneously fit stellar sources in all available images. This required the creation of a masterlist including all the stars that have to be fitted. Using \texttt{DAOMATCH} and \texttt{DAOMASTER}, we generated, for each filter, a list of stars with their average magnitudes and associated errors, determined from the dispersion of individual measurements. Only sources measured in at least half of the images +1 were included. The lists from the different filters were then combined into the final masterlist, again using \texttt{DAOMATCH} and \texttt{DAOMASTER}.

The WFC3 images are characterized by significant geometrical distortions across the FoV. Therefore, we corrected the instrumental positions of the first dataset with the equations by \cite{Bellini2009} for the filter F814W, while we used the coefficients of the F225W for the second dataset.
We then transformed the instrumental coordinates to the absolute coordinate systems by cross-matching the WFC3 catalogue with the Gaia DR3 dataset \citep{Gaia2023}. Instrumental magnitudes were reported to the VEGAMAG photometric system either by using stars in common with the catalogues by \cite{Dalessandro2013}, \cite{Piotto2015} and \cite{Nardiello2018} for bands in common, or by applying appropriate equations and zeropoints for the remaining filters (which are the F225W and F435W), as reported on the \textit{HST} website \footnote{\url{https://www.stsci.edu/hst/instrumentation/wfc3/data-analysis/photometric-calibration/uvis-photometric-calibration}} \citep[see also][]{Sirianni2005}.

Finally, the catalogues derived from the photometric analysis of the two WFC3 datasets were cross-matched and combined, producing the final WFC3 catalogue. This comprehensive catalogue includes all stars detected in at least one band across both WFC3 observation datasets and contains over 130000 sources.

\subsection{ACS/HRC data}\label{hrcdata}
The WFC3 catalogue has been supplemented with additional images obtained with the ACS HRC.
The HRC images used in this work were taken on 2004 December 06 as part of the proposal 10401 (PI: Chandar). This dataset comprises 16 exposures, each with $t_{\mathrm{exp}}=85$ s, in the filter F435W. These high-resolution images, despite their limited FoV, are a valuable resource for our work as they encompass a significant portion (22 out of 33) of the X-ray sources identified by \cite{Kumawat2024}, as shown in Fig.~\ref{map_zoom}. In fact, their inclusion is primarily motivated by the need for enhanced resolution in the cluster's most densely populated region. Additionally, this dataset allowed us to study the variability of our counterparts over a longer temporal baseline (see Sect.~\ref{variability}).

In this analysis we employed the basic fully pipeline-calibrated individual exposures (\texttt{$\_$flt} images). We processed each image using the most recent PAM images from the \textit{HST} website. Following the same procedure as for the WFC3 images, we used \texttt{DAOPHOT} for the photometric analysis and selected approximately 100 bright isolated stars to construct the PSF model. We employed a Penny model for all frames. The PSF model was applied using \texttt{ALLSTAR} and \texttt{ALLFRAME}, and the star positions and magnitudes were then combined using \texttt{DAOMATCH} and \texttt{DAOMASTER}, requiring sources to be measured in a minimum of nine out of 16 images. We corrected for the geometrical distortions across the FoV using the equations by \citet{meurer2004}. The instrumental positions have been transformed to the absolute coordinate system using the same Gaia DR3 reference catalogue mentioned above and the instrumental magnitudes have been converted to the VEGAMAG photometric system following the guidelines and zero points provided in the "ACS Calibration and Zeropoints" Web site \footnote{\url{https://acszeropoints.stsci.edu}}. The final HRC catalogue contains $\sim 11000$ sources.

\subsection{ACS/WFC data and final catalogue construction}\label{acsdata}
Finally, we include ACS/WFC photometric information, mostly from the public ACS/WFC catalogue from the `Hubble Space Telescope Proper Motion (HSTPROMO) Catalogs of Galactic Globular Cluster' \citep{Libralato2018,Libralato2022}. This includes measurements in the F606W and F814W bands, derived from images taken under proposal ID 10775 (PI: Sarajedini). Additionally, we included the $m_{\rm F435W}$, $m_{\rm F625W}$ and $m_{\rm F658N}$ magnitudes (from proposal ID 10005) provided by \cite{Libralato2018,Libralato2022} via private communication. As outlined in Sect.~\ref{halpha}, this allowed us to analyse the H$\alpha$ emission of our candidate counterparts.

\begin{figure*}
    \centering
    \includegraphics[width=\textwidth]{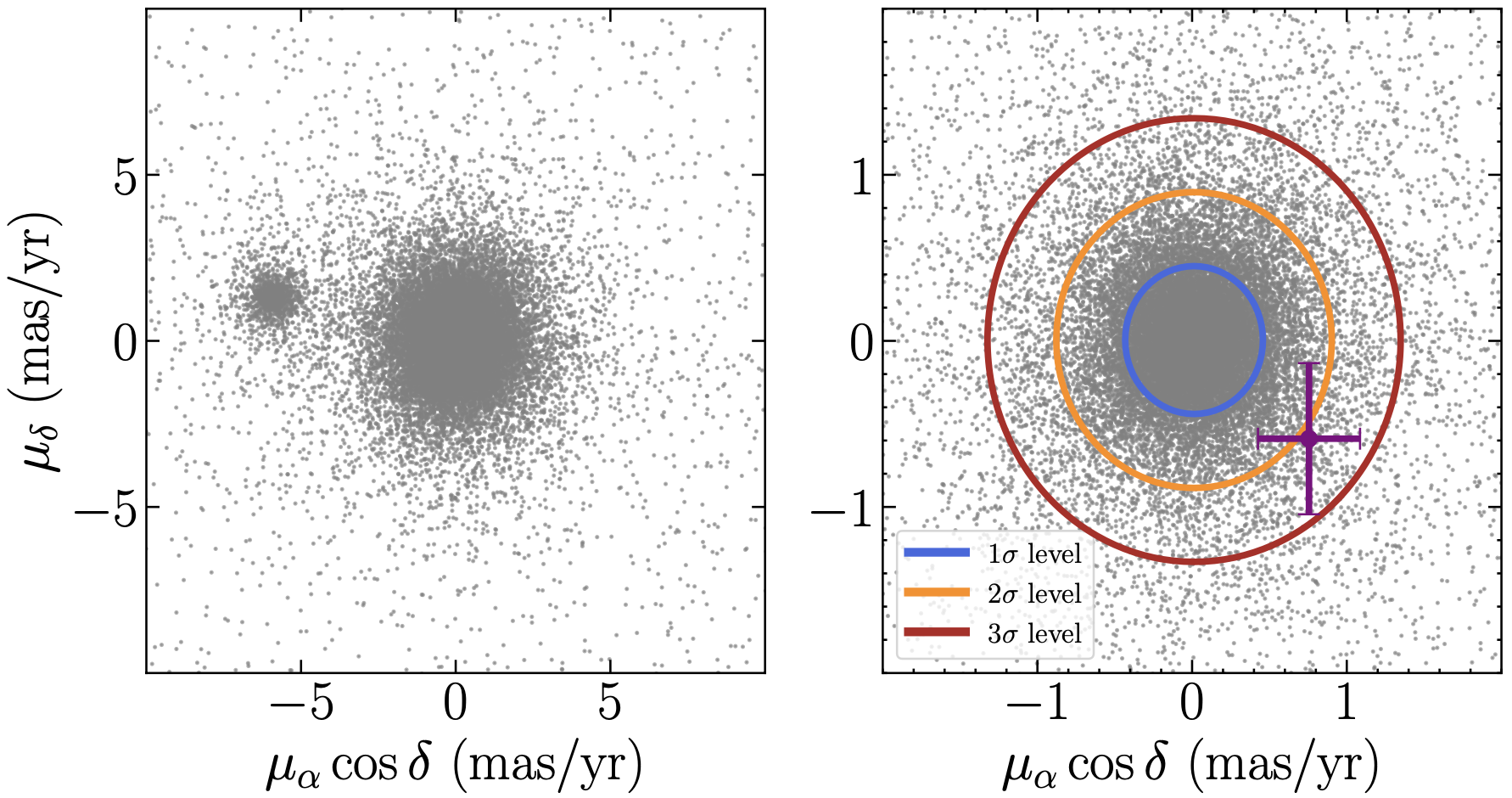}
    \caption{Left-hand panel: full proper motion VPD from the HSTPROMO catalogue. Right-hand panel: Proper motion VPD for stars within the F606W magnitude bin that includes one of the candidate counterparts of X5 (the object shown in blue in Fig.~\ref{finding}). The purple point indicates the object's position in the VPD along with its associated errors. The blue, orange and red circles represent the $1\sigma$, $2\sigma$ and $3\sigma$ contours, respectively.}
    \label{propermotion}
\end{figure*}

By cross-matching and combining our WFC3 and HRC catalogues with the public HSTPROMO catalogue, we constructed a comprehensive dataset containing more than 154000 sources. This dataset includes relative proper motion measurements for stars in common between our catalogues and the HSTPROMO catalogue, which will be used for investigating the cluster membership probabilities.
To assess membership, we excluded stars associated with the Small Magellanic Cloud (SMC) by selecting only stars with proper motions within a 3 mas/yr radius circle centered at ($\mu_{\alpha}\cos\delta=0$  mas/yr, $\mu_{\delta}=0$ mas/yr) in the proper motion vector-point diagram (VPD). In fact, in addition to the cluster population centered in the VPD, the diagram reveals another sub-population, at ($\mu_{\alpha}\cos\delta = -6$  mas/yr, $\mu_{\delta} = 1.5$ mas/yr) which corresponds to SMC stars (see left-hand panel of Fig.~\ref{propermotion}). 
Subsequently, we divided the catalogue into seven bins based on F606W magnitude.  We performed a 2D Gaussian fit on the VPD for each bin to determine the $1\sigma$, $2\sigma$, and $3\sigma$ contours. Stars falling outside the $3\sigma$ contour, even after accounting for proper motion uncertainties, were classified as non-members. 
An example of this analysis is shown in the right-hand panel of Fig.~\ref{propermotion}, where we highlight the position of one of the candidate counterpart of X5 (the object marked in blue in Fig.~\ref{finding}).

\begin{figure}
    \centering
    \includegraphics[width=\hsize]{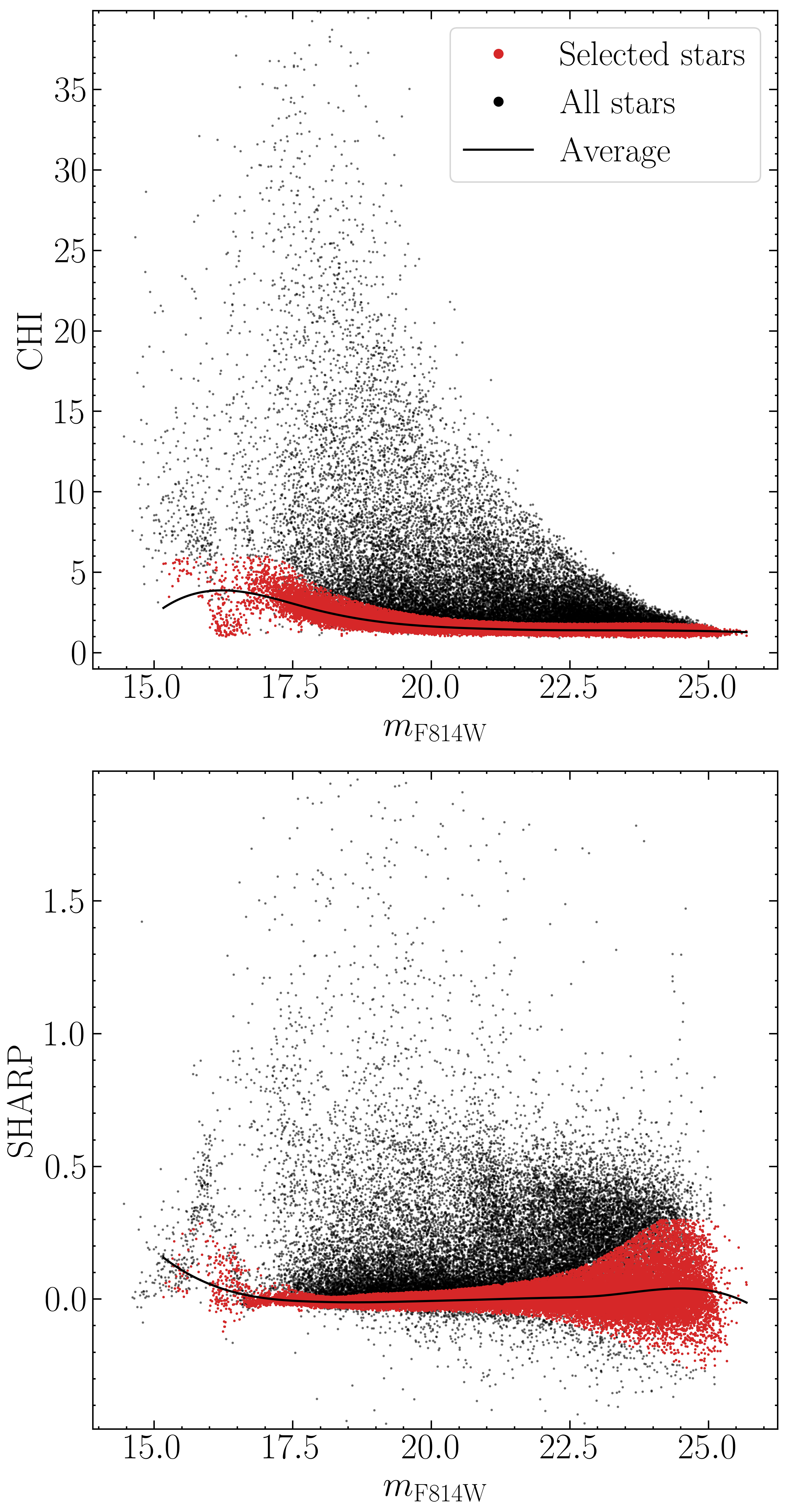}
    \caption{Plot of CHI parameter (top panel) and SHARP parameter (bottom panel) as function of $m_{\rm F814W}$ magnitude. Objects in red represent selected stars following the $3\sigma$-rejection method, while black points are rejected stars. The solid black line represents the average CHI (top panel) and average SHARP (lower panel). It can be observed that both CHI and SHARP display an irregular pattern at brighter magnitudes, particularly for $m_{\rm F814W} < 17$ . This irregularity occurs due to the saturation affecting the magnitudes of these stars. Consequently, the average and standard deviation within these bins are affected, leading to the observed behaviour.}
    \label{chisharp}
\end{figure}

\begin{figure*}
    \centering
    \includegraphics[width=\textwidth]{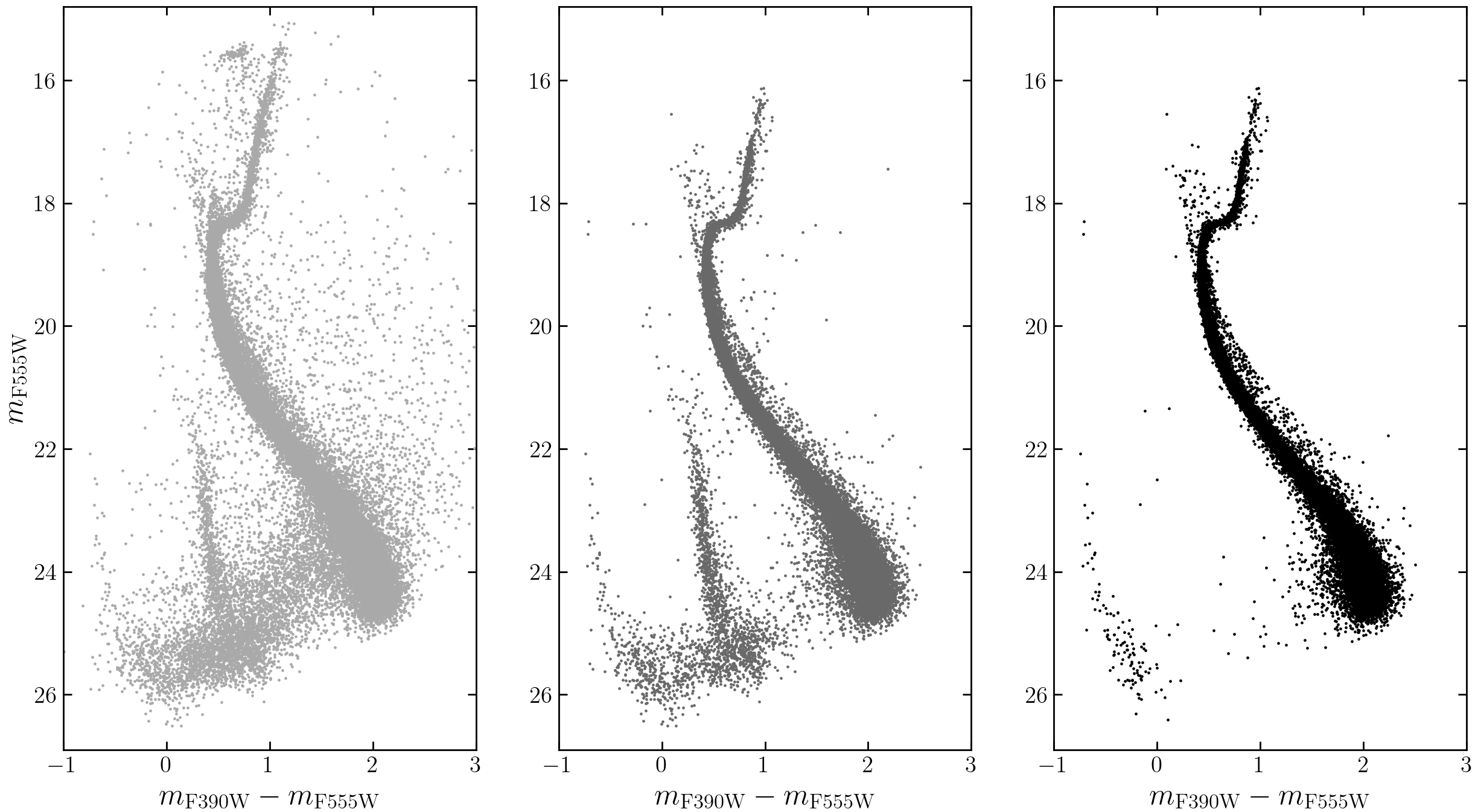}
    \caption{($m_{\rm F555W}$, $m_{\rm F390W}$ - $m_{\rm F555W}$) CMDs schematically illustrating the data selection process. The left panel shows the observed CMD, including all detected stars. The middle panel shows the cleaned CMD after photometric quality cuts. The fraction of stars removed according to photometric quality cuts is $0.53$. It can be observed that the stars in the upper part of the RGB and those on the Horizontal Branch (HB) are excluded after the photometric quality cuts, which is a result of saturation. The right panel shows the final CMD after membership selection. The fraction of stars removed according to membership evaluation with respect to the complete catalogue shown in the first panel is $0.05$.
    }
    \label{cmd}
\end{figure*}

Additionally, we assigned a photometric quality flag to each star in the catalogue. For the magnitudes of the HSTPROMO catalogue, we used the diagnostic parameters described in \cite{Libralato2018}, applying the same quality cuts as outlined in their appendix B. 
For the WFC3/UVIS and ACS/HRC magnitudes, we identified well-measured stars performing a selection based on the \texttt{CHI} and \texttt{SHARP} photometric parameters derived for each filter by \texttt{DAOPHOT}. To this aim, we divided our sample into ten magnitude bins, we computed the mean and performed the selection using an iterative $3\sigma$-rejection method. An example is shown in Fig.~\ref{chisharp} for the F814W filter. 
In Fig.~\ref{cmd} we present a sequence of ($m_{\rm F555W}$, $m_{\rm F390W}$ - $m_{\rm F555W}$) CMDs illustrating the data refinement process.
The first panel displays the observed CMD, including all stars of the catalogue. In the second panel, we display the CMD after removing the stars with bad photometric quality. Finally, the last, rightmost panel shows the final CMD, refined by applying photometric quality cuts and membership selection, as outlined above.
 The vertical sequence in the first two CMDs at ($m_{\rm F390W}$ - $m_{\rm F555W}$) $\sim 0.7$ is the MS of the SMC.  From the cleaned diagram, it is clear that all evolutionary sequences are narrow and well-defined. A key feature of our catalogue is its depth, extending up to 8 magnitudes below the turn-off.

\section{Identification of optical counterparts}\label{methods}
The primary goal of this work is to identify and characterize the optical counterparts of the 33 X-ray sources identified in NGC 362 by \cite{Kumawat2024}. To this aim we will follow a 4-step approach which accounts for i) positional coincidence between X-ray observations and \textit{HST}, ii) stellar position in the CMDs, iii) photometric variability, iv) H$\alpha$  excess.

In the following, we will refer to the X-ray sources using the same names adopted by \cite{Kumawat2024}, and we will mark the candidate optical counterparts with a "c" as a superscript. For X-ray sources with multiple candidate counterparts, we distinguish among different candidates by appending a capital letter to their names, starting from "A".

\subsection{Positional coincidence and CMD location}\label{selection}
As a first step, we searched for optical counterparts for every X-ray source within the X-ray positional $95\%$ confidence uncertainty radius ($\mathrm{unc_X}$) using the WFC3 catalogue. An example of a finding chart is shown in Fig.~\ref{finding} for the X-ray source X5. All the other finding charts are shown in the Appendix~\ref{Appendix}. Then, all candidates were placed in the optical CMDs, using different filter combinations, to investigate their position. Particular attention was given to objects that displayed unusual positions, deviating from the expected evolutionary paths. 
Indeed, an anomalous position in the CMD could be indicative of multiple sources of light in the system or a perturbed state of the companion star. For instance, an accretion disk can shift a CV to the blue of the main sequence, or a second star can shift an AB to the red of the main sequence, while heating, mass loss, or tidal deformation can also alter the CMD position \citep[see for example][]{Cocozza2008,Pallanca2010}.

\begin{figure}
    \centering
    \includegraphics[width=0.9\hsize]{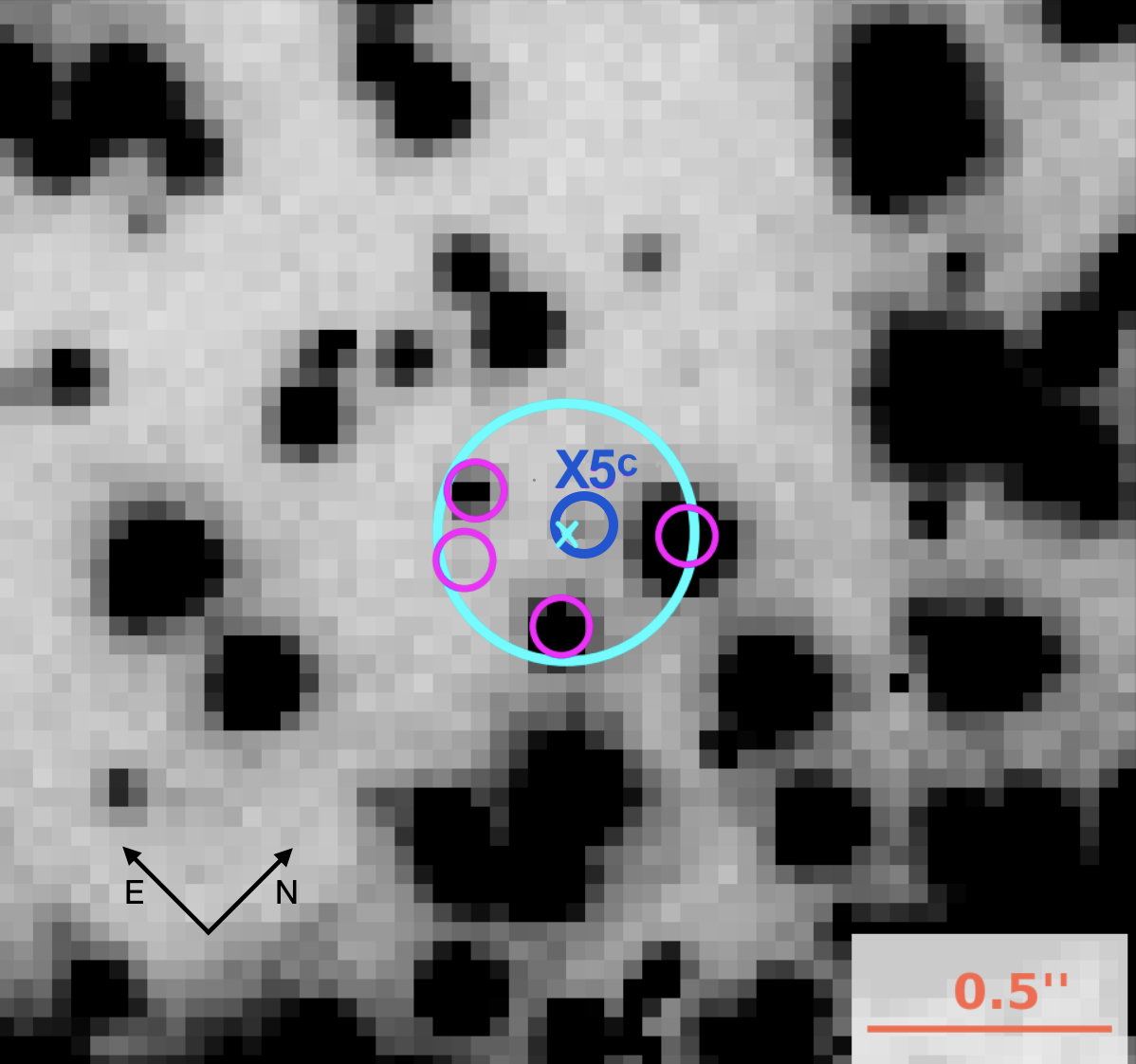}
    \caption{Finding chart for source X5 in the F390W WFC3 band, covering an area of $2.4\arcsec$ on each side. The cyan circle is centered on the X-ray position, indicated by a cyan cross, and has a radius equal to the X-ray position uncertainty $\mathrm{unc_X}$. The magenta circles denote the positions of the candidate counterparts. We mark in blue the candidate counterpart $\mathrm{X5^{c}}$, which is the same as the one presented in Fig.~\ref{propermotion} and Fig.~\ref{variaX5}.}
    \label{finding}
\end{figure}

\subsection{Counterparts variability}\label{variability}
One of the most important additions of our work is the possibility to study the variability of the optical counterparts. This is a crucial point when searching for companions of exotic objects. For example, in interacting binaries we expect photometric variability due to different processes as irradiation or tidal distortions. Moreover, in the case of MSPs, there is strong evidence confirming the optical counterpart as the companion to the MSP if the pulsar's orbital period and ascending node time (e.g. phase), derived from radio observations, match those of the optical counterpart \citep[e.g.][]{Edmonds2002,Pallanca2010}.

To identify variable stars, we used the Stetson variability index \citep[$J$ as defined by][]{Stetson1996}.  This index is computed in the photometric reduction performed with \texttt{DAOPHOT} and it examines the reliability of successive changes in the star's brightness, taking into account the uncertainties associated with the measurements. Specifically, we plotted the $J$ index as a function of magnitude for all the stars in the catalogue, for each filter. We divided the sample into ten magnitude bins, calculated the median for each bin, and obtained a list of variable stars, considering as variable those sources that were more than 2 sigma from the median in at least two filters. 
We then checked if any of our candidate counterparts were present in the list of all variable stars and visualized their light curves. In addition to this, we also visually inspected the light curves of all those candidate counterparts displaying unusual positions in the CMDs.

To study the variability of our sources, we used both WFC3 and HRC data. The temporal baseline spans approximately 7 hours for the F390W, F555W, and F814W images, with overlapping observation times. The F225W images cover a time range of nearly 6 hours, while the F275W frames span more than 3 hours. Also in this case, the images overlap in time. On the other side, the temporal baseline of HRC images spans 2 hours and 27 minutes. To get an estimate of the period of the flux modulation in each filter we applied the Lomb-Scargle Periodogram method \citep{Lomb1976,Scargle1982} implemented in \texttt{astropy} \citep{astropy2013,Price-Whelan2018A}. 
Afterwards, we opted to merge all the available measurements, possibly broadening our temporal baseline and allowing us to capture objects with longer periods. 
To do so, we chose one reference magnitude and, for each available filter, we computed the shift between its average magnitude and the average reference magnitude. We thus obtained the combined global light curve and performed the Lomb-Scargle analysis on it. Moreover, we computed the False Alarm Probability (FAP) using the method described by \cite{Baluev2008} and implemented in \texttt{astropy}. 
\begin{figure}
    \centering
    \includegraphics[width=\hsize]{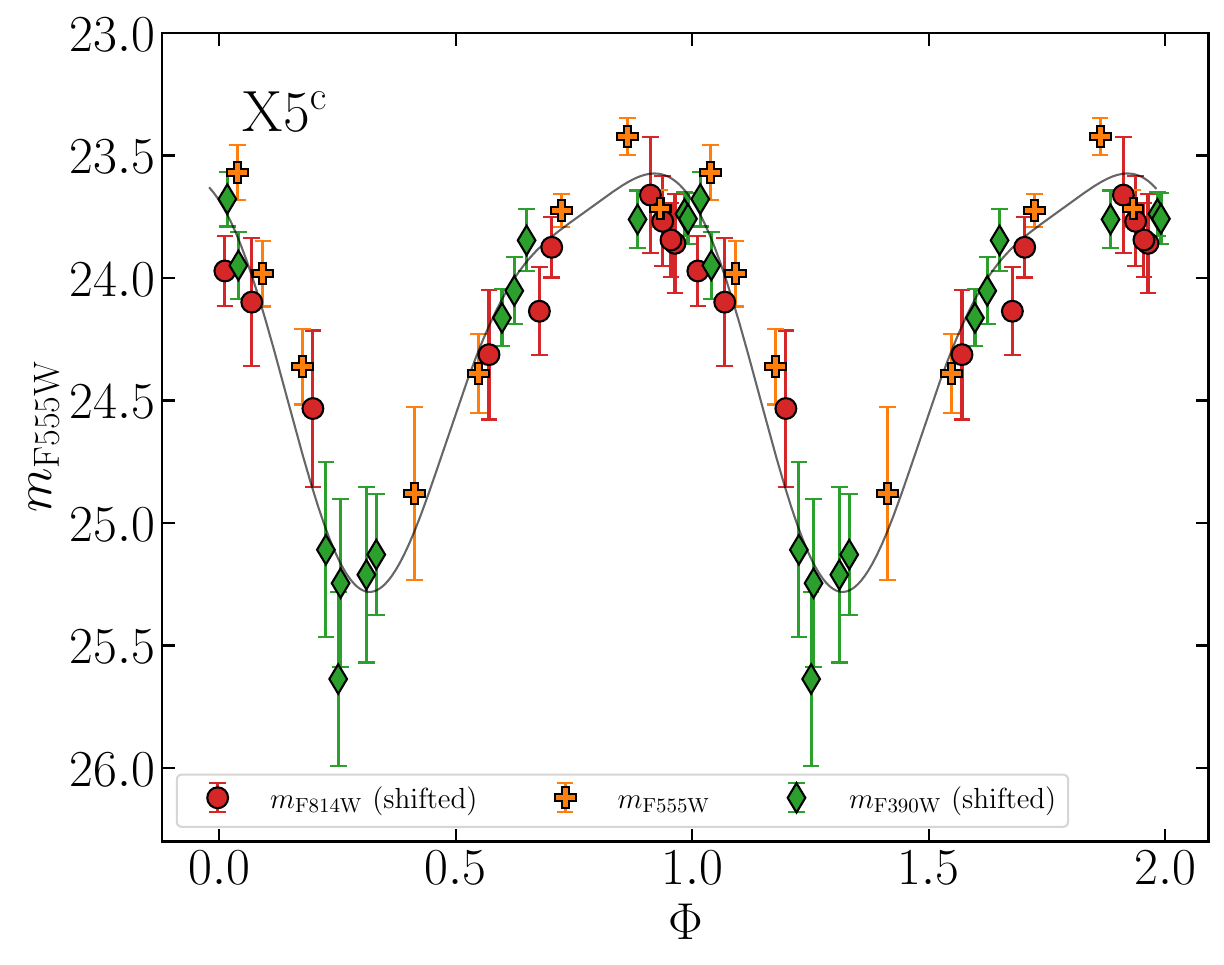}
    \caption{Global light curve of one candidate counterpart to the X5 source (object marked in blue in Fig.~\ref{finding}), obtained combining the data points from F390W, F555W and F814W WFC3 images (in green, orange and red respectively). The light curve is folded with a period of 5.08 hours, obtained from the Lomb-Scargle analysis. The grey solid line, a combination of sines and cosines fitted to the data points. It has no physical meaning, it is plotted only to facilitate the visualization of the curve.}
    \label{variaX5}
\end{figure}
As an example, we plot in Fig.~\ref{variaX5} the global light curve of one of the candidate counterparts for the X-ray source X5. This object is the one marked with a blue circle in Fig.~\ref{finding}, closest to the X-ray position, and is also shown in the VPD in Fig.~\ref{propermotion}. This is the light curve obtained combining all the filters of the first WFC3 dataset, taking $m_{\rm F555W}$ as reference magnitude, and folding the measured data points with the period obtained from the Lomb-Scargle method. The light curve shows a very broad peak and varies within a range of over 2 magnitudes (for more details see Sect.~\ref{result2}).

\subsection{Selection of H$\alpha$ emitter candidates}\label{halpha}
As mentioned previously, employing the $m_{\rm F658N}$ catalogue \citep{Libralato2018,Libralato2022}, we are able to identify the objects showing H$\alpha$ excess among the X-ray candidate counterparts. In fact, when a binary system harbors a compact object, mass transfer phenomena may occur. This can generate substantial X-ray and UV radiation in addition to emission lines like H$\alpha$. Especially when we have multiple potential counterparts for an X-ray source, which is common when using positional coincidence as a selection criterion in dense environments like the cores of GCs, the presence of H$\alpha$ excess in one of them would identify it as the most likely counterpart among the candidates. For example, \cite{Beccari2013} demonstrated the effectiveness of combining broadband \textit{V} and \textit{I} with narrowband $H\alpha$ images to detect CVs. Moreover, this method has been successfully employed in GCs, as demonstrated by \cite{Pallanca2017} in NGC 6397. For this reason, studying the H$\alpha$ emission provides valuable information that we can incorporate to identify these counterparts and obtain more insights about accretion processes or coronal activity occurring in these systems.

Here we used the ($m_{\rm F555W}$$-$$m_{\rm F658N}$) versus ($m_{\rm F555W}$$-$$m_{\rm F814W}$) colour-colour diagram as diagnostic to identify H$\alpha$ emitters. Before proceeding, we decided to clean the catalogue, both in terms of membership and photometric quality, following the procedure described at the end of Sect.~\ref{acsdata}. The main point of this analysis is that the majority of the stars in the cluster are expected to have a negligible H$\alpha$ emission and they define a relatively narrow sequence in the colour-colour diagram, while the objects showing H$\alpha$ emission are located above this sequence. As the influence of the H$\alpha$ line on the $m_{\rm F555W}$ magnitude is negligible, the H$\alpha$ excess can be derived computing the distance of the object from the main sequence of the colour-colour diagram. In this work we decided to follow the same kind of approach adopted by \cite{DeMarchi2010}, \cite{Beccari2013} and \cite{Pallanca2013,Pallanca2017}. We computed the median of the ($m_{\rm F555W}$$-$$m_{\rm F658N}$) colour using only stars with intrinsic photometric error (defined as $\sigma_{m_{\rm F555W}- m_{\rm F658N}}=\sqrt{\sigma_{m_{\rm F555W}}^2+\sigma_{m_{\rm F658N}}^2}$) smaller than 0.1 mag. The median defines the reference line for stars without H$\alpha$ excess. For each star of the catalogue we then computed the difference between the intrinsic ($m_{\rm F555W}$$-$$m_{\rm F658N}$) and the value of ($m_{\rm F555W}$$-$$m_{\rm F658N}$) obtained projecting on the median line at the source ($m_{\rm F555W}$$-$$m_{\rm F814W}$). We call this difference $\Delta\mathrm{H}\alpha$. If a star shows $\Delta\mathrm{H}\alpha > 5\sigma_{(m_{\rm F555W} - m_{\rm F658N})}$, then it is classified as an H$\alpha$ emitter. This approach allowed us to select the objects with a true colour excess and discard those with large intrinsic photometric error. 
Following this approach we identified 1366 candidate stars showing H$\alpha$ excess out of more than 45000 selected stars. While we acknowledge that our approach is more conservative in terms of the number of objects classified as candidate H$\alpha$ emitters, it was chosen to ensure the inclusion of weak emitters, such as ABs and MSPs. A more restrictive 3$\sigma$-based method was also considered, where $\sigma$ was defined as vertical width of the sequence, but this approach posed the risk of excluding such objects. The effectiveness of our method is supported by cases like the $\mathrm{X20^{c}}$ counterpart, which was spectroscopically confirmed as an H$\alpha$ emitter by \citet[][see Sect.~\ref{result3}]{Gottgens2019}.

\begin{figure}
    \centering
    \includegraphics[width=\hsize]{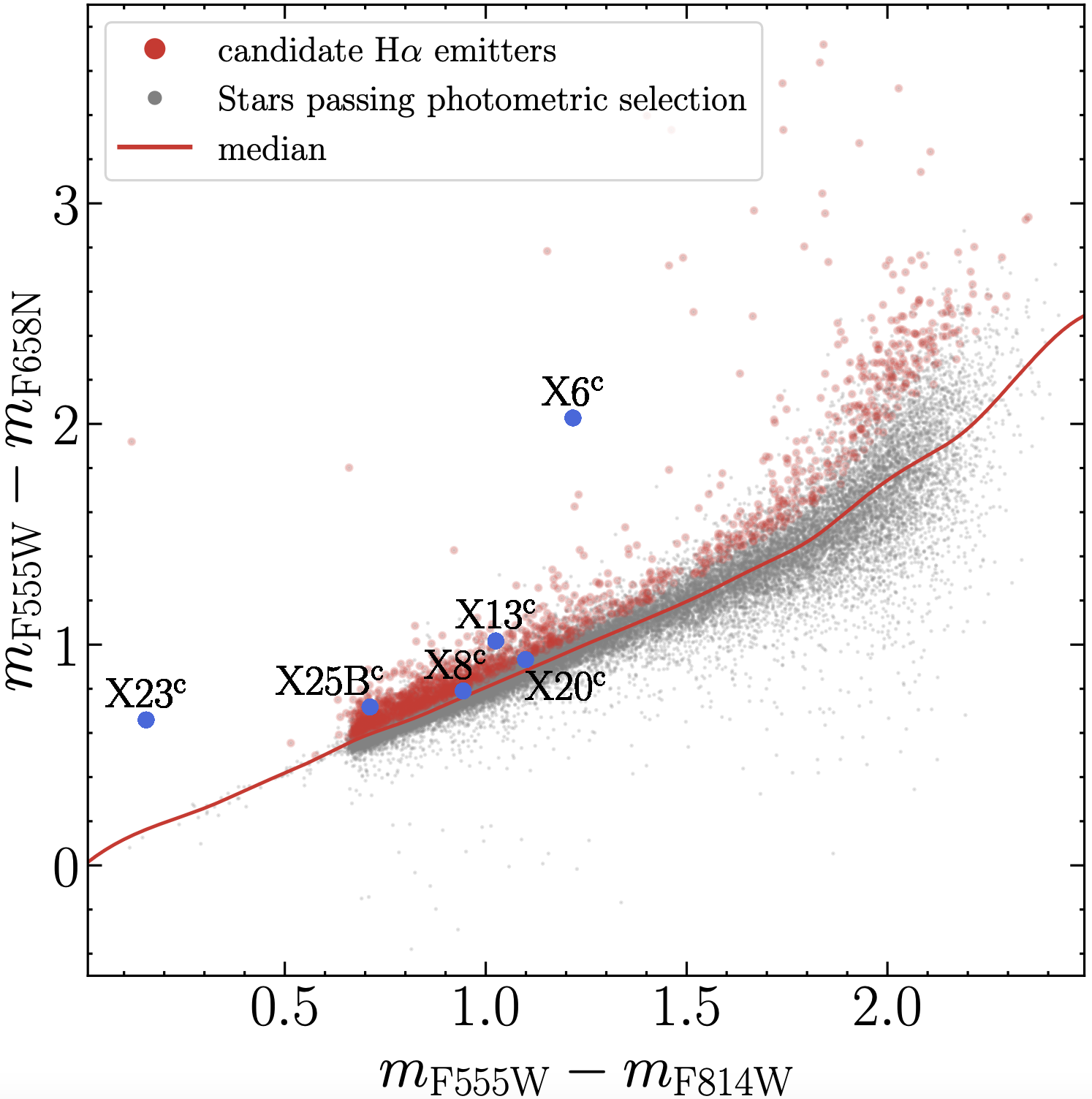}
    \caption{($m_{\rm F555W}$$-$$m_{\rm F658N}$) versus ($m_{\rm F555W}$$-$$m_{\rm F814W}$) colour-colour diagram. Gray points are stars passing the photometric and membership selection.
    Candidate H$\alpha$ emitters are represented in red. We plot in blue the candidate optical counterparts that we found to have an H$\alpha$ excess (see Sect.~\ref{result3}).}
    \label{halphafig}
\end{figure}

For the selected candidate emitters we also computed the photometric equivalent width (pEW) of the H$\alpha$ emission line adopting the equation 4 in \cite{DeMarchi2010}:
\begin{equation}
    \mathrm{pEW} = \mathrm{RW} \times [1 - 10^{(-0.4\,\times\,\Delta\mathrm{H}\alpha)
}]
\end{equation}
where RW is the rectangular width of the filter in \r{A} units. In this context, the pEW is a very useful parameter for distinguishing between different types of objects. 
In Fig.~\ref{halphafig} we plot the colour-colour diagram and we highlight in red the median line and the candidate H$\alpha$ emitters. We also plot in blue the candidate counterparts showing an H$\alpha$ excess (see Sect.~\ref{result3}).

\section{Results}\label{results}
In this section we summarise the main results obtained in our analysis. In Sect.~\ref{result1} we present the most interesting counterparts in terms of CMD position. Section~\ref{result2} lists the counterparts showing coherent and variable light curves, while the candidate counterparts for which we found H$\alpha$ excess are presented in Sect.~\ref{result3}. As a final outcome of this analysis we build a list of `high-confidence' optical counterparts to the X-ray sources in NGC 362. This sample includes stars that meet at least one of the following criteria: interesting CMD position, magnitude variability and presence of H$\alpha$ excess. Additionally, in cases where only one counterpart is found within the X-ray uncertainty radius, we included it among the high-confidence counterparts.

\subsection{Counterparts with unusual CMD position}\label{result1}
\begin{figure*}
    \centering
    \includegraphics[width=\textwidth]{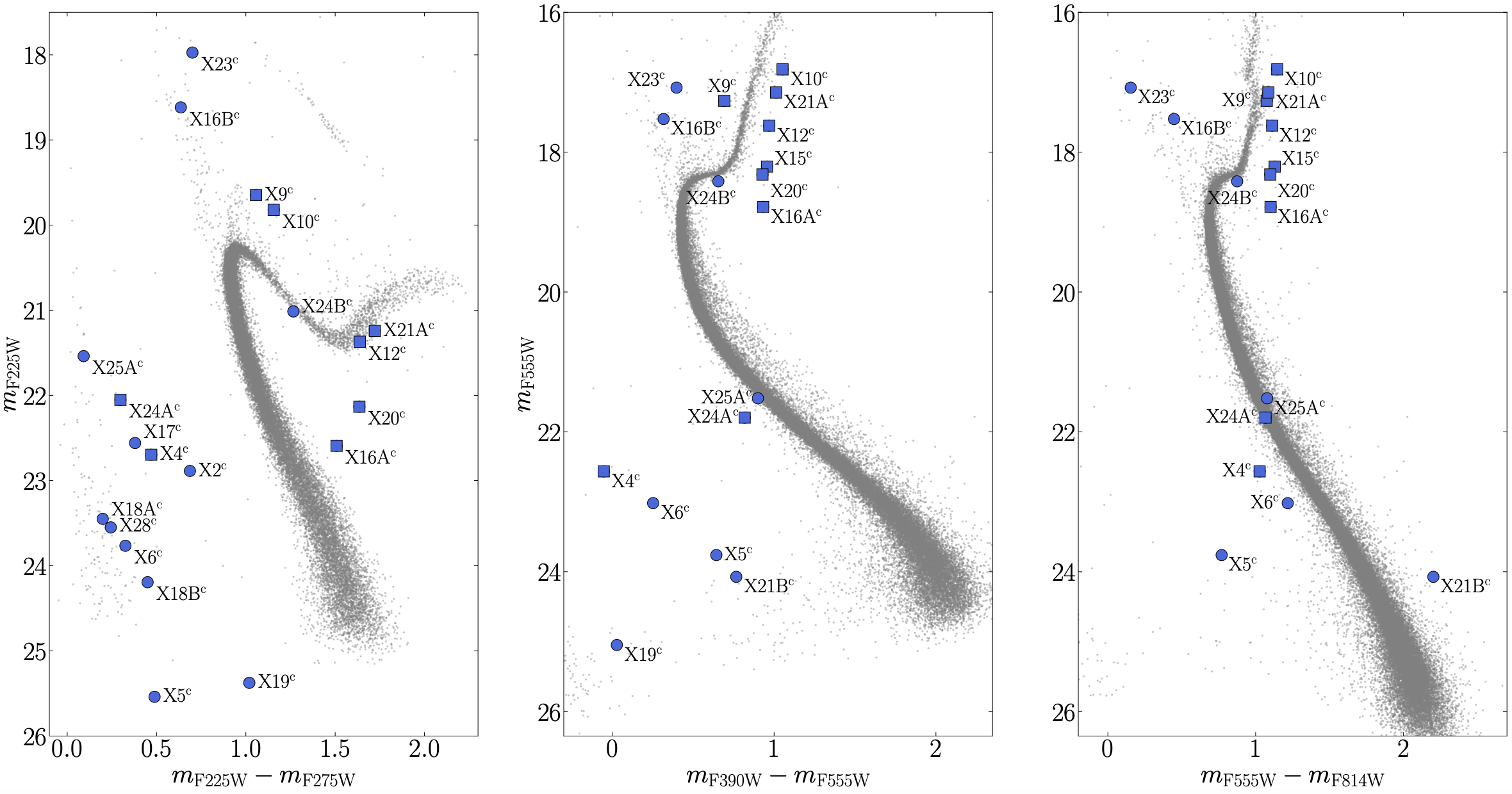}
    \caption{From left to right: ($m_{\rm F225W}$, $m_{\rm F225W}$$-$$m_{\rm F275W}$), ($m_{\rm F555W}$, $m_{\rm F390W}$$-$$m_{\rm F555W}$) and ($m_{\rm F555W}$, $m_{\rm F555W}$$-$$m_{\rm F814W}$) CMDs. Counterparts showing peculiar CMD position are plotted in blue. Objects that become bluer in the $m_{\rm F390W}$$-$$m_{\rm F555W}$ CMD might have an accretion disk and/or a WD. Squares indicate those counterparts that were identified also by \cite{Kumawat2024}. The step observed along the RGB at $m_{\mathrm {F555W}}$ $\sim 17$ in the ($m_{\rm F555W}$, $m_{\rm F555W}$$-$$m_{\rm F814W}$) CMD is due to issues related to saturation in the F814W filter.}
    \label{cmd_int}
\end{figure*}

As anticipated in Sect.~\ref{selection}, the necessary condition for an object to be considered as a candidate counterpart is that its location falls within the uncertainty radius of the X-ray position. Using our catalogue, we found at least one candidate optical counterpart for each X-ray source. When the uncertainty radius is small, we found a limited number of counterparts for a single X-ray source, while for larger uncertainty radii we found up to a maximum of 45 candidate counterparts. The final result is a list of 530 candidate counterparts for all the 33 X-ray sources, including all the 15 candidate counterparts identified by \cite{Kumawat2024}. Among these 530 candidates, ten have only measurements in the F435W ACS/HRC data. Unfortunately, the lack of measurements in other filters for these ten candidates prevented their visualization within the CMDs. Additionally, variability analysis of these candidates did not reveal any evidence of variable light curves, likely due to the limited temporal coverage of the HRC images (see Sect.~\ref{variability}). All the other candidates were placed in different CMDs in order to investigate their position within the evolutionary sequences, using different combinations of filters. In Fig.~\ref{cmd_int} we show the location of candidate counterparts falling out of the main evolutionary sequences.

Four objects ($\mathrm{X9^{c}}$, $\mathrm{X10^{c}}$, $\mathrm{X12^{c}}$, $\mathrm{X21A^{c}}$) are located on the red side of the RGB in the ($m_{\rm F555W}$, $m_{\rm F555W}$$-$$m_{\rm F814W}$) CMD. However, in the ($m_{\rm F225W}$, $m_{\rm F225W}$$-$$m_{\rm F275W}$) and ($m_{\rm F555W}$, $m_{\rm F390W}$$-$$m_{\rm F555W}$) CMDs, $\mathrm{X9^{c}}$ shifts towars bluer colours, indicating the possible presence of an accretion disk and/or a blue component, such as a WD. A similar behaviour is observed for $\mathrm{X10^{c}}$ in the ($m_{\rm F225W}$, $m_{\rm F225W}$$-$$m_{\rm F275W}$) CMD. Two counterparts ($\mathrm{X16B^{c}}$ and $\mathrm{X23^{c}}$) are found in the blue straggler stars (BSS) sequence. In the ($m_{\rm F555W}$, $m_{\rm F390W}$$-$$m_{\rm F555W}$) CMD, $\mathrm{X23^{c}}$ is slightly redder than the BSS sequence, but interestingly, it becomes bluer than the bulk of BSS stars in the ($m_{\rm F555W}$,$m_{\rm F555W}$$-$$m_{\rm F814W}$) CMD. Following our membership evaluation, all these objects appear to be members of the cluster. In this context, the study conducted by \cite{Rozyczka2016} gives us insight into the nature of these systems. Specifically, they monitored the NGC 362 field to identify variable stars and obtained light curves for 151 periodic variable stars. Two of the most interesting objects they identified are the so-called V20 and V24. The latter coincides with our $\mathrm{X10^{c}}$, as also pointed out by \cite{Kumawat2024}. The period measured for this object is 8.1 days, and the light curve suggests that this could be a semi-detached binary system, formed by two $\sim 0.8 \, \mathrm{M_\odot}$ stars, where the primary giant is filling its Roche Lobe and the blue companion could be either a BSS or an accreting WD. It is worth mentioning that the variability of $\mathrm{X10^{c}}$ (see also the case of $\mathrm{X23^{c}}$  as discussed in Sect.~\ref{result2}) could account for their rather unusual colours.

$\mathrm{X15^{c}}$, $\mathrm{X16A^{c}}$, $\mathrm{X20^{c}}$ and $\mathrm{X24B^{c}}$ are located in the sub-subgiant region. All of them are classified as cluster members according to our analysis of the proper motions. Within the scope of this work, the sub-subgiant region is of particular interest, since it is a typical CMD position where exotic binaries as SSGs are often found. In general, objects located in the SSG region have experienced unusual binary evolution, such as mass transfer from a star  evolving up the subgiant branch \citep[SGB;][]{Leiner2017}. Moreover, RB MSPs can also populate this region, as found, for example, by \cite{Ferraro2001}, \cite{Cocozza2008} and \cite{Bogdanov2010,Zhang2022} for COM-6397A, COM-6266B and COM-6397B, respectively. Another example of exotic object found in this CMD region is the CV AKO 9 found in 47 Tucanae by \cite{Knigge2003}.

Another interesting CMD region is the area, at faint magnitudes, between the cluster's MS and the WD sequence. Typically, this region hosts BW systems, as in the case of COM-M5C in \cite{Pallanca2014} and COM-M71A in \cite{Cadelano2015}, CVs, as CV4 and CV7 in \cite{Pallanca2017} and RBs, as 47 Tuc W found by \cite{Edmonds2002}. This region also includes qLMXBs \citep[e.g. the X5 source found in 47 Tucanae by][]{Edmonds2002} and background galaxies with AGN \citep[e.g. CX2 in M4;][]{Bassa2005}.
In the ($m_{\rm F555W}$, $m_{\rm F390W}$$-$$m_{\rm F555W}$) CMD, this area contains six counterparts ($\mathrm{X4^{c}}$, $\mathrm{X5^{c}}$, $\mathrm{X6^{c}}$, $\mathrm{X19^{c}}$, $\mathrm{X21B^{c}}$, and $\mathrm{X24A^{c}}$). However, in the ($m_{\rm F555W}$, $m_{\rm F555W}$$-$$m_{\rm F814W}$) CMD, 
$\mathrm{X24A^{c}}$ moves towards the MS and $\mathrm{X21B^{c}}$ shows a shift towards redder colours relative to the MS.  
In the ($m_{\rm F225W}$, $m_{\rm F225W}$$-$$m_{\rm F275W}$) CMD $\mathrm{X6^{c}}$ lies on the WD cooling sequence, possibly suggesting the presence of a WD in this system. On the other side, $\mathrm{X5^{c}}$ is missing from the (F225W, F275W) catalogue, as it is fainter than the detection threshold reached in our photometric reduction of the second WFC3 dataset. 
We decided to recover its $m_{\rm F225W}$ and $m_{\rm F275W}$ magnitudes by combining all the images using the \texttt{MONTAGE} command in \texttt{DAOPHOT} \citep{Stetson1987},  and performing the same photometric reduction described in Sect.~\ref{wfc3data} up to the \texttt{ALLSTAR} step on the combined images. This process allowed us to retrieve the $m_{\rm F225W}$ and $m_{\rm F275W}$ magnitudes for $\mathrm{X5^{c}}$ and to display it in the ($m_{\rm F225W}$, $m_{\rm F225W}$$-$$m_{\rm F275W}$) CMD as well.
Finally, there are two other stars that populate this area in the ($m_{\rm F225W}$, $m_{\rm F225W}$$-$$m_{\rm F275W}$) CMD. These objects are potential counterparts for the X17 and X2 X-ray sources. The $\mathrm{X2^{c}}$ candidate has no measurements in the other WFC3 filters due to saturation from a nearby star, whereas $\mathrm{X17^{c}}$ is more unusual, as it is missing from the other WFC3 images, as detailed further in Sect.~\ref{result2}.
As for the membership, only $\mathrm{X5^{c}}$ and $\mathrm{X6^{c}}$ fall within the $3\sigma$ contours on their proper motion diagram when errors are considered. On the other side, $\mathrm{X4^{c}}$, $\mathrm{X24A^{c}}$ and $\mathrm{X21B^{c}}$ do not appear to be cluster members, while no proper motion information is available for $\mathrm{X17^{c}}$ and $\mathrm{X2^{c}}$. 
The counterpart $\mathrm{X19^{c}}$, is located near the WD sequence of the ($m_{\rm F555W}$, $m_{\rm F390W}$$-$$m_{\rm F555W}$) CMD, while it is absent in the ($m_{\rm F555W}$, $m_{\rm F555W}$$-$$m_{\rm F814W}$) CMD because it lacks the $m_{\rm F814W}$ magnitude. 
In the ($m_{\rm F225W}$, $m_{\rm F225W}$$-$$m_{\rm F275W}$) CMD this source shifts towards redder colours. However, we should note here that to recover the $m_{\rm F225W}$ and $m_{\rm F275W}$ magnitudes of $\mathrm{X19^{c}}$, we applied the same procedure used for $\mathrm{X5^{c}}$. As a consequence, the derived magnitudes may be affected by relatively large uncertainties. Also in this case no proper motion information is available.

Finally, in the ($m_{\rm F225W}$, $m_{\rm F225W}$$-$$m_{\rm F275W}$) CMD, the WD cooling sequence includes four objects ($\mathrm{X18A^{c}}$, $\mathrm{X18B^{c}}$, $\mathrm{X25A^{c}}$ and $\mathrm{X28^{c}}$). The counterpart $\mathrm{X25A^{c}}$ appears redder in the other two CMDs shown in Fig.~\ref{cmd_int}. Specifically, it aligns with the MS in both the ($m_{\rm F555W}$, $m_{\rm F390W}$$-$$m_{\rm F555W}$) and ($m_{\rm F555W}$, $m_{\rm F555W}$$-$$m_{\rm F814W}$) CMDs. This behaviour could be due to the presence of an accretion disk or a blue companion, such as a WD, which becomes prominent in the UV filters.
Lastly, the candidate counterpart for X28, which lies along the WD cooling sequence in the ($m_{\rm F225W}$, $m_{\rm F225W}$$-$$m_{\rm F275W}$) CMD, lacks $m_{\rm F390W}$ and $m_{\rm F555W}$ measurements due to being below the detection threshold. For the same reason, both $\mathrm{X18A^{c}}$ and $\mathrm{X18B^{c}}$ are detected only in the F225W and F275W filters. As a result, the position of these three counterparts in the other two CMDs in Fig.~\ref{cmd_int} could not be investigated. According to the VPD analysis, $\mathrm{X25A^{c}}$ and $\mathrm{X28^{c}}$ are not cluster members, while no information is available for $\mathrm{X18A^{c}}$ and $\mathrm{X18B^{c}}$. Nevertheless, the fact that these four objects align with the WD cooling sequence in the UV CMD makes them interesting candidate counterparts. For example, they could be CVs, where the UV emission is dominated by the WD.

\subsection{Counterparts showing variability}\label{result2}
As previously mentioned in Sect.~\ref{variability}, the second step of our analysis consists in studying the flux modulation of all candidate counterparts that displayed unusual positions in the CMD (see Sect.~\ref{result1}) and / or were classified as variables following our analysis of the Stetson variability index $J$ \citep{Stetson1996}, as detailed in Sect.~\ref{variability}.

\begin{figure*}
    \centering
    \includegraphics[width=\textwidth]{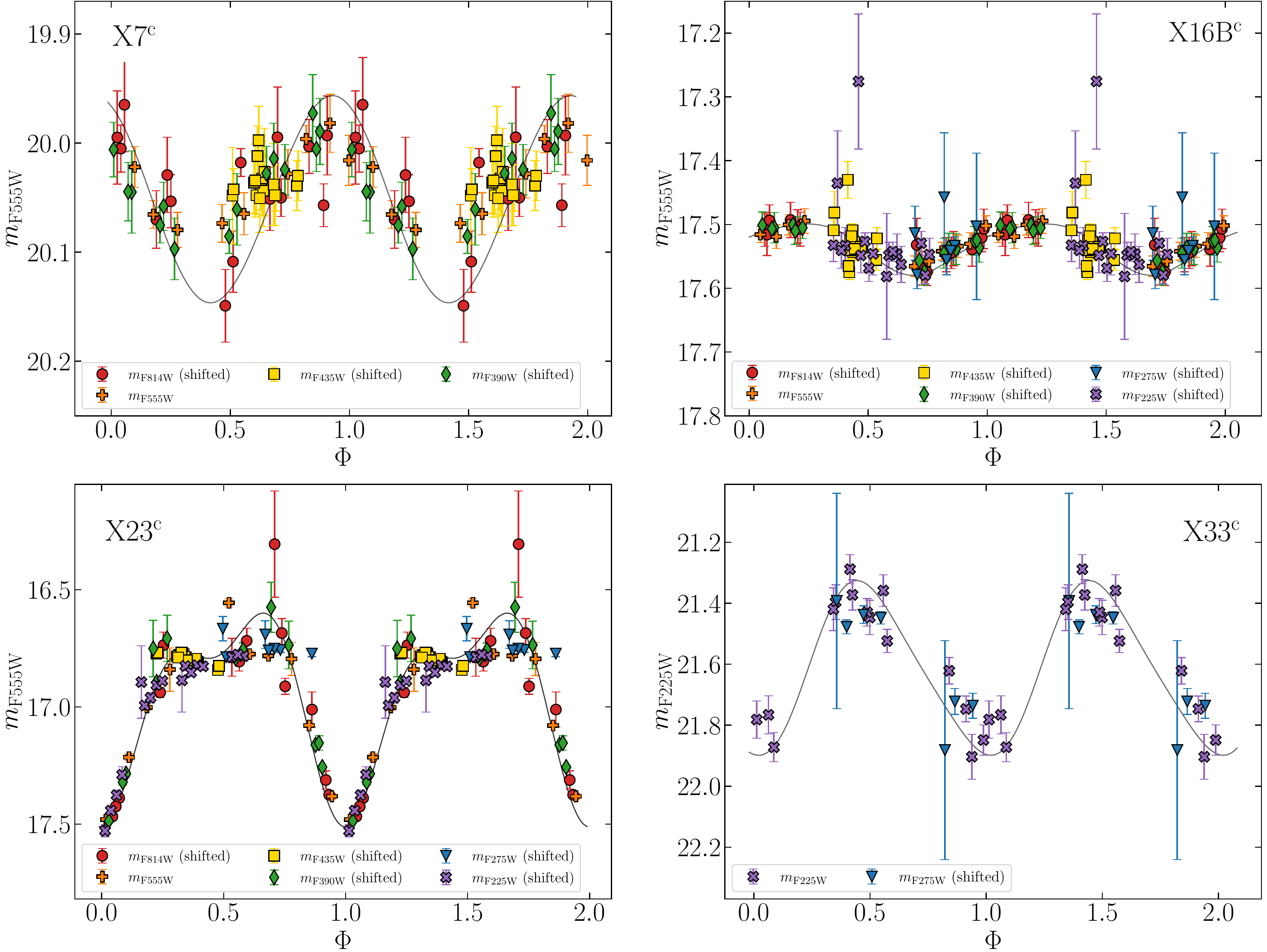}
    \caption{From left to right, top to bottom: Global light curves of $\mathrm{X7^{c}}$, $\mathrm{X16B^{c}}$, $\mathrm{X23^{c}}$, and $\mathrm{X33^{c}}$. The light curve of $\mathrm{X7^{c}}$ does not show $m_{\rm F225W}$ and $m_{\rm F275W}$ measurements for clarity. The curves are folded using periods of 8.89 hours, 13.44 hours, 9.53 hours and 3.10 hours respectively, determined from Lomb-Scargle analysis performed on the global curves across all available filters. In each panel, the grey solid line is a combination of sines and cosines fitted to the data points with the only purpose of helping the visualization of the light curve.}
    \label{multipanel}
\end{figure*}

Among all the candidates, only five \footnote{Our dataset spans a quite limited temporal baseline, allowing us to detect variability only in objects with relatively short periods.} objects display clear variability and consistent light curves across all filters: $\mathrm{X5^{c}}$, $\mathrm{X7^{c}}$, $\mathrm{X16B^{c}}$, $\mathrm{X23^{c}}$ and $\mathrm{X33^{c}}$. 
The light curve of $\mathrm{X5^{c}}$ was shown as an example in Fig.~\ref{variaX5}, while the light curves of the other four candidate counterparts are presented in Fig.~\ref{multipanel}. We overlay a grey solid line representing a combination of sines and cosines. This line has no physical meaning; it is purely meant to facilitate the reading of the light curve.
We anticipate that the FAP computed for these five sources, as detailed in Sect.~\ref{variability}, are very low, and their specific values are provided in the following paragraphs. 

\begin{figure}
    \centering
    \includegraphics[width=0.9\hsize]{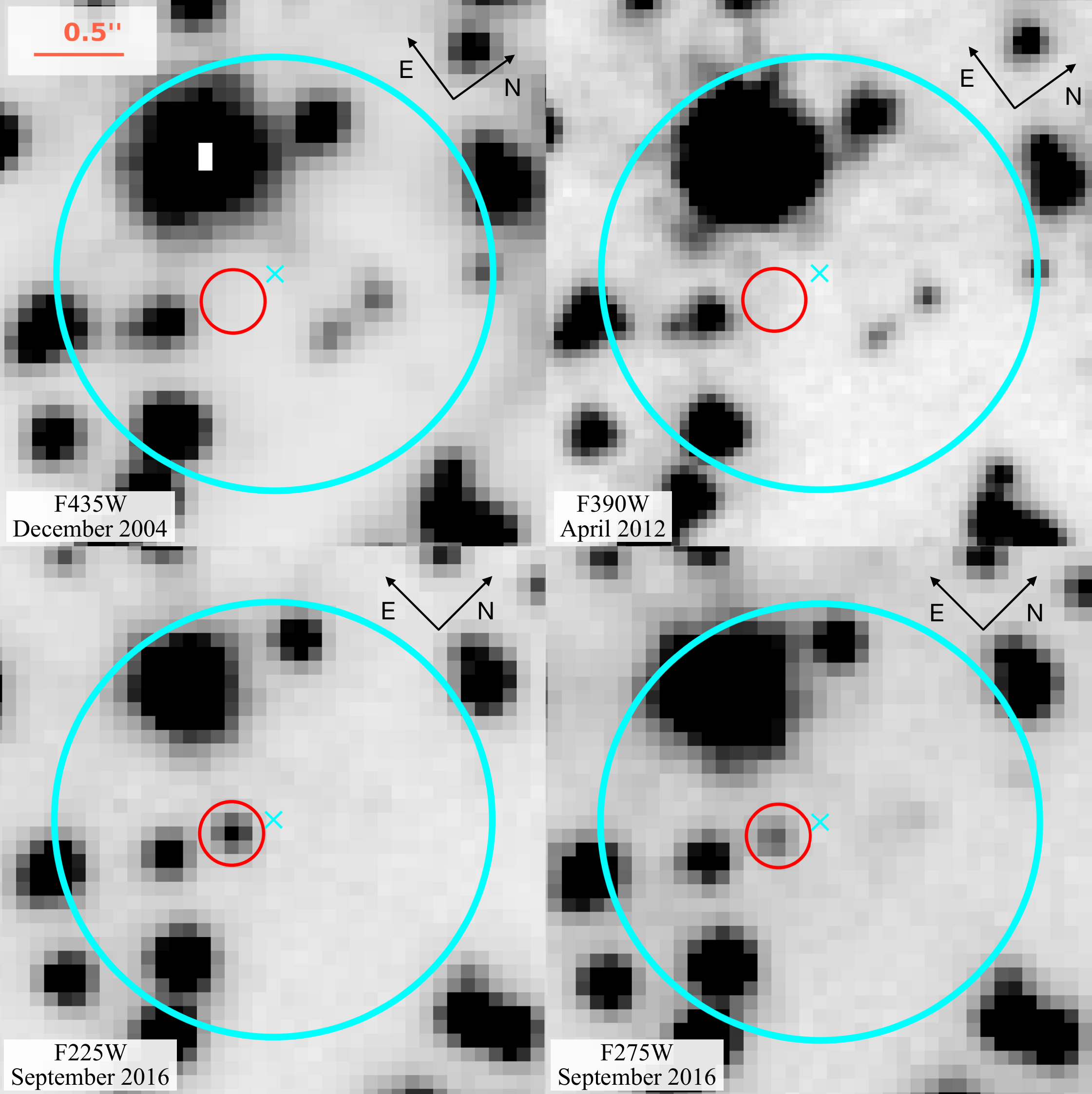}
    \caption{Four \textit{HST} combined and calibrated images (.drz file for the F435W filter and .drc files for the other three filters) centered on the position of the X-ray source X17. In the top-left panel, the F435W image from the December 2004 ACS/HRC observation is shown. The image in the top-right panel was taken in the F390W filter as part of the WFC3/UVIS observations of April 2012. The bottom panels, from left to right, display the F225W and F275W WFC3/UVIS images from September 2016. In each panel, the X-ray source position is indicated by a cyan cross, along with a cyan circle whose radius represents the uncertainty in the X-ray position, $\mathrm{unc_{X}}$. The red circle marks the location of the counterpart $\mathrm{X17^{c}}$.}
    \label{figx17}
\end{figure}

\paragraph*{$\mathbf{X5^{c}}$}
This is one of the five objects falling inside the uncertainty radius of the X5 X-ray source. It is located at only $0\arcsec\!.03$ from the X-ray position and it has a very blue position in the CMDs, as detailed in the previous section. As also shown in Fig.~\ref{propermotion}, the VPD suggests that $\mathrm{X5^{c}}$ is a likely cluster member. The FAP for this source is $3.29 \,\cdot\,10^{-7}$. The light curve of this candidate counterpart is depicted in Fig.~\ref{variaX5}. In this case, we were unable to supplement the light curve with HRC data because the star is absent from the final catalogue obtained with HRC images. In fact, $\mathrm{X5^{c}}$ is located right at the edge of HRC’s FoV, and due to the slight offset applied to each exposure, the source appears in fewer than half of the images, thus not fulfilling the criteria adopted during the data-reduction procedure. Additionally, individual measurements in the F225W and F275W filters are not available for this object either. As described in the previous section, retrieving the $m_{\rm F225W}$ and $m_{\rm F275W}$ magnitudes for $\mathrm{X5^{c}}$ required photometric reduction on the combined images. As a consequence, it was not possible to obtain individual magnitudes. 
After aligning the three light curves to the mean $m_{\rm F555W}$ magnitude and using the Lomb-Scargle periodogram method, we obtained a period of 5.08 hours, which is the period we used to fold the light curve in Fig.~\ref{variaX5}. The light curve shows a broad maximum, with an amplitude of more than 2 magnitudes. This pattern resembles typical BW light curves, known for their characteristic singular minimum and maximum structure, which is commonly attributed to the heating of the companion side exposed to the pulsar's flux. An illustrative example is the light curve derived for the BW system PSR B1957+20 by \cite{Reynolds2007}. 
Interestingly, the period nicely aligns with typical BW periods.
The hypothesis that this object could be a BW is also supported by its position in the CMD, which resembles that of COM-M71 \citep{Cadelano2015} and COM-M5C \citep{Pallanca2014}. An alternative hypothesis is that this could be a CV. 
\paragraph*{$\mathbf{X7^{c}}$}
This object is the only candidate for the X7 X-ray source surpassing at least one criterion for being included in our list of high-confidence counterparts. It is located in the binary sequence at $m_{\rm F555W}$ $\sim 20$ (see Fig.~\ref{cmd_finale}). Notably, most X-ray sources with counterparts detected in the binary sequence are thought to be coronally active binaries, or BY Dra systems \citep{Edmonds2003,Bassa2004}. 
No membership information is available for this source, as it is not present in the HSTPROMO catalogue \citep{Libralato2018,Libralato2022}. The FAP of $\mathrm{X7^{c}}$ amounts to $1.62 \,\cdot\,10^{-7}$ 
As $\mathrm{X7^{c}}$ falls within the HRC FoV and it is retrieved in the final HRC catalogue, we had the opportunity to study its variability with HRC images.  
Due to the fact that the temporal baseline of HRC observations is considerably shorter compared to WFC3, we did not detect any variability using HRC alone. However, we used HRC data to complement WFC3 measurements and perform the period analysis on a much wider baseline, by shifting all the magnitudes to the $m_{\rm F555W}$ one. The period derived from the global light curve is 8.89 hours, and it is used to fold the combined curve shown in the top-left panel of Fig.~\ref{multipanel}.
Although the $m_{\rm F225W}$ and $m_{\rm F275W}$ data points were used to derive the period from the combined light curve, we chose not to display them for clarity reasons, as their larger errors compared to the other filters would make less clear the visualization of the light curve. With an amplitude of roughly 0.2 magnitudes, the variability of $\mathrm{X7^{c}}$ is less pronounced than that of $\mathrm{X5^{c}}$. However, this object is still pretty interesting, particularly considering that also exotic objects like RB systems typically exhibit small variability amplitudes. 
\paragraph*{$\mathbf{X16B^{c}}$}
This is, with the SSG $\mathrm{X16A^{c}}$, one of the two counterparts with interesting CMD position that we identified for the X-ray source. It aligns precisely with the BSS sequence across all filter combinations and it is confirmed as a cluster member based on the proper motion analysis. This object was detected both in the WFC3 and HRC images. 
By combining all available measurements using $m_{\rm F555W}$ as the reference magnitude, we derived a period of 13.44 hours. This period was used to fold the combined light curve, shown in the top-right panel of Fig.~\ref{multipanel}. The FAP is equal to $2.66 \,\cdot\,10^{-8}$. 
As in the previous case, the light curve displays small fluctuations, within 0.1 mag. The light curve, the CMD position and the period suggest that this star might be a WUMa star, defined as a semi-detached binary system with ongoing mass transfer. Four WUMa stars were found in NGC 362 by \cite{Dalessandro2013}. $\mathrm{X16B^{c}}$ is not among these stars, however it could be another possible candidate.
Moreover, using the equation 1 in \cite{Rucinski2000} for the $\rm{M}_V=\rm{M}_V(\log P,V-I)$ calibration for WUMa systems, we get a fairly good agreement between the $\rm{{M}}^{VI}_V = 2.59$ computed from the equation using the period inferred from our analysis and the $\rm{{M}}_V = 2.50$ computed for $\mathrm{X16B^{c}}$ assuming E(B$-$V) = 0.05 \citep{Harris1996} and a distance of 8.8 kpc \citep{Dotter2010}. This further indicates that this system might indeed be a WUMa. Finally, WUMa stars are closely linked to X-ray emission, as noted by \citet{Heinke2005}, who highlights that 11 out of 15 WUMa binaries in 47 Tucanae were detected as \textit{Chandra} sources. In fact, due to their rapid rotation, with periods ranging from 0.3 to 0.6 days, WUMa stars are expected to exhibit the highest coronal activity relative to their surface area.
\paragraph*{$\mathbf{X23^{c}}$}
This object was already included in the list of objects showing peculiar CMD position. Indeed, in the ($m_{\rm F390W}$$-$$m_{\rm F555W}$) CMD, it is close to the BSS sequence but exhibits a slightly redder colour. On the other side, when plotted in a ($m_{\rm F555W}$-$m_{\rm F814W}$) CMD, the star shifts to a bluer colour. As anticipated in Sect.~\ref{result1}, this object is classified as a cluster member. Similar to $\mathrm{X5^{c}}$, we observe consistent light curves across the filters. The FAP corresponding to this source is the lowest among our five variable counterparts, being $7.29 \,\cdot\,10^{-25}$.
As in the case of $\mathrm{X7^{c}}$, the HRC data alone were not helpful due to their limited temporal range. However, by combining the WFC3 and HRC data, we determined a period of 9.53 hours, which is used to fold the light curve shown in the bottom-left panel of Fig.~\ref{multipanel}.
The curve displays an amplitude of $\sim 1$ mag and it features an asymmetric light profile with two minima, one of which is about 0.7 magnitudes deeper than the other. As anticipated in Sect.~\ref{result1}, one of the most interesting variables found by \cite{Rozyczka2016} is V20. In the (B-V) CMD this star exhibit the same position as $\mathrm{X23^{c}}$, being slightly redder than the BSS sequence. At the same time, the light curve plotted in their fig. 4 is very similar to the curve we derived for $\mathrm{X23^{c}}$. According to \cite{Rozyczka2016} this object could be an eclipsing binary BSS with a period of 9.6 hours. In this instance, we matched our catalogue with the position of the variable stars listed in their table 1 and table 2. While we found a match between V24 and $\mathrm{X10^{c}}$, as already reported in Sect.~\ref{result1}, we did not find any match with V20, as the latter is $\sim 30 \arcsec$ away from $\mathrm{X23^{c}}$. However, the similarity in terms of CMD position, variability and period would suggest that this object is likely a close eclipsing binary BSS. This X-ray emission is likely due to coronal activity.
\paragraph*{$\mathbf{X33^{c}}$}
This object stands out as the only candidate counterpart to the X-ray source X33 with distinctive features. Given that the positional uncertainty of X33 is $1\arcsec\!.00$, this object is one of 45 potential counterparts for this X-ray source. $\mathrm{X33^{c}}$ is detected only in the $m_{\rm F225W}$ and $m_{\rm F275W}$ WFC3 catalogue, while in all other images from our dataset, its detection was restricted by the proximity of a saturated star. In the ($m_{\rm F225W}$, $m_{\rm F225W}$$-$$m_{\rm F275W}$) CMD (see Fig.~\ref{cmd_finale}), the object lies along the binary sequence, suggesting that it could be an AB with the X-ray emission powered by coronal activity, as these are typically found on this sequence. 
Using the $m_{\rm F225W}$ as reference magnitude, we constructed the combined light curve, whose analysis determined a period of 3.10 hours, subsequently used to fold the global curve shown in the bottom-right panel of Fig.~\ref{multipanel}. The FAP of $\mathrm{X33^{c}}$ amounts to $1.62 \,\cdot\,10^{-6}$.  The light curve exhibits photometric variability with a modulation amplitude of approximately 0.6 magnitudes.
If this object is an AB, the period derived from our analysis could be unreliable, as an object located at this position in the CMD (consistent with a $0.8 \, M_\odot$ star) cannot physically sustain a 3.2-hour orbit. In addition, it is worth stressing that, since the temporal baseline available for this source to derive its period is only $\sim8$ hours, it is likely that we have not identified the true period.
An alternative interpretation of this source, assuming it is the true optical counterpart of X33, is that it could be a RB MSP. Both its period and light curve remind the typical characteristics of such systems, where this pattern is generally attributed to the tidal distortion caused by the pulsar on the companion star. 
\paragraph*{\textbf{The peculiar case of the $\mathbf{X17^{c}}$ optical counterpart}}
We selected this object as a candidate counterpart for the X17 X-ray source due to its unusual position in the ($m_{\rm F225W}$, $m_{\rm F225W}$–$m_{\rm F275W}$) CMD, where it appears significantly offset from the MS, closer to the WD cooling sequence, as can be seen from the first panel of Fig.~\ref{cmd_int}. However, this star lacks $m_{\rm F390W}$, $m_{\rm F435W}$, $m_{\rm F555W}$, and $m_{\rm F814W}$ magnitudes and it is also missing from the public catalogue by \cite{Libralato2018, Libralato2022}.
To understand why this source only has $m_{\rm F225W}$ and $m_{\rm F275W}$ magnitudes, we performed a visual inspection of all the images in the archive available for NGC 362, confirming that none of the images showed a clear detection of $\mathrm{X17^{c}}$, with the only exception of the F225W and F275W images that we used in this work. As an example, Fig.~\ref{figx17} illustrates four \textit{HST} combined and calibrated images (\texttt{$\_$drz} and \texttt{$\_$drc} files) centered on the position of the X-ray source X17, captured by two different instruments over three distinct epochs. In the top-left panel, the image from the 2004 ACS/HRC observation of NGC 362 in the F435W filter is shown; this dataset was used to construct our HRC catalogue and is the second entry in Table~\ref{dataset}. The image in the top-right panel was taken in the F390W filter as part of the WFC3/UVIS observations of April 2012, which were used to construct our WFC3 catalog, and corresponds to the fifth entry in Table~\ref{dataset}. Finally, the bottom panels, from left to right, display the F225W and F275W WFC3/UVIS images from 2016 which were used in this work (last two entries of Table~\ref{dataset}). In each panel, the X-ray source position is indicated by a cyan cross, along with a cyan circle whose radius represents the uncertainty in the X-ray position, $\mathrm{unc_{X}}$. The red circle marks the location of the counterpart $\mathrm{X17^{c}}$. As can be seen from the figure, there is a clear detection of this source in both the F225W and F275W images (bottom panels), whereas the object is completely absent in the other two frames. We note that this is not an issue of detection threshold, as the F435W and F390W frames are deeper than the F225W and F275W frames. Figures ~\ref{cmd} and \ref{cmd_int} demonstrate that many stars of similar colour, and fainter than $\mathrm{X17^{c}}$, such as $\mathrm{X4^{c}}$ and the WD sequence, are detected in the ($m_{\rm F555W}$, $m_{\rm 390W}$–$m_{\rm F555W}$) CMD.
This object appears to show more extreme behaviour in terms of photometric variability than the other sources described in this section, suggesting that it may be a system experiencing an outburst, similar to the object discovered by \cite{Pallanca2013} in the GC M28. However, further information on its cluster membership or possible H$\alpha$ excess is currently lacking, as it is absent from \cite{Libralato2018,Libralato2022} catalogues, and its light curve analysis in the F225W and F275W filters does not indicate short-term variability. Regarding the X-ray emission, the \textit{HST} archival images temporally closest to the \textit{Chandra} detection from January 2004 are those from program ID 10005 (PI: Lewin), captured in December 2003 (first row of Table~\ref{dataset}); however, $\mathrm{X17^{c}}$ is not present in these images. In contrast, no X-ray emission was detected around September 2016, despite the source being clearly visible in the optical images. Therefore, associating the X-ray emission with this candidate counterpart is not straightforward. However, it is possible that by chance, no \textit{HST} observations captured the object during its outburst, except for the F225W and F275W images. It is likely that this object is associated with the cluster, since most of the stars in this direction are associated with the cluster, and neither AGN nor foreground/background stars are substantially more likely to show eruptions in the UV than cluster stars. If the outburst were from a LMXB hosting a neutron star or black hole, the implied increase in accretion would generate an X-ray luminosity of $>10^{36}$ erg/s, which would almost certainly have been detected by the MAXI all-sky monitor \citep{Negoro2016}. Thus it seems more likely that this was a dwarf nova outburst from a CV, as have been seen repeatedly in globular clusters \citep{Shara2005,Pietrukowicz2008,Modiano2020}.

\subsection{Counterparts showing H$\alpha$ emission}\label{result3}
\begin{figure}
    \centering
    \includegraphics[width=\hsize]{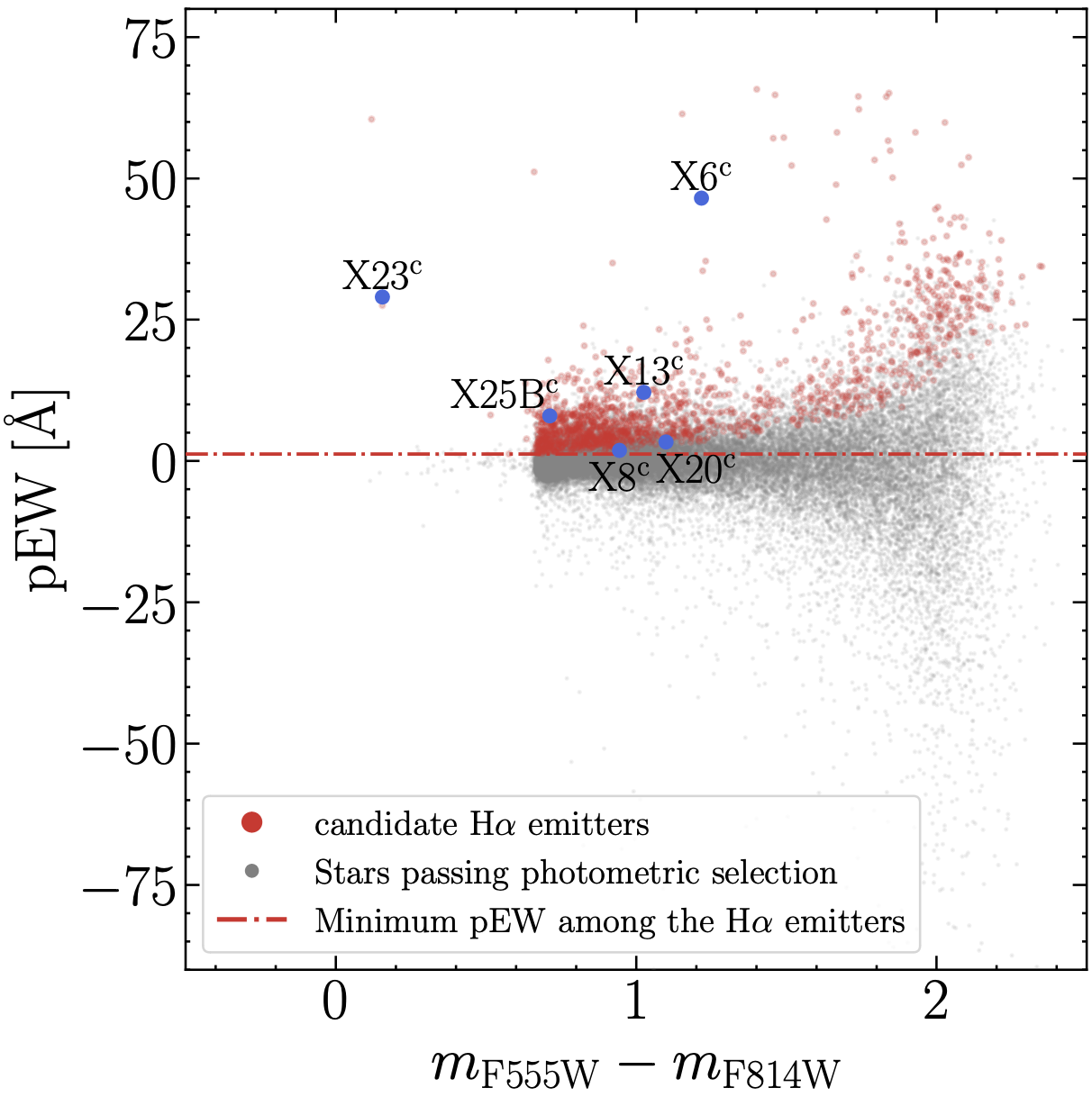}
    \caption{Photometric equivalent width (pEW) as function of the ($m_{\rm F555W}$$-$$m_{\rm F814W}$) colour. Grey points are all the objects that passed the photometric selection, while red points are the candidate H$\alpha$ emitters. The dotted line marks the minimum pEW of the candidate H$\alpha$ emitters. We plot in blue the candidate optical counterpart that we found to have an H$\alpha$ excess.}
    \label{pew}
\end{figure}

In this section we summarise the results obtained from our photometric study of candidate H$\alpha$ emitters in NGC 362. 
In Fig.~\ref{halphafig} we show the ($m_{\rm F555W}$$-$$m_{\rm F658N}$) versus ($m_{\rm F555W}$$-$$m_{\rm F814W}$) colour-colour diagram and we highlight in red the 1366 candidate H$\alpha$ emitters we found following the method described in Sect.~\ref{halpha}. 
In Fig.~\ref{pew} we plot the pEW as function of the ($m_{\rm F555W}$$-$$m_{\rm F814W}$) colour and we highlight in red the H$\alpha$ emitters. The pEW of the H$\alpha$ emitters selected with this method ranges between 1.22 \r{A} and 65.8 \r{A}. 
Since our focus is on the optical counterparts of the 33 X-ray sources, we examined whether any of the 1366 emitters are included among our 530 candidate optical counterparts. We identified five objects, $\mathrm{X8^{c}}$, $\mathrm{X13^{c}}$, $\mathrm{X20^{c}}$, $\mathrm{X25B^{c}}$ and $\mathrm{X23^{c}}$, showing H$\alpha$ excess. These objects are represented by blue points in Fig.~\ref{halphafig} and Fig.~\ref{pew}. We also found another counterpart, $\mathrm{X6^{c}}$, which exhibits a significant ($m_{\rm F555W}$$-$$m_{\rm F658N}$) excess and satisfies the requirement to be considered an H$\alpha$ emitter, but fails the $m_{\rm F814W}$ photometric quality selection. However, as illustrated in Fig.~\ref{halphafig}, $\mathrm{X6^{c}}$ exhibits such a significant excess that it would require a ($m_{\rm F555W}$$-$$m_{\rm F814W}$) shift of more than $1$ magnitude to no longer be classified as an emitter. Therefore, even if it does not pass the F814W-band photometric quality selection, we included it among the list of counterparts showing H$\alpha$ excess.

\begin{figure}
    \centering
    \includegraphics[width=0.85\hsize]{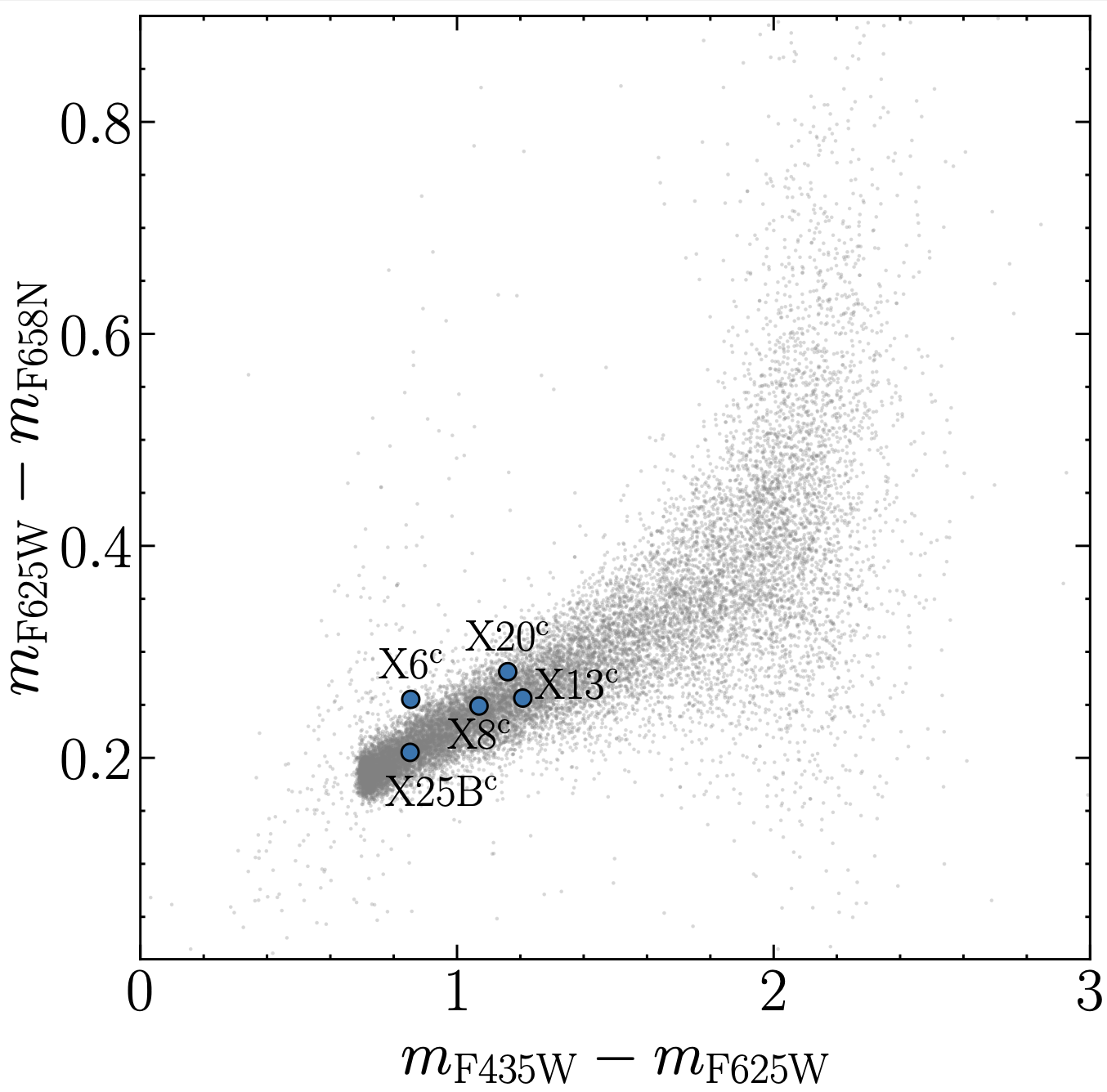}
    \caption{($m_{\rm F555W}$$-$$m_{\rm F658N}$) versus ($m_{\rm F555W}$$-$$m_{\rm F814W}$) colour-colour diagram. Gray points are stars passing the photometric selection by \cite{Libralato2018} and membership selection. The blue dots represent our candidate counterparts for which we found an H$\alpha$ excess with the method described in Sect.~\ref{halpha}.}
    \label{cmdhalpha}
\end{figure}

It should be noticed that in our H$\alpha$ analysis we use coeval F555W and F814W measurements, while those in the F658N band were collected at a different epoch. However, many objects, especially CVs, are variable at optical wavelengths. Thus, taking F658N data from a different epoch than the comparison broadband data could lead to incorrect measurements of H$\alpha$ strength. In principle, since proposal ID 10005 includes F435W and F625W images alongside F658N, one should prefer using the magnitudes in these three filters to study the H$\alpha$ excess. However, since the \textit{R} band (F625W) is over an order of magnitude wider than the H$\alpha$ filter and contains the H$\alpha$ feature itself, it is not an appropriate choice to retrieve a precise estimate of the continuum level and only gives a rough estimate of it, as also pointed out by \citet{DeMarchi2010}. 
Moreover, as mentioned in Sect.~\ref{halpha}, for this type of study we use a two-colour diagram, where the colours are indicative of the H$\alpha$ excess (on the y-axis) and the stellar effective temperature (on the x-axis). Hence, since we also need a useful colour index for temperature, we cannot rely on the F435W and F658N images only. On the other side, an accurate determination of the continuum can be achieved using a combination of \textit{V}, \textit{I}, and \textit{H}$\alpha$ magnitudes, as the contribution of the H$\alpha$ line to the \textit{V} and \textit{I} bands is negligible. In addition, the \textit{V}$-$\textit{I} is a useful colour index for determining the effective temperature, and it accounts for the variation of the stellar continuum below the H$\alpha$ line across different spectral types.
Moreover, a solid estimate of the level of the stellar continuum inside the H$\alpha$ band is needed to measure the pEW. As anticipated in Sect.~\ref{halpha},  the latter is an important piece of information in this context, as it allows a classification of the identified emitters. 
Therefore, we chose the ($m_{\rm F555W}-m_{\rm F814W}$, $m_{\rm F555W}-m_{\rm F658N}$) colour-colour diagram, shown in Fig.~\ref{halphafig}, as the primary diagnostic tool for identifying candidate H$\alpha$ emitters. Nonetheless, we also used the coeval F435W, F625W, and F658N images to check on the counterparts of our candidate H$\alpha$ emitters identified with this method. In fact, we plotted our candidate H$\alpha$ emitter counterparts on a ($m_{\rm F555W}$$-$$m_{\rm F658N}$) versus ($m_{\rm F555W}$$-$$m_{\rm F814W}$) colour-colour diagram, presented in Fig.~\ref{cmdhalpha}. In this diagram, we can plot five of the six counterparts presented in Sect.~\ref{result3}, as $\mathrm{X23^{c}}$ lacks the ${m_{\rm{F625W}}}$ measurement.  We notice that our candidate emitters show an excess which is broadly consistent with what we observe in Fig.~\ref{halphafig}.
The only true exception is  $\mathrm{X6^{c}}$, which shows such a significant excess in our colour-colour plane that is not reflected in the {($m_{\rm F555W}$$-$$m_{\rm F658N}$)} versus ($m_{\rm F555W}$$-$$m_{\rm F814W}$) colour-colour diagram.
To investigate further, we analysed the light curve of $\mathrm{X6^{c}}$ but found no evidence of variability. We believe that, in this case, the H$\alpha$ excess might result from either long-term variability or a transient event, such as a disk flare, that temporarily enhanced the H$\alpha$ emission.
Table~\ref{halphatable} summarises the X-ray source candidate counterparts for which we found H$\alpha$ excess, while a more detailed description of each object is provided in the following paragraphs.
\begin{table*}[h]
\caption{\label{halphatable}Candidate optical counterparts showing H$\alpha$ excess.}
\centering
\begin{tabular}{lccccccc}
\hline\hline
Object name & R.A. & Dec. & $m_{\rm F555W}$ & $m_{\rm F814W}$ & $m_{\rm F658N}$ & pEW\\
&  [$^\circ$] & [$^\circ$] & [mag]& [mag]& [mag] & [\r{A}]\\
\hline
$\mathrm{X6^{c}}$  & 15.8111520 & $-70.8482723$& 23.019 & 21.802 & 20.992 &  46.5\\
$\mathrm{X8^{c}}$  & 15.8058206 & $-70.8503698$ & 17.934 & 16.990 & 17.144 & 1.86\\
$\mathrm{X13^{c}}$ & 15.8175106 & $-70.8513261$ & 21.439 & 20.414 & 20.422 & 12.1\\
$\mathrm{X20^{c}}$  & 15.8033216 & $-70.8549707$ & 18.319 & 17.220 & 17.387 &  3.35\\
$\mathrm{X23^{c}}$  & 15.8085802 & $-70.8488293$ & 17.077 & 16.922 & 16.418 & 27.5\\
$\mathrm{X25B^{c}}$ & 15.8081627 & $-70.8517508$ & 19.704 & 18.992 & 18.986 & 7.97 \\
\hline
\end{tabular}
\tablefoot{From left to right: the name of the object in our list of counterparts, the right ascension and declination (in degrees), the $m_{\rm F555W}$, $m_{\rm F814W}$ and $m_{\rm F658N}$ magnitudes and the pEW of the H$\alpha$ line in \r{A}.}
\end{table*}
\paragraph{$\mathbf{X6^{c}}$}
This star is one of the four objects falling inside the uncertainty radius for the X-ray source $\mathrm{X6}$. Its position in the ($m_{\rm F555W}$, $m_{\rm F390W}$$-$$m_{\rm F555W}$) CMD resembles that of $\mathrm{X5^{c}}$, as it is located between the MS and the WD cooling sequence, while it shifts towards the WD cooling sequence in the ($m_{\rm F225W}$, $m_{\rm F225W}$$-$$m_{\rm F275W}$) CMD, suggesting that this could be a system containing a WD. The proper motion diagram suggests that this object is a cluster member. Figure~\ref{halphafig}, clearly shows that this star, has a significant ($m_{\rm F555W}$$-$$m_{\rm F658N}$) colour excess. Moreover, its pEW of 46.5 \r{A} is the highest among the X-ray counterparts with H$\alpha$ emission. As a reference, in the analysis performed by \cite{Pallanca2017} on the H$\alpha$ emitters of NGC 6397, the objects with the highest H$\alpha$ excess are 7 CVs, with pEW ranging from a minimum of 15.2 \r{A} to a maximum of 90.4 \r{A}, as reported in their table 1. The pEW of $\mathrm{X6^{c}}$ falls precisely within this range, hence, both the CMD position and the prominent H$\alpha$ pEW suggest that this object may be a CV.
On the other side, as anticipated before and shown in Fig.~\ref{cmdhalpha}, this excess seems to be absent in the CMD constructed with coeval \textit{H$\alpha$} and \textit{R} images. As previously noted, the H$\alpha$ excess detected in our analysis could potentially result from significant long-term variability and/or a transient event, such as a flare in the accretion disk.

\paragraph{$\mathbf{X8^{c}}$}
$\mathrm{X8^{c}}$ is one of the seven counterparts found for its X-ray source. In the CMD it is located along the RGB and according to \citep{Libralato2018} proper motions is a cluster member.  The ($m_{\rm F555W}$$-$$m_{\rm F658N}$) colour excess of $\mathrm{X8^{c}}$ is not as remarkable as the one of $\mathrm{X6^{c}}$, and its pEW of 1.86 Å is the lowest among the identified H$\alpha$ emitters. Given the very low value of the pEW and its CMD position, this system might be an AB with a red giant component. 
\paragraph{$\mathbf{X20^{c}}$}
This is one of the three sub-subgiant counterparts we found in this work, together with $\mathrm{X15^{c}}$ and $\mathrm{X16A^{c}}$. Given its peculiar CMD position, this object was already included in our list of high-confidence counterparts. 
In this case, the presence of an H$\alpha$ emission together with the peculiar CMD position suggests even more strongly that this could be the real counterpart to the X-ray source X20. Its emission is also confirmed by the survey on 26 GCs performed by \citep{Gottgens2019} with MUSE (see Sect.~\ref{gottgens}). The value of the pEW we found for $\mathrm{X20^{c}}$ is 3.35 \r{A}, the second smallest among the H$\alpha$ emitters. 
The census of H$\alpha$ emitters in NGC 6397 performed by \cite{Pallanca2017} revealed that the MSPs companions COM-6397A and COM-6397B, are among the H$\alpha$ emitters of NGC 6397, with a pEW of 3.2 \r{A} and 4.3 \r{A}, respectively. Both the values of the pEW and the CMD position of COM-6397B are consistent with our $\mathrm{X20^{c}}$ candidate counterpart. Moreover, its light curve does not show any variability; hence, it is possible that, if variable, either this object has a longer period than our temporal baseline or the inclination of the system is such that we do not see variability. Given these characteristics, it is possible that this counterpart might represent an example of spider MSP. On the other side, this object is also consistent in being an AB, both in terms its CMD position and moderate H$\alpha$ excess. At this time, we cannot dismiss either of the two possibilities.
\paragraph{$\mathbf{X23^{c}}$}
We previously discussed this object in Sect.~\ref{result1} and Sect.~\ref{result2} due to its peculiar CMD position and distinct light curve, exhibiting clear variability among the three WFC3 filters. Based on the comparison with the V20 source in \cite{Rozyczka2016}, we interpret this object as a close eclipsing binary BSS, where the X-ray emission could be explained by coronal activity.
Furthermore, this object displays a significant H$\alpha$ excess with a pEW of 27.5 \r{A}. In this case, the H$\alpha$ excess might be related to an ongoing mass transfer process. On the other side, it is worth noticing that, the remarkable variability of this source might also affect the single-epoch measurement of the H$\alpha$ magnitude.
\paragraph{\textbf{$\mathbf{X13^{c}}$ and $\mathbf{X25B^{c}}$}}
These two objects lie on the cluster's MS, both in the ($m_{\rm F390W}$$-$$m_{\rm F555W}$) and ($m_{\rm F555W}$$-$$m_{\rm F814W}$) CMDs. Both $\mathrm{X13^{c}}$ and $\mathrm{X25B^{c}}$ are classified as cluster members according to the proper motion analysis.
They are characterized by moderate values for the pEW, being 12.1 and 7.97 \r{A} for $\mathrm{X13^{c}}$ and $\mathrm{X25B^{c}}$, respectively. However, the fact that their CMD position remains unchanged across different filter combinations suggests the absence of a blue component in this system. Furthermore, considering the moderate H$\alpha$ excess, it is possible that these two objects are ABs.
None of the other counterparts of $\mathrm{X13^{c}}$ show any peculiarities, apart from the proximity to the X-ray source, either in their CMD position or light curve. Therefore, the presence of H$\alpha$ emission in the $\mathrm{X13^{c}}$  suggests that it could be the most probable counterpart for this X-ray source. On the other hand, $\mathrm{X25B^{c}}$ is not the only candidate counterpart we identified for the X25 X-ray source. In fact, in addition to $\mathrm{X25B^{c}}$, we identified another potential counterpart, $\mathrm{X25A^{c}}$, which is more likely to be the correct identification due to its clear UV excess, suggesting a CV nature.

\section{Clues from spectroscopy}\label{spectroscopy}
\subsection{H$\alpha$ analysis by G{\"o}ttgens et al. (2019)}\label{gottgens}
\cite{Gottgens2019} identified 156 H$\alpha$  emitters in 27 Galactic GCs by using MUSE. Within this sample, 7 stars belong to NGC 362. 
Using the coordinates listed in their Table A.2, we matched them with our catalogue to determine whether any of these 7 objects correspond to candidate counterparts in our list. We were able to identify three of them. 
As anticipated in Sect.~\ref{result3}, our $\mathrm{X20^{c}}$ corresponds to the first object listed for NGC 362 in table A.2 of \cite{Gottgens2019}. In our analysis we found for this object an H$\alpha$ excess with a pEW of 3.34 \r{A}. Interestingly, \cite{Gottgens2019} found a variable H$\alpha$ emission for this source. This additional information provides interesting clues about the nature of this object. As anticipated before, $\mathrm{X20^{c}}$ is in the sub-subgiant region, which is populated by close binaries, such as close, coronally emitting ABs, or more exotic objects such as RB MSPs. The variability of this star did not show any peculiarity, however the presence of a variable emission in H$\alpha$ might suggest that this could be an example of MSP where the emission is likely produced by a combination of material stripped from the companion and a strong intrabinary shock. The fifth object associated with NGC 362 as listed in \cite{Gottgens2019} corresponds to the optical counterpart to X12. As $\mathrm{X20^{c}}$, this counterpart is also characterized by a variable H$\alpha$ emission.  In the CMDs shown in Fig.~\ref{cmd_int}, $\mathrm{X12^{c}}$ lies significantly redwards of the RGB. Given its CMD position and the variable H$\alpha$ emission found by \cite{Gottgens2019}, X12 could be interpreted as an AB system, such as an RSS. In our study, we did not find any photometric evidence of H$\alpha$ emission for it. Finally, the last object with variable emission listed in \cite{Gottgens2019} corresponds to our counterpart $\mathrm{X10^{c}}$. As anticipated in Sect.~\ref{result1}, we also found a match between $\mathrm{X10^{c}}$ and the V24 variable star in \cite{Rozyczka2016}. Here, this system is interpreted as an eclipsing semidetached binary system, formed by two $\sim 0.8 \, \mathrm{M_\odot}$ stars, where the primary giant is filling its Roche Lobe and the secondary star is a blue object, either a BSS or a WD. Moreover, in our analysis, $\mathrm{X10^{c}}$ satisfies the requirement for being considered an H$\alpha$ emitter, but fails the $m_{\rm F814W}$ photometric selection criterion, both in \texttt{CHI} and \texttt{SHARP}. Although it has a pEW of 13.0 \r{A}, $\mathrm{X10^{c}}$'s H$\alpha$ excess is not significant enough to confidently include it among the confirmed emitters as we did instead for $\mathrm{X6^{c}}$. Nonetheless, the H$\alpha$ emission detected by \cite{Gottgens2019}, would suggest that accretion is taking place in this system, probably through Roche lobe overflow, as suggested by \cite{Rozyczka2016}.

\subsection{MUSE radial velocities}
Multi-epoch RV measurements are particularly useful  for understanding the motion of celestial bodies. In particular, the orbital motion of binary system components causes their RVs to vary periodically. Therefore, detecting regular changes in an object's RV curve is a strong indication that it may belong to a binary system, with its mass ratio, eccentricity, and inclination shaping the radial velocity modulation. Multi-epoch RVs analysis provide a powerful complementary information to magnitude variability to identify and characterize binary systems.

In order to gain deeper insights into the nature of our high-confidence counterparts, we examined their RVs across multiple observation epochs. The starting point of this analysis is the survey of 27 Galactic GCs (PI: S. Kamann, formerly S. Dreizler) performed with the integral field spectrograph MUSE at the VLT \citep{Kamann2018}. Here we used observations of NGC 362 obtained in Wide Field Mode (WFM) in three main epochs: November 22, 2014; October 8-13, 2015; and November 2, 2017. The RV measurements were produced by combining individual exposures from each pointing within each 1-hour observing block (OB). Typically, a single OB included observations at four distinct pointings, with each pointing observed three times, resulting in four data cubes per OB. For each star we discarded the measurements with reliability $R < 0.8$ or flagged as outliers. The methods for reliability assessment and outlier detection are described in section 3.2 of \cite{Giesers2019}. 

We found that eight out of our 28 high-confidence counterparts (see Sect.~\ref{conclusions}) have multi-epoch RV measurements. These include the brightest counterparts: $\mathrm{X8^{c}}$, $\mathrm{X9^{c}}$, $\mathrm{X10^{c}}$, $\mathrm{X12^{c}}$, $\mathrm{X16B^{c}}$, $\mathrm{X20^{c}}$, $\mathrm{X21A^{c}}$, and $\mathrm{X23^{c}}$. As a first step, to determine whether our high-confidence counterparts exhibited any variation in RV, we applied the method recently proposed by \cite{Giesers2019} in their study of binary systems in NGC 3201. To detect RV variations in a given star with $m$ measurements, they compute the $\chi^2$ for the set of measurements $x_{j}$ with uncertainties $\sigma_j$. This determines how consistent the measurements are with a constant weighted mean $\overline x$ (null hypothesis) of the data \citep[see eq. 2 in][]{Giesers2019}.
\begin{figure}
    \centering
    \includegraphics[width=\hsize]{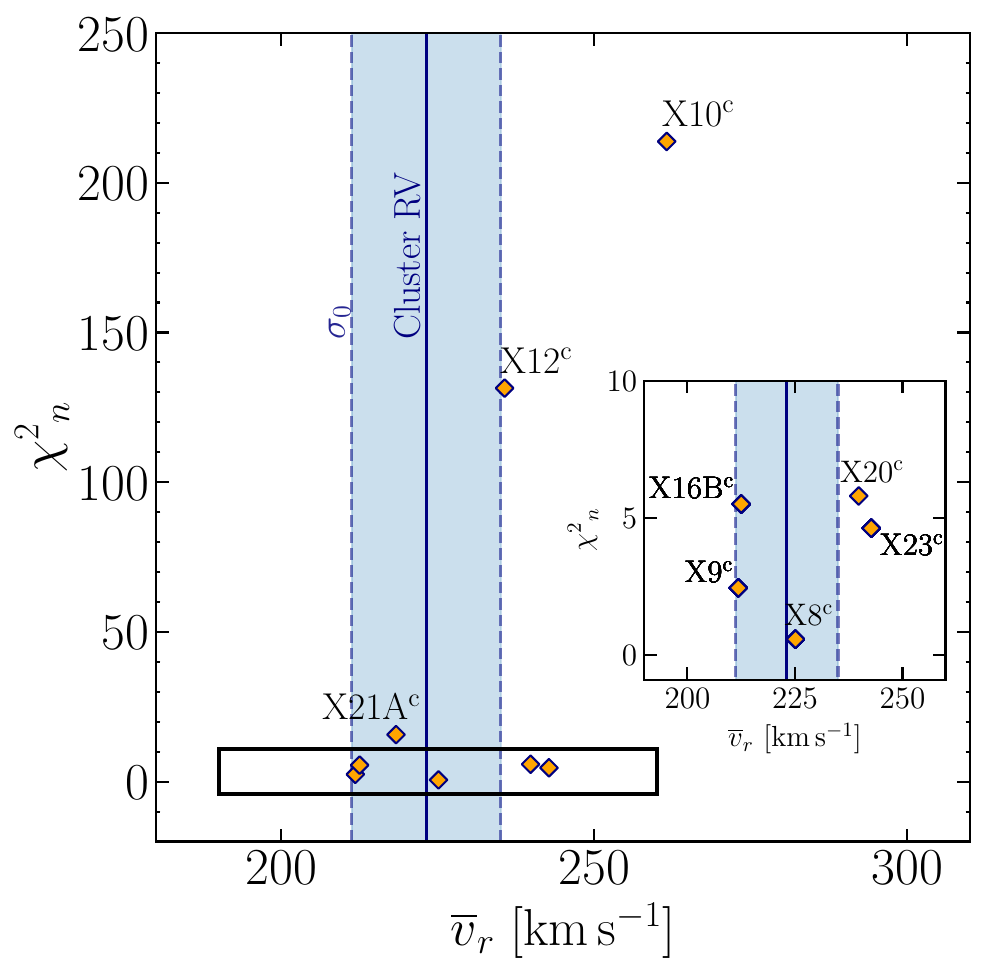}
    \caption{Reduced $\chi^2$ as function of the weighted mean of the RVs for our eight high-confidence counterparts with available multi-epoch RV measurements (orange diamonds). The solid vertical blue line is the heliocentric RV of NGC 362, while the dashed lines represent the central 1D velocity dispersion $\sigma_{0}$. The inset on the right is a zoom-in of the area highlighted by the black rectangle.}
    \label{chiRV}
\end{figure}
We computed the $\chi^2$ for each of the eight counterparts and normalized it by dividing by $m-1$. 
In Fig.~\ref{chiRV} we plot the reduced chi-square ${\chi^{2}}_{n}$  as a function of the weighted mean of the RV measurements for each star ($\overline v_r$). The eight high-confidence counterparts are highlighted with orange diamonds, and their names are labelled. The solid vertical blue line indicates the heliocentric RV of NGC 362, while the dashed lines represent the central 1D velocity dispersion $\sigma_{0}$, whose values are taken from the catalogue provided by \citet{Baumgardt2020} \footnote{The catalogue is publicly available at \url{https://people.smp.uq.edu.au/HolgerBaumgardt/globular/}. Values used in this work are from the 4th version of the catalogue updated in March 2023}.
The black rectangle highlights the area of the plot with $\chi^2 <10$, which is shown in more detail in the zoomed-in inset on the right.

Most of the counterparts shown in the figure have an average RV close to the systemic velocity of the cluster, which supports the membership of these eight objects, as already suggested by the study of their proper motions (see Sect.~\ref{result1}). Specifically, all objects fall within $3 \,\sigma_{0}$ of the cluster’s velocity, with the only exception of $\mathrm{X10^{c}}$. However, based on proper motions, $\mathrm{X10^{c}}$ appears to be a member of the cluster. On the other hand, it should be noted that $\mathrm{X10^{c}}$ has the highest ${\chi^{2}}_{n}$ value among all, exceeding 200, which may indicate significant variability. This variability could, in turn, affect the assessment of its membership when considering the resulting average RV. In general, as also mentioned by \cite{Giesers2019}, a star with $\chi^2>1$ typically exhibits RV variations exceeding the associated uncertainties. This condition holds for all the counterparts shown here, except for $\mathrm{X8^{c}}$, which has a ${\chi^{2}}_{n}$ of 0.58.

\begin{figure*}
    \centering
    \includegraphics[width=\textwidth]{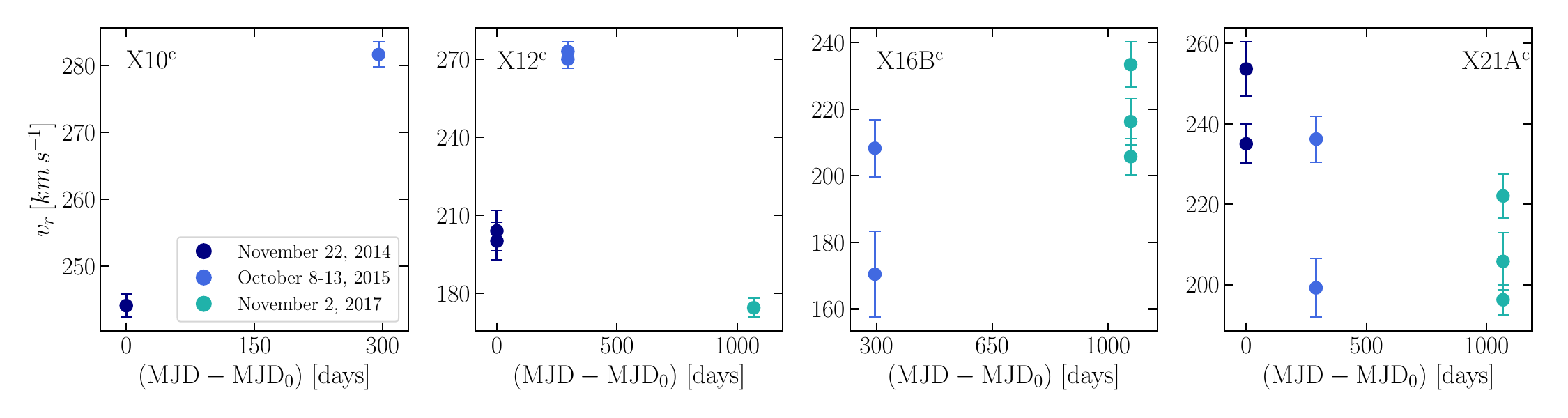}
    \caption{From left to right: RV curves for the optical counterparts $\mathrm{X10^{c}}$, $\mathrm{X12^{c}}$, $\mathrm{X16B^{c}}$, and $\mathrm{X21A^{c}}$ obtained from MUSE observations of NGC 362 in November 22, 2014; October 8-13, 2015; and November 2, 2017. The data points in the plots are colour-coded according to the macro-epoch to which they belong. The x-axis represents the $\mathrm{MJD-MJD_{0}}$, where $\mathrm{MJD_{0}} = 2456983.5075578704$ is the MJD of the first observation from the November 22, 2014 dataset.}
    \label{RVcurves}
\end{figure*}

In addition to analyzing the $\chi^2$ value for each high-confidence counterpart with available multi-epoch RV measurements, we also opted to visually inspect the RV curves for these objects, selecting the ones that displayed the clearest indications of significant variability. By combining the ${\chi^{2}}_{n}$ analysis with a visual inspection of the RV curves, we identified clear variability in the RV of four out of the eight counterparts with available measurements. Specifically, this applies to $\mathrm{X10^{c}}$, $\mathrm{X12^{c}}$, $\mathrm{X16B^{c}}$, and $\mathrm{X21A^{c}}$, whose RV curves are presented in Fig.~\ref{RVcurves}. For clarity, the x-axis represents the $\mathrm{MJD-MJD_{0}}$, where $\mathrm{MJD_{0}} = 2456983.5076$, which corresponds to the epoch of the first observation from the November 2014 dataset. The data points in the plots are colour-coded according to the macro-epoch to which they belong. In fact, as mentioned above, the available RV measurements were obtained from observations grouped into three macro-epochs. Within the same epoch, however, the measurements are taken in close succession, with intervals ranging from a minimum of $\sim3$ minutes for the November 2014 observations, $\sim10$ minutes for the November 2017 observations, up to approximately one day for the October 2015 observations. Consequently, for each counterpart, there are at most three RV measurements available that are separated by a reasonable temporal interval. This limitation prevented us from conducting period studies on RV curves.

Nonetheless, we can confidently state that $\mathrm{X10^{c}}$, $\mathrm{X12^{c}}$, $\mathrm{X16B^{c}}$, and $\mathrm{X21A^{c}}$ show significant variability in their RV curves, with amplitudes ranging from $\sim 38 \, \mathrm{km \, s^{-1}}$ in the case of $\mathrm{X10^{c}}$ up to $98 \, \mathrm{km \, s^{-1}}$ for $\mathrm{X12^{c}}$, showing that these four objects belong to binary systems.

Finally, it is worth noting that for $\mathrm{X16B^{c}}$, we also observed consistent variability in its photometric curve (see Sect.~\ref{result2}), identifying a period of 13.44 hours. Due to the reasons outlined earlier, we could not determine an estimate of the period from the analysis of RV curves, hence we cannot make a comparison between the results obtained from photometry and spectroscopy. Nevertheless, this additional analysis supports the hypothesis proposed in Sect.~\ref{result2} that this object likely resides in a binary system. Conversely, no signs of photometric modulation were detected for the other three objects exhibiting RV variability. This absence may be attributed to their significantly longer periods compared to the temporal coverage of our \textit{HST} observations. This is indeed the case for the $\mathrm{X10^{c}}$, for which \cite{Rozyczka2016} found a period of 8.1 days.
\section{Conclusions}\label{conclusions}
\begin{figure*}
    \centering
    \includegraphics[width=\hsize]{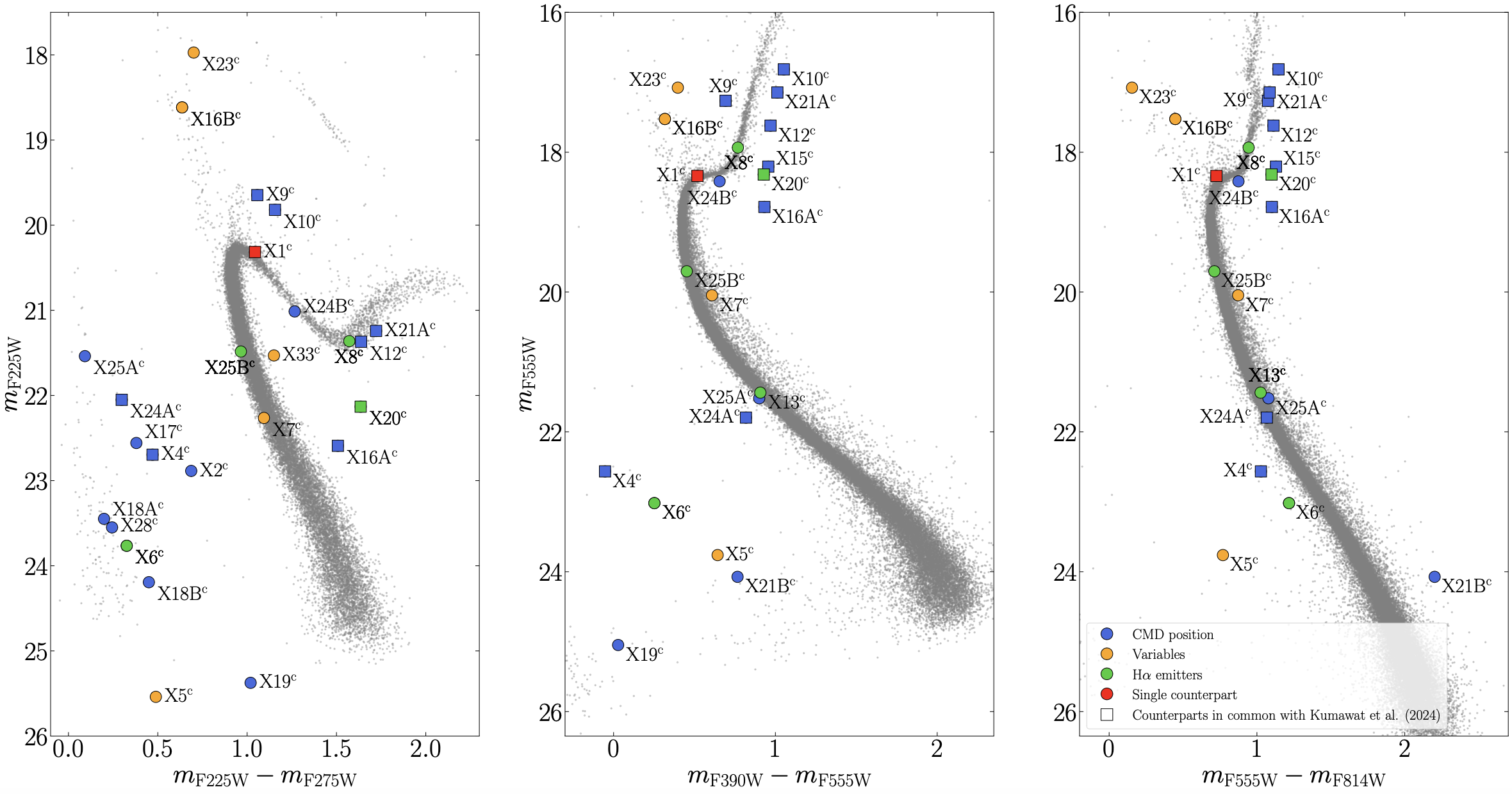}
    \caption{(From left to right: ($m_{\rm F225W}$, $m_{\rm F225W}$$-$$m_{\rm F275W}$), ($m_{\rm F555W}$, $m_{\rm F390W}$$-$$m_{\rm F555W}$) and ($m_{\rm F555W}$, $m_{\rm F555W}$$-$$m_{\rm F814W}$) CMDs containing the 28 high-confidence optical counterparts found in this study. Orange points indicate counterparts exhibiting evident variability, while green points denote objects for which we found an H$\alpha$ emission. Additionally, blue counterparts indicate those occupying interesting positions in the CMD. Red counterparts are the only counterparts found within the uncertainty radius of their respective X-ray source. Squares indicate those counterparts that we find in common with \cite{Kumawat2024}.}
    \label{cmd_finale}
\end{figure*}

The photometric search of optical counterparts to X-ray sources in the GC NGC 362 performed in this work yielded 28 high-confidence candidates. Figure~\ref{cmd_finale} shows their location in the ($m_{\rm F225W}$, $m_{\rm F225W}$$-$$m_{\rm F275W}$), ($m_{\rm F555W}$, $m_{\rm F390W}$$-$$m_{\rm F555W}$) and ($m_{\rm F555W}$, $m_{\rm F555W}$$-$$m_{\rm F814W}$) CMDs. We opted to use different colours to denote the distinct criteria leading to the inclusion of counterparts in the final list of the most interesting ones. Orange points indicate counterparts exhibiting evident variability, while green points denote objects for which we found H$\alpha$ emission. Additionally, blue counterparts indicate those occupying peculiar positions in the CMD. X1 is the only case for which we found only one counterpart within the uncertainty radius and it is highlighted in red. Ten of these high-confidence counterparts are in common with those suggested by \cite{Kumawat2024} and they are highlighted as squares in Fig.~\ref{cmd_finale}. Among the 28 high-confidence counterparts identified in this study, 17 were selected based solely on their peculiar positions in the CMD, five due to their photometric variability, and six for their H$\alpha$ excess. Notably, four objects satisfy two of these criteria: $\mathrm{X5^{c}}$ and $\mathrm{X16B^{c}}$ display both variability and an unusual CMD position, while $\mathrm{X20^{c}}$ and $\mathrm{X6^{c}}$ combine an anomalous CMD position with H$\alpha$ excess. Remarkably, $\mathrm{X23^{c}}$ fulfills all the criteria established in this analysis. These findings indicate that these five objects are the most intriguing and promising counterparts identified in our work, pointing to their potential association with interacting binary systems, MSPs, or CVs. In \citet{Kumawat2024}, only two of these five objects are identified, specifically $\mathrm{X16B^{c}}$ and $\mathrm{X20^{c}}$. This is reasonable, as our identification of the remaining three objects was primarily based on variability and H$\alpha$ analyses, which were not performed by \citep{Kumawat2024}. Additionally, the integration of our photometric analysis with MUSE multi-epoch RV measurements seems to indicate that $\mathrm{X10^{c}}$, $\mathrm{X12^{c}}$, $\mathrm{X16B^{c}}$, and $\mathrm{X21A^{c}}$ reside in binary systems.
All the high confidence counterparts proposed in our work are summarised in Table~\ref{summarytable} in Appendix \ref{summaryt}.

\cite{Kumawat2024} suggests that the X-ray spectrum of X1 is well-fitted by a NS hydrogen atmosphere and interpret this object as a qLMXB.  On the other hand, no qLMXB counterpart has ever been found on the SGB of a cluster. To explain this aspect, \cite{Kumawat2024} suggests that either the qLMXB accretion disc is faint and not showing up against a relatively bright star, or the real qLMXB companion could be a fainter star. In this regard, as visible in fig. A1 of \cite{Kumawat2024}, there is a fainter star close to the position of X1 that lacks HUGS photometry. We also could not find it in the WFC3 images, where we only recovered the same counterpart reported in \cite{Kumawat2024}. However, thanks to the HRC database, we were able to obtain the F435W photometry for the fainter star. Its magnitude is $\sim 1.5$ mag fainter with respect to the SGB counterpart; however, we could not assess its position in the CMD since we only have one magnitude available. In addition, it does not show any evidence of variability. Nevertheless, our photometry differs from the HUGS photometry used by \cite{Kumawat2024}, which places $\mathrm{X1^{c}}$ in the sub-subgiant region, whereas our photometry situates it on the SGB. This shift makes the star a less likely candidate counterpart.

We identified the same optical counterparts as \cite{Kumawat2024} for X9 and X10, which lie redward of the RGB in the ($m_{\rm F555W}$, $m_{\rm F555W}$$-$$m_{\rm F814W}$) CMD, but lie substantially blueward of the RGB in the ($m_{\rm F225W}$, $m_{\rm F225W}$$-$$m_{\rm F275W}$), suggesting very blue and red light sources. Additionally, the match we performed with the stars analysed by \cite{Gottgens2019} reveals that $\mathrm{X10^{c}}$ exhibits variable H$\alpha$ emission and its RV curve suggests that this object might be in a binary system.

Standing out among the promising counterparts identified by \cite{Kumawat2024} are $\mathrm{X15^{c}}$, $\mathrm{X16A^{c}}$, and $\mathrm{X20^{c}}$. All three are classified as sub-subgiants and are located below the MSTO in the CMD. As previously mentioned, we also included these three objects because of their CMD position. Moreover, for $\mathrm{X20^{c}}$ we also found evidence of H$\alpha$ emission which is also spectroscopically confirmed by \citet[][see Sect.~\ref{gottgens}]{Gottgens2019}.

Concerning their counterparts proposed for the X12 and X21A sources, we agree with the hypothesis that, given their CMD position, these appear to be RSSs, with X-rays driven by chromospheric activity. Moreover, for both of them we found RV curves showing strong variability, suggesting that they could be residing in binary systems. Finally, the H$\alpha$ emission found by \cite{Gottgens2019} for $\mathrm{X12^{c}}$, together with the X-ray emission, corroborates the hypothesis that this system is characterized by intense chromospheric activity.

Among the other candidate ABs found by \cite{Kumawat2024} there are five counterparts that they called X21B, X23A, X23B, X23C and X26. Since these objects do not meet our criteria, we did not include them among our final list of high confidence counterparts. 

Finally, \cite{Kumawat2024} identified X4 and X24 as two background galaxies. We confirmed their identification of the same counterparts for both X4 and X24 sources within their uncertainty radius and included them due to their peculiar CMD position.

Of the remaining 18 high-confidence counterparts, ten are not present in the HUGS catalogue used by \cite{Kumawat2024}, namely $\mathrm{X2^{c}}$  $\mathrm{X6^{c}}$ $\mathrm{X7^{c}}$, $\mathrm{X17^{c}}$, $\mathrm{X18A^{c}}$, $\mathrm{X18B^{c}}$, $\mathrm{X19^{c}}$, $\mathrm{X21B^{c}}$, $\mathrm{X25A^{c}}$ and $\mathrm{X33^{c}}$. Of the remaining eight candidates, $\mathrm{X5^{c}}$, $\mathrm{X16B^{c}}$, and $\mathrm{X23^{c}}$ are included in our list due to their variability, while $\mathrm{X8^{c}}$, $\mathrm{X13^{c}}$, and $\mathrm{X25B^{c}}$ exhibit H$\alpha$ emission. Lastly, we included $\mathrm{X24B^{c}}$ and $\mathrm{X28^{c}}$ due to their peculiar positions in the CMDs. Finally, it is worth noticing that \cite{Kumawat2024} did not find any CV candidate, although they calculated that some CV systems are expected to be present in NGC 362, as suggested by extrapolation of the stellar encounter rate from 47 Tucanae. They attribute this aspect to the optical crowding in the core of the cluster. 
On the other hand, our identification of several CV candidates ($
\mathrm{X2^{c}}$,$\mathrm{X5^{c}}$, $\mathrm{X6^{c}}$,$\mathrm{X17^{c}}
$, $\mathrm{X19^{c}}$, $\mathrm{X18A^{c}}$, $\mathrm{X18B^{c}}$, $
\mathrm{X25A^{c}}$and $\mathrm{X28^{c}}$) demonstrates that our photometry is more effective at identifying faint, blue stars compared to the HUGS photometry. We attribute this to the fact that the HUGS photometry relies on the F555W and F814W star catalogs for its master list. As shown in our CMDs, many faint blue stars that are clearly visible in bluer filters remain undetectable in F555W and F814W. This is either due to the presence of nearby bright stars in the redder filters or because the stars are too blue and faint to be captured in these redder bands.

In conclusion, the comprehensive methodology used in this study—integrating multi-wavelength, multi-epoch, and multi-instrument data—along with the synergy between optical and X-ray observations, has been essential in advancing our understanding on the nature of these systems. Moreover, future timing studies of MSPs in NGC 362 will be pivotal in confirming our candidates by linking our variability measures and period estimates to precise orbital periods as derived from radio observations. In addition, the proposed MSPs optical counterparts identified in this work will contribute to ongoing radio analyses with MeerKAT, facilitating the identification and comprehensive characterization of MSPs in NGC 362.

\begin{acknowledgements}
The research activities described in this paper were carried out with contribution of the Next Generation EU funds within the National Recovery and Resilience Plan (PNRR), Mission 4 - Education and Research, Component 2 - From Research to Business (M4C2), Investment Line 3.1 - Strengthening and creation of Research Infrastructures, Project IR0000034 – ``STILES - Strengthening the Italian Leadership in ELT and SKA''.
This work has been funded using resources from the INAF Large Grant 2022 “GCjewels” (P.I. Andrea Possenti) approved with the Presidential Decree 30/2022.
S.D. acknowledges the support of the Deutsche Forschungsgemeinschaft (DFG) through project DR 281/41-1. S.D. acknowledges the support by the BMBF from the ErUM program through grants 05A14MGA, 05A17MGA, 05A20MGA. C.O.H. is supported by NSERC Discovery Grant RGPIN-2023-04264. The authors thank the anonymous referee for their suggestions and comments, which improved the clarity of this work. G.E. is grateful to A. Della Croce for the useful discussions and comments.
\end{acknowledgements}
\bibliographystyle{aa} 
\bibliography{biblio} 

\begin{thebibliography}{104}
\expandafter\ifx\csname natexlab\endcsname\relax\def\natexlab#1{#1}\fi

\bibitem[{{Archibald} {et~al.}(2009){Archibald}, {Stairs}, {Ransom}, {Kaspi},
  {Kondratiev}, {Lorimer}, {McLaughlin}, {Boyles}, {Hessels}, {Lynch}, {van
  Leeuwen}, {Roberts}, {Jenet}, {Champion}, {Rosen}, {Barlow}, {Dunlap}, \&
  {Remillard}}]{Archibald2009}
{Archibald}, A.~M., {Stairs}, I.~H., {Ransom}, S.~M., {et~al.} 2009, Science,
  324, 1411

\bibitem[{{Astropy Collaboration} {et~al.}(2018){Astropy Collaboration},
  {Price-Whelan}, {Sip{\H{o}}cz}, {G{\"u}nther}, {Lim}, {Crawford}, {Conseil},
  {Shupe}, {Craig}, {Dencheva}, {Ginsburg}, {VanderPlas}, {Bradley},
  {P{\'e}rez-Su{\'a}rez}, {de Val-Borro}, {Aldcroft}, {Cruz}, {Robitaille},
  {Tollerud}, {Ardelean}, {Babej}, {Bach}, {Bachetti}, {Bakanov}, {Bamford},
  {Barentsen}, {Barmby}, {Baumbach}, {Berry}, {Biscani}, {Boquien}, {Bostroem},
  {Bouma}, {Brammer}, {Bray}, {Breytenbach}, {Buddelmeijer}, {Burke},
  {Calderone}, {Cano Rodr{\'\i}guez}, {Cara}, {Cardoso}, {Cheedella}, {Copin},
  {Corrales}, {Crichton}, {D'Avella}, {Deil}, {Depagne}, {Dietrich}, {Donath},
  {Droettboom}, {Earl}, {Erben}, {Fabbro}, {Ferreira}, {Finethy}, {Fox},
  {Garrison}, {Gibbons}, {Goldstein}, {Gommers}, {Greco}, {Greenfield},
  {Groener}, {Grollier}, {Hagen}, {Hirst}, {Homeier}, {Horton}, {Hosseinzadeh},
  {Hu}, {Hunkeler}, {Ivezi{\'c}}, {Jain}, {Jenness}, {Kanarek}, {Kendrew},
  {Kern}, {Kerzendorf}, {Khvalko}, {King}, {Kirkby}, {Kulkarni}, {Kumar},
  {Lee}, {Lenz}, {Littlefair}, {Ma}, {Macleod}, {Mastropietro}, {McCully},
  {Montagnac}, {Morris}, {Mueller}, {Mumford}, {Muna}, {Murphy}, {Nelson},
  {Nguyen}, {Ninan}, {N{\"o}the}, {Ogaz}, {Oh}, {Parejko}, {Parley}, {Pascual},
  {Patil}, {Patil}, {Plunkett}, {Prochaska}, {Rastogi}, {Reddy Janga},
  {Sabater}, {Sakurikar}, {Seifert}, {Sherbert}, {Sherwood-Taylor}, {Shih},
  {Sick}, {Silbiger}, {Singanamalla}, {Singer}, {Sladen}, {Sooley},
  {Sornarajah}, {Streicher}, {Teuben}, {Thomas}, {Tremblay}, {Turner},
  {Terr{\'o}n}, {van Kerkwijk}, {de la Vega}, {Watkins}, {Weaver}, {Whitmore},
  {Woillez}, {Zabalza}, \& {Astropy Contributors}}]{Price-Whelan2018A}
{Astropy Collaboration}, {Price-Whelan}, A.~M., {Sip{\H{o}}cz}, B.~M., {et~al.}
  2018, \aj, 156, 123

\bibitem[{{Astropy Collaboration} {et~al.}(2013){Astropy Collaboration},
  {Robitaille}, {Tollerud}, {Greenfield}, {Droettboom}, {Bray}, {Aldcroft},
  {Davis}, {Ginsburg}, {Price-Whelan}, {Kerzendorf}, {Conley}, {Crighton},
  {Barbary}, {Muna}, {Ferguson}, {Grollier}, {Parikh}, {Nair}, {Unther},
  {Deil}, {Woillez}, {Conseil}, {Kramer}, {Turner}, {Singer}, {Fox}, {Weaver},
  {Zabalza}, {Edwards}, {Azalee Bostroem}, {Burke}, {Casey}, {Crawford},
  {Dencheva}, {Ely}, {Jenness}, {Labrie}, {Lim}, {Pierfederici}, {Pontzen},
  {Ptak}, {Refsdal}, {Servillat}, \& {Streicher}}]{astropy2013}
{Astropy Collaboration}, {Robitaille}, T.~P., {Tollerud}, E.~J., {et~al.} 2013,
  \aap, 558, A33

\bibitem[{{Bacon} {et~al.}(2010){Bacon}, {Accardo}, {Adjali}, {Anwand},
  {Bauer}, {Biswas}, {Blaizot}, {Boudon}, {Brau-Nogue}, {Brinchmann},
  {Caillier}, {Capoani}, {Carollo}, {Contini}, {Couderc}, {Daguis{\'e}},
  {Deiries}, {Delabre}, {Dreizler}, {Dubois}, {Dupieux}, {Dupuy}, {Emsellem},
  {Fechner}, {Fleischmann}, {Fran{\c{c}}ois}, {Gallou}, {Gharsa}, {Glindemann},
  {Gojak}, {Guiderdoni}, {Hansali}, {Hahn}, {Jarno}, {Kelz}, {Koehler},
  {Kosmalski}, {Laurent}, {Le Floch}, {Lilly}, {Lizon}, {Loupias}, {Manescau},
  {Monstein}, {Nicklas}, {Olaya}, {Pares}, {Pasquini}, {P{\'e}contal-Rousset},
  {Pell{\'o}}, {Petit}, {Popow}, {Reiss}, {Remillieux}, {Renault}, {Roth},
  {Rupprecht}, {Serre}, {Schaye}, {Soucail}, {Steinmetz}, {Streicher}, {Stuik},
  {Valentin}, {Vernet}, {Weilbacher}, {Wisotzki}, \& {Yerle}}]{Bacon2010}
{Bacon}, R., {Accardo}, M., {Adjali}, L., {et~al.} 2010, in Society of
  Photo-Optical Instrumentation Engineers (SPIE) Conference Series, Vol. 7735,
  Ground-based and Airborne Instrumentation for Astronomy III, ed. I.~S.
  {McLean}, S.~K. {Ramsay}, \& H.~{Takami}, 773508

\bibitem[{{Bailyn}(1995)}]{Bailyn1995}
{Bailyn}, C.~D. 1995, \araa, 33, 133

\bibitem[{{Baluev}(2008)}]{Baluev2008}
{Baluev}, R.~V. 2008, \mnras, 385, 1279

\bibitem[{{Bassa} {et~al.}(2004){Bassa}, {Pooley}, {Homer}, {Verbunt},
  {Gaensler}, {Lewin}, {Anderson}, {Margon}, {Kaspi}, \& {van der
  Klis}}]{Bassa2004}
{Bassa}, C., {Pooley}, D., {Homer}, L., {et~al.} 2004, \apj, 609, 755

\bibitem[{{Bassa} {et~al.}(2005){Bassa}, {Pooley}, {Homer}, {Verbunt},
  {Gaensler}, {Lewin}, {Anderson}, {Margon}, {Kaspi}, \& {van der
  Klis}}]{Bassa2005}
{Bassa}, C., {Pooley}, D., {Homer}, L., {et~al.} 2005, \apj, 619, 1189

\bibitem[{{Baumgardt} {et~al.}(2020){Baumgardt}, {Sollima}, \&
  {Hilker}}]{Baumgardt2020}
{Baumgardt}, H., {Sollima}, A., \& {Hilker}, M. 2020, \pasa, 37, e046

\bibitem[{{Beccari} {et~al.}(2014){Beccari}, {De Marchi}, {Panagia}, \&
  {Pasquini}}]{Beccari2013}
{Beccari}, G., {De Marchi}, G., {Panagia}, N., \& {Pasquini}, L. 2014, \mnras,
  437, 2621

\bibitem[{{Bellazzini} {et~al.}(1995){Bellazzini}, {Pasquali}, {Federici},
  {Ferraro}, \& {Pecci}}]{Bellanzini1995}
{Bellazzini}, M., {Pasquali}, A., {Federici}, L., {Ferraro}, F.~R., \& {Pecci},
  F.~F. 1995, \apj, 439, 687

\bibitem[{{Bellini} \& {Bedin}(2009)}]{Bellini2009}
{Bellini}, A. \& {Bedin}, L.~R. 2009, \pasp, 121, 1419

\bibitem[{{Benvenuto} {et~al.}(2014){Benvenuto}, {De Vito}, \&
  {Horvath}}]{Benvenuto2014}
{Benvenuto}, O.~G., {De Vito}, M.~A., \& {Horvath}, J.~E. 2014, \apjl, 786, L7

\bibitem[{{Benvenuto} {et~al.}(2015){Benvenuto}, {De Vito}, \&
  {Horvath}}]{Benvenuto2015}
{Benvenuto}, O.~G., {De Vito}, M.~A., \& {Horvath}, J.~E. 2015, \apj, 798, 44

\bibitem[{{Bianchini} {et~al.}(2018){Bianchini}, {Webb}, {Sills}, \&
  {Vesperini}}]{Bianchini2018}
{Bianchini}, P., {Webb}, J.~J., {Sills}, A., \& {Vesperini}, E. 2018, \mnras,
  475, L96

\bibitem[{{Bogdanov} {et~al.}(2006){Bogdanov}, {Grindlay}, {Heinke}, {Camilo},
  {Freire}, \& {Becker}}]{Bogdanov2006}
{Bogdanov}, S., {Grindlay}, J.~E., {Heinke}, C.~O., {et~al.} 2006, \apj, 646,
  1104

\bibitem[{{Bogdanov} {et~al.}(2010){Bogdanov}, {van den Berg}, {Heinke},
  {Cohn}, {Lugger}, \& {Grindlay}}]{Bogdanov2010}
{Bogdanov}, S., {van den Berg}, M., {Heinke}, C.~O., {et~al.} 2010, \apj, 709,
  241

\bibitem[{{Bond} {et~al.}(2002){Bond}, {White}, {Becker}, \&
  {O'Brien}}]{Bond2002}
{Bond}, H.~E., {White}, R.~L., {Becker}, R.~H., \& {O'Brien}, M.~S. 2002,
  \pasp, 114, 1359

\bibitem[{{Cadelano} {et~al.}(2020){Cadelano}, {Chen}, {Pallanca}, {Istrate},
  {Ferraro}, {Lanzoni}, {Freire}, \& {Salaris}}]{Cadelano2020}
{Cadelano}, M., {Chen}, J., {Pallanca}, C., {et~al.} 2020, \apj, 905, 63

\bibitem[{{Cadelano} {et~al.}(2019){Cadelano}, {Ferraro}, {Istrate},
  {Pallanca}, {Lanzoni}, \& {Freire}}]{Cadelano2019}
{Cadelano}, M., {Ferraro}, F.~R., {Istrate}, A.~G., {et~al.} 2019, \apj, 875,
  25

\bibitem[{{Cadelano} {et~al.}(2015){Cadelano}, {Pallanca}, {Ferraro}, {Stairs},
  {Ransom}, {Dalessandro}, {Lanzoni}, {Hessels}, \& {Freire}}]{Cadelano2015}
{Cadelano}, M., {Pallanca}, C., {Ferraro}, F.~R., {et~al.} 2015, \apj, 807, 91

\bibitem[{{Carretta} {et~al.}(2009){Carretta}, {Bragaglia}, {Gratton},
  {D'Orazi}, \& {Lucatello}}]{Carretta2009}
{Carretta}, E., {Bragaglia}, A., {Gratton}, R., {D'Orazi}, V., \& {Lucatello},
  S. 2009, \aap, 508, 695

\bibitem[{{Chatterjee} {et~al.}(2007){Chatterjee}, {Gaensler}, {Melatos},
  {Brisken}, \& {Stappers}}]{Chatterjee2007}
{Chatterjee}, S., {Gaensler}, B.~M., {Melatos}, A., {Brisken}, W.~F., \&
  {Stappers}, B.~W. 2007, \apj, 670, 1301

\bibitem[{{Chen} {et~al.}(2013){Chen}, {Chen}, {Tauris}, \& {Han}}]{Chen2013}
{Chen}, H.-L., {Chen}, X., {Tauris}, T.~M., \& {Han}, Z. 2013, \apj, 775, 27

\bibitem[{{Clark}(1975)}]{Clark1975}
{Clark}, G.~W. 1975, \apjl, 199, L143

\bibitem[{{Cocozza} {et~al.}(2008){Cocozza}, {Ferraro}, {Possenti}, {Beccari},
  {Lanzoni}, {Ransom}, {Rood}, \& {D'Amico}}]{Cocozza2008}
{Cocozza}, G., {Ferraro}, F.~R., {Possenti}, A., {et~al.} 2008, \apjl, 679,
  L105

\bibitem[{{Cohn} {et~al.}(2010){Cohn}, {Lugger}, {Couch}, {Anderson}, {Cool},
  {van den Berg}, {Bogdanov}, {Heinke}, \& {Grindlay}}]{Cohn2010}
{Cohn}, H.~N., {Lugger}, P.~M., {Couch}, S.~M., {et~al.} 2010, \apj, 722, 20

\bibitem[{{Cool} {et~al.}(1995){Cool}, {Grindlay}, {Cohn}, {Lugger}, \&
  {Slavin}}]{Cool1995}
{Cool}, A.~M., {Grindlay}, J.~E., {Cohn}, H.~N., {Lugger}, P.~M., \& {Slavin},
  S.~D. 1995, \apj, 439, 695

\bibitem[{{Dalessandro} {et~al.}(2013){Dalessandro}, {Ferraro}, {Massari},
  {Lanzoni}, {Miocchi}, {Beccari}, {Bellini}, {Sills}, {Sigurdsson},
  {Mucciarelli}, \& {Lovisi}}]{Dalessandro2013}
{Dalessandro}, E., {Ferraro}, F.~R., {Massari}, D., {et~al.} 2013, \apj, 778,
  135

\bibitem[{{Dalessandro} {et~al.}(2008{\natexlab{a}}){Dalessandro}, {Lanzoni},
  {Ferraro}, {Rood}, {Milone}, {Piotto}, \& {Valenti}}]{dalessandro2008b}
{Dalessandro}, E., {Lanzoni}, B., {Ferraro}, F.~R., {et~al.}
  2008{\natexlab{a}}, \apj, 677, 1069

\bibitem[{{Dalessandro} {et~al.}(2008{\natexlab{b}}){Dalessandro}, {Lanzoni},
  {Ferraro}, {Vespe}, {Bellazzini}, \& {Rood}}]{dalessandro2008a}
{Dalessandro}, E., {Lanzoni}, B., {Ferraro}, F.~R., {et~al.}
  2008{\natexlab{b}}, \apj, 681, 311

\bibitem[{{De Marchi} {et~al.}(2010){De Marchi}, {Panagia}, \&
  {Romaniello}}]{DeMarchi2010}
{De Marchi}, G., {Panagia}, N., \& {Romaniello}, M. 2010, \apj, 715, 1

\bibitem[{{de Martino} {et~al.}(2015){de Martino}, {Papitto}, {Belloni},
  {Burgay}, {De Ona Wilhelmi}, {Li}, {Pellizzoni}, {Possenti}, {Rea}, \&
  {Torres}}]{DeMartino2015}
{de Martino}, D., {Papitto}, A., {Belloni}, T., {et~al.} 2015, \mnras, 454,
  2190

\bibitem[{{De Vito} {et~al.}(2020){De Vito}, {Benvenuto}, \&
  {Horvath}}]{Devito2020}
{De Vito}, M.~A., {Benvenuto}, O.~G., \& {Horvath}, J.~E. 2020, \mnras, 493,
  2171

\bibitem[{{Dempsey} {et~al.}(1997){Dempsey}, {Linsky}, {Fleming}, \&
  {Schmitt}}]{Dempsey1997}
{Dempsey}, R.~C., {Linsky}, J.~L., {Fleming}, T.~A., \& {Schmitt}, J.~H.~M.~M.
  1997, \apj, 478, 358

\bibitem[{{Dickson} {et~al.}(2024){Dickson}, {Smith}, {H{\'e}nault-Brunet},
  {Gieles}, \& {Baumgardt}}]{Dickson2024}
{Dickson}, N., {Smith}, P.~J., {H{\'e}nault-Brunet}, V., {Gieles}, M., \&
  {Baumgardt}, H. 2024, \mnras, 529, 331

\bibitem[{{Dotter} {et~al.}(2010){Dotter}, {Sarajedini}, {Anderson},
  {Aparicio}, {Bedin}, {Chaboyer}, {Majewski}, {Mar{\'\i}n-Franch}, {Milone},
  {Paust}, {Piotto}, {Reid}, {Rosenberg}, \& {Siegel}}]{Dotter2010}
{Dotter}, A., {Sarajedini}, A., {Anderson}, J., {et~al.} 2010, \apj, 708, 698

\bibitem[{{Edmonds} {et~al.}(2002){Edmonds}, {Gilliland}, {Camilo}, {Heinke},
  \& {Grindlay}}]{Edmonds2002}
{Edmonds}, P.~D., {Gilliland}, R.~L., {Camilo}, F., {Heinke}, C.~O., \&
  {Grindlay}, J.~E. 2002, \apj, 579, 741

\bibitem[{{Edmonds} {et~al.}(2003){Edmonds}, {Gilliland}, {Heinke}, \&
  {Grindlay}}]{Edmonds2003}
{Edmonds}, P.~D., {Gilliland}, R.~L., {Heinke}, C.~O., \& {Grindlay}, J.~E.
  2003, \apj, 596, 1177

\bibitem[{{Ferraro} {et~al.}(2009){Ferraro}, {Beccari}, {Dalessandro},
  {Lanzoni}, {Sills}, {Rood}, {Pecci}, {Karakas}, {Miocchi}, \&
  {Bovinelli}}]{Ferraro2009}
{Ferraro}, F.~R., {Beccari}, G., {Dalessandro}, E., {et~al.} 2009, \nat, 462,
  1028

\bibitem[{{Ferraro} {et~al.}(2012){Ferraro}, {Lanzoni}, {Dalessandro},
  {Beccari}, {Pasquato}, {Miocchi}, {Rood}, {Sigurdsson}, {Sills}, {Vesperini},
  {Mapelli}, {Contreras}, {Sanna}, \& {Mucciarelli}}]{Ferraro2012}
{Ferraro}, F.~R., {Lanzoni}, B., {Dalessandro}, E., {et~al.} 2012, \nat, 492,
  393

\bibitem[{{Ferraro} {et~al.}(2015){Ferraro}, {Pallanca}, {Lanzoni}, {Cadelano},
  {Massari}, {Dalessandro}, \& {Mucciarelli}}]{Ferraro2015b}
{Ferraro}, F.~R., {Pallanca}, C., {Lanzoni}, B., {et~al.} 2015, \apjl, 807, L1

\bibitem[{{Ferraro} {et~al.}(2001){Ferraro}, {Possenti}, {D'Amico}, \&
  {Sabbi}}]{Ferraro2001}
{Ferraro}, F.~R., {Possenti}, A., {D'Amico}, N., \& {Sabbi}, E. 2001, \apjl,
  561, L93

\bibitem[{{Freire} {et~al.}(2008){Freire}, {Ransom}, {B{\'e}gin}, {Stairs},
  {Hessels}, {Frey}, \& {Camilo}}]{Freire2008}
{Freire}, P. C.~C., {Ransom}, S.~M., {B{\'e}gin}, S., {et~al.} 2008, \apj, 675,
  670

\bibitem[{{Gaia Collaboration} {et~al.}(2023){Gaia Collaboration}, {Vallenari},
  {Brown}, {Prusti}, {de Bruijne}, {Arenou}, {Babusiaux}, {Biermann},
  {Creevey}, {Ducourant}, {Evans}, {Eyer}, {Guerra}, {Hutton}, {Jordi},
  {Klioner}, {Lammers}, {Lindegren}, {Luri}, {Mignard}, {Panem}, {Pourbaix},
  {Randich}, {Sartoretti}, {Soubiran}, {Tanga}, {Walton}, {Bailer-Jones},
  {Bastian}, {Drimmel}, {Jansen}, {Katz}, {Lattanzi}, {van Leeuwen}, {Bakker},
  {Cacciari}, {Casta{\~n}eda}, {De Angeli}, {Fabricius}, {Fouesneau},
  {Fr{\'e}mat}, {Galluccio}, {Guerrier}, {Heiter}, {Masana}, {Messineo},
  {Mowlavi}, {Nicolas}, {Nienartowicz}, {Pailler}, {Panuzzo}, {Riclet}, {Roux},
  {Seabroke}, {Sordo}, {Th{\'e}venin}, {Gracia-Abril}, {Portell}, {Teyssier},
  {Altmann}, {Andrae}, {Audard}, {Bellas-Velidis}, {Benson}, {Berthier},
  {Blomme}, {Burgess}, {Busonero}, {Busso}, {C{\'a}novas}, {Carry}, {Cellino},
  {Cheek}, {Clementini}, {Damerdji}, {Davidson}, {de Teodoro}, {Nu{\~n}ez
  Campos}, {Delchambre}, {Dell'Oro}, {Esquej}, {Fern{\'a}ndez-Hern{\'a}ndez},
  {Fraile}, {Garabato}, {Garc{\'\i}a-Lario}, {Gosset}, {Haigron}, {Halbwachs},
  {Hambly}, {Harrison}, {Hern{\'a}ndez}, {Hestroffer}, {Hodgkin}, {Holl},
  {Jan{\ss}en}, {Jevardat de Fombelle}, {Jordan}, {Krone-Martins}, {Lanzafame},
  {L{\"o}ffler}, {Marchal}, {Marrese}, {Moitinho}, {Muinonen}, {Osborne},
  {Pancino}, {Pauwels}, {Recio-Blanco}, {Reyl{\'e}}, {Riello}, {Rimoldini},
  {Roegiers}, {Rybizki}, {Sarro}, {Siopis}, {Smith}, {Sozzetti}, {Utrilla},
  {van Leeuwen}, {Abbas}, {{\'A}brah{\'a}m}, {Abreu Aramburu}, {Aerts},
  {Aguado}, {Ajaj}, {Aldea-Montero}, {Altavilla}, {{\'A}lvarez}, {Alves},
  {Anders}, {Anderson}, {Anglada Varela}, {Antoja}, {Baines}, {Baker},
  {Balaguer-N{\'u}{\~n}ez}, {Balbinot}, {Balog}, {Barache}, {Barbato},
  {Barros}, {Barstow}, {Bartolom{\'e}}, {Bassilana}, {Bauchet}, {Becciani},
  {Bellazzini}, {Berihuete}, {Bernet}, {Bertone}, {Bianchi}, {Binnenfeld},
  {Blanco-Cuaresma}, {Blazere}, {Boch}, {Bombrun}, {Bossini}, {Bouquillon},
  {Bragaglia}, {Bramante}, {Breedt}, {Bressan}, {Brouillet}, {Brugaletta},
  {Bucciarelli}, {Burlacu}, {Butkevich}, {Buzzi}, {Caffau}, {Cancelliere},
  {Cantat-Gaudin}, {Carballo}, {Carlucci}, {Carnerero}, {Carrasco},
  {Casamiquela}, {Castellani}, {Castro-Ginard}, {Chaoul}, {Charlot}, {Chemin},
  {Chiaramida}, {Chiavassa}, {Chornay}, {Comoretto}, {Contursi}, {Cooper},
  {Cornez}, {Cowell}, {Crifo}, {Cropper}, {Crosta}, {Crowley}, {Dafonte},
  {Dapergolas}, {David}, {David}, {de Laverny}, {De Luise}, {De March}, {De
  Ridder}, {de Souza}, {de Torres}, {del Peloso}, {del Pozo}, {Delbo},
  {Delgado}, {Delisle}, {Demouchy}, {Dharmawardena}, {Di Matteo}, {Diakite},
  {Diener}, {Distefano}, {Dolding}, {Edvardsson}, {Enke}, {Fabre}, {Fabrizio},
  {Faigler}, {Fedorets}, {Fernique}, {Fienga}, {Figueras}, {Fournier},
  {Fouron}, {Fragkoudi}, {Gai}, {Garcia-Gutierrez}, {Garcia-Reinaldos},
  {Garc{\'\i}a-Torres}, {Garofalo}, {Gavel}, {Gavras}, {Gerlach}, {Geyer},
  {Giacobbe}, {Gilmore}, {Girona}, {Giuffrida}, {Gomel}, {Gomez},
  {Gonz{\'a}lez-N{\'u}{\~n}ez}, {Gonz{\'a}lez-Santamar{\'\i}a},
  {Gonz{\'a}lez-Vidal}, {Granvik}, {Guillout}, {Guiraud},
  {Guti{\'e}rrez-S{\'a}nchez}, {Guy}, {Hatzidimitriou}, {Hauser}, {Haywood},
  {Helmer}, {Helmi}, {Sarmiento}, {Hidalgo}, {Hilger}, {H{\l}adczuk}, {Hobbs},
  {Holland}, {Huckle}, {Jardine}, {Jasniewicz}, {Jean-Antoine Piccolo},
  {Jim{\'e}nez-Arranz}, {Jorissen}, {Juaristi Campillo}, {Julbe}, {Karbevska},
  {Kervella}, {Khanna}, {Kontizas}, {Kordopatis}, {Korn}, {K{\'o}sp{\'a}l},
  {Kostrzewa-Rutkowska}, {Kruszy{\'n}ska}, {Kun}, {Laizeau}, {Lambert},
  {Lanza}, {Lasne}, {Le Campion}, {Lebreton}, {Lebzelter}, {Leccia}, {Leclerc},
  {Lecoeur-Taibi}, {Liao}, {Licata}, {Lindstr{\o}m}, {Lister}, {Livanou},
  {Lobel}, {Lorca}, {Loup}, {Madrero Pardo}, {Magdaleno Romeo}, {Managau},
  {Mann}, {Manteiga}, {Marchant}, {Marconi}, {Marcos}, {Marcos Santos},
  {Mar{\'\i}n Pina}, {Marinoni}, {Marocco}, {Marshall}, {Martin Polo},
  {Mart{\'\i}n-Fleitas}, {Marton}, {Mary}, {Masip}, {Massari},
  {Mastrobuono-Battisti}, {Mazeh}, {McMillan}, {Messina}, {Michalik}, {Millar},
  {Mints}, {Molina}, {Molinaro}, {Moln{\'a}r}, {Monari}, {Mongui{\'o}},
  {Montegriffo}, {Montero}, {Mor}, {Mora}, {Morbidelli}, {Morel}, {Morris},
  {Muraveva}, {Murphy}, {Musella}, {Nagy}, {Noval}, {Oca{\~n}a}, {Ogden},
  {Ordenovic}, {Osinde}, {Pagani}, {Pagano}, {Palaversa}, {Palicio},
  {Pallas-Quintela}, {Panahi}, {Payne-Wardenaar}, {Pe{\~n}alosa Esteller},
  {Penttil{\"a}}, {Pichon}, {Piersimoni}, {Pineau}, {Plachy}, {Plum}, {Poggio},
  {Pr{\v{s}}a}, {Pulone}, {Racero}, {Ragaini}, {Rainer}, {Raiteri}, {Rambaux},
  {Ramos}, {Ramos-Lerate}, {Re Fiorentin}, {Regibo}, {Richards}, {Rios Diaz},
  {Ripepi}, {Riva}, {Rix}, {Rixon}, {Robichon}, {Robin}, {Robin}, {Roelens},
  {Rogues}, {Rohrbasser}, {Romero-G{\'o}mez}, {Rowell}, {Royer}, {Ruz Mieres},
  {Rybicki}, {Sadowski}, {S{\'a}ez N{\'u}{\~n}ez}, {Sagrist{\`a} Sell{\'e}s},
  {Sahlmann}, {Salguero}, {Samaras}, {Sanchez Gimenez}, {Sanna},
  {Santove{\~n}a}, {Sarasso}, {Schultheis}, {Sciacca}, {Segol}, {Segovia},
  {S{\'e}gransan}, {Semeux}, {Shahaf}, {Siddiqui}, {Siebert}, {Siltala},
  {Silvelo}, {Slezak}, {Slezak}, {Smart}, {Snaith}, {Solano}, {Solitro},
  {Souami}, {Souchay}, {Spagna}, {Spina}, {Spoto}, {Steele},
  {Steidelm{\"u}ller}, {Stephenson}, {S{\"u}veges}, {Surdej}, {Szabados},
  {Szegedi-Elek}, {Taris}, {Taylor}, {Teixeira}, {Tolomei}, {Tonello}, {Torra},
  {Torra}, {Torralba Elipe}, {Trabucchi}, {Tsounis}, {Turon}, {Ulla}, {Unger},
  {Vaillant}, {van Dillen}, {van Reeven}, {Vanel}, {Vecchiato}, {Viala},
  {Vicente}, {Voutsinas}, {Weiler}, {Wevers}, {Wyrzykowski}, {Yoldas}, {Yvard},
  {Zhao}, {Zorec}, {Zucker}, \& {Zwitter}}]{Gaia2023}
{Gaia Collaboration}, {Vallenari}, A., {Brown}, A.~G.~A., {et~al.} 2023, \aap,
  674, A1

\bibitem[{{Geller} {et~al.}(2017){Geller}, {Leiner}, {Bellini}, {Gleisinger},
  {Haggard}, {Kamann}, {Leigh}, {Mathieu}, {Sills}, {Watkins}, \&
  {Zurek}}]{Geller2017}
{Geller}, A.~M., {Leiner}, E.~M., {Bellini}, A., {et~al.} 2017, \apj, 840, 66

\bibitem[{{Giesers} {et~al.}(2019){Giesers}, {Kamann}, {Dreizler}, {Husser},
  {Askar}, {G{\"o}ttgens}, {Brinchmann}, {Latour}, {Weilbacher}, {Wendt}, \&
  {Roth}}]{Giesers2019}
{Giesers}, B., {Kamann}, S., {Dreizler}, S., {et~al.} 2019, \aap, 632, A3

\bibitem[{{G{\"o}ttgens} {et~al.}(2019){G{\"o}ttgens}, {Husser}, {Kamann},
  {Dreizler}, {Giesers}, {Kollatschny}, {Weilbacher}, {Roth}, \&
  {Wendt}}]{Gottgens2019}
{G{\"o}ttgens}, F., {Husser}, T.-O., {Kamann}, S., {et~al.} 2019, \aap, 631,
  A118

\bibitem[{{G{\"u}del}(2004)}]{Gudel2004}
{G{\"u}del}, M. 2004, \aapr, 12, 71

\bibitem[{{Harris}(1996)}]{Harris1996}
{Harris}, W.~E. 1996, \aj, 112, 1487

\bibitem[{{Heinke} {et~al.}(2005){Heinke}, {Grindlay}, {Edmonds}, {Cohn},
  {Lugger}, {Camilo}, {Bogdanov}, \& {Freire}}]{Heinke2005}
{Heinke}, C.~O., {Grindlay}, J.~E., {Edmonds}, P.~D., {et~al.} 2005, \apj, 625,
  796

\bibitem[{{Heinke} {et~al.}(2003){Heinke}, {Grindlay}, {Lugger}, {Cohn},
  {Edmonds}, {Lloyd}, \& {Cool}}]{Heinke2003}
{Heinke}, C.~O., {Grindlay}, J.~E., {Lugger}, P.~M., {et~al.} 2003, \apj, 598,
  501

\bibitem[{{Hertz} \& {Grindlay}(1983)}]{Hertz1983}
{Hertz}, P. \& {Grindlay}, J.~E. 1983, \apjl, 267, L83

\bibitem[{{Hui} {et~al.}(2010){Hui}, {Cheng}, \& {Taam}}]{Hui2010}
{Hui}, C.~Y., {Cheng}, K.~S., \& {Taam}, R.~E. 2010, \apj, 714, 1149

\bibitem[{{Ivanova} {et~al.}(2006){Ivanova}, {Heinke}, {Rasio}, {Taam},
  {Belczynski}, \& {Fregeau}}]{Ivanova2006}
{Ivanova}, N., {Heinke}, C.~O., {Rasio}, F.~A., {et~al.} 2006, \mnras, 372,
  1043

\bibitem[{{Kamann} {et~al.}(2018){Kamann}, {Husser}, {Dreizler}, {Emsellem},
  {Weilbacher}, {Martens}, {Bacon}, {den Brok}, {Giesers}, {Krajnovi{\'c}},
  {Roth}, {Wendt}, \& {Wisotzki}}]{Kamann2018}
{Kamann}, S., {Husser}, T.~O., {Dreizler}, S., {et~al.} 2018, \mnras, 473, 5591

\bibitem[{{K{\i}z{\i}ltan} {et~al.}(2017){K{\i}z{\i}ltan}, {Baumgardt}, \&
  {Loeb}}]{Kiziltan2017}
{K{\i}z{\i}ltan}, B., {Baumgardt}, H., \& {Loeb}, A. 2017, \nat, 542, 203

\bibitem[{{Knigge}(2012)}]{Knigge2012}
{Knigge}, C. 2012, \memsai, 83, 549

\bibitem[{{Knigge} {et~al.}(2003){Knigge}, {Zurek}, {Shara}, {Long}, \&
  {Gilliland}}]{Knigge2003}
{Knigge}, C., {Zurek}, D.~R., {Shara}, M.~M., {Long}, K.~S., \& {Gilliland},
  R.~L. 2003, \apj, 599, 1320

\bibitem[{{Kumawat} {et~al.}(2024){Kumawat}, {Heinke}, {Cohn}, \&
  {Lugger}}]{Kumawat2024}
{Kumawat}, G., {Heinke}, C.~O., {Cohn}, H.~N., \& {Lugger}, P.~M. 2024, \mnras,
  530, 82

\bibitem[{{Lanzoni} {et~al.}(2016){Lanzoni}, {Ferraro}, {Alessandrini},
  {Dalessandro}, {Vesperini}, \& {Raso}}]{Lanzoni2016}
{Lanzoni}, B., {Ferraro}, F.~R., {Alessandrini}, E., {et~al.} 2016, \apjl, 833,
  L29

\bibitem[{{Leiner} {et~al.}(2017){Leiner}, {Mathieu}, \& {Geller}}]{Leiner2017}
{Leiner}, E., {Mathieu}, R.~D., \& {Geller}, A.~M. 2017, \apj, 840, 67

\bibitem[{{Libralato} {et~al.}(2018){Libralato}, {Bellini}, {van der Marel},
  {Anderson}, {Watkins}, {Piotto}, {Ferraro}, {Nardiello}, \&
  {Vesperini}}]{Libralato2018}
{Libralato}, M., {Bellini}, A., {van der Marel}, R.~P., {et~al.} 2018, \apj,
  861, 99

\bibitem[{{Libralato} {et~al.}(2022){Libralato}, {Bellini}, {Vesperini},
  {Piotto}, {Milone}, {van der Marel}, {Anderson}, {Aparicio}, {Barbuy},
  {Bedin}, {Borsato}, {Cassisi}, {Dalessandro}, {Ferraro}, {King}, {Lanzoni},
  {Nardiello}, {Ortolani}, {Sarajedini}, \& {Sohn}}]{Libralato2022}
{Libralato}, M., {Bellini}, A., {Vesperini}, E., {et~al.} 2022, \apj, 934, 150

\bibitem[{{Lomb}(1976)}]{Lomb1976}
{Lomb}, N.~R. 1976, \apss, 39, 447

\bibitem[{{Meurer} {et~al.}(2003){Meurer}, {Lindler}, {Blakeslee}, {Cox},
  {Martel}, {Tran}, {Bouwens}, {Ford}, {Clampin}, {Hartig}, {Sirianni}, \& {De
  Marchi}}]{meurer2004}
{Meurer}, G.~R., {Lindler}, D.~J., {Blakeslee}, J., {et~al.} 2003, in Society
  of Photo-Optical Instrumentation Engineers (SPIE) Conference Series, Vol.
  4854, Future EUV/UV and Visible Space Astrophysics Missions and
  Instrumentation., ed. J.~C. {Blades} \& O.~H.~W. {Siegmund}, 507--514

\bibitem[{{Modiano} {et~al.}(2020){Modiano}, {Parikh}, \&
  {Wijnands}}]{Modiano2020}
{Modiano}, D., {Parikh}, A.~S., \& {Wijnands}, R. 2020, \aap, 634, A132

\bibitem[{{Moffat}(1969)}]{Moffat1969}
{Moffat}, A.~F.~J. 1969, \aap, 3, 455

\bibitem[{{Nardiello} {et~al.}(2018){Nardiello}, {Libralato}, {Piotto},
  {Anderson}, {Bellini}, {Aparicio}, {Bedin}, {Cassisi}, {Granata}, {King},
  {Lucertini}, {Marino}, {Milone}, {Ortolani}, {Platais}, \& {van der
  Marel}}]{Nardiello2018}
{Nardiello}, D., {Libralato}, M., {Piotto}, G., {et~al.} 2018, \mnras, 481,
  3382

\bibitem[{{Negoro} {et~al.}(2016){Negoro}, {Kohama}, {Serino}, {Saito},
  {Takahashi}, {Miyoshi}, {Ozawa}, {Suwa}, {Asada}, {Fukushima}, {Eguchi},
  {Kawai}, {Kennea}, {Mihara}, {Morii}, {Nakahira}, {Ogawa}, {Sugawara},
  {Tomida}, {Ueno}, {Ishikawa}, {Isobe}, {Kawamuro}, {Kimura}, {Masumitsu},
  {Nakagawa}, {Nakajima}, {Sakamoto}, {Shidatsu}, {Sugizaki}, {Sugimoto},
  {Suzuki}, {Takagi}, {Tanaka}, {Tsuboi}, {Tsunemi}, {Ueda}, {Yamaoka},
  {Yamauchi}, {Yoshida}, \& {Matsuoka}}]{Negoro2016}
{Negoro}, H., {Kohama}, M., {Serino}, M., {et~al.} 2016, \pasj, 68, S1

\bibitem[{{Pallanca} {et~al.}(2017){Pallanca}, {Beccari}, {Ferraro},
  {Pasquini}, {Lanzoni}, \& {Mucciarelli}}]{Pallanca2017}
{Pallanca}, C., {Beccari}, G., {Ferraro}, F.~R., {et~al.} 2017, \apj, 845, 4

\bibitem[{{Pallanca} {et~al.}(2013){Pallanca}, {Dalessandro}, {Ferraro},
  {Lanzoni}, \& {Beccari}}]{Pallanca2013}
{Pallanca}, C., {Dalessandro}, E., {Ferraro}, F.~R., {Lanzoni}, B., \&
  {Beccari}, G. 2013, \apj, 773, 122

\bibitem[{{Pallanca} {et~al.}(2010){Pallanca}, {Dalessandro}, {Ferraro},
  {Lanzoni}, {Rood}, {Possenti}, {D'Amico}, {Freire}, {Stairs}, {Ransom}, \&
  {B{\'e}gin}}]{Pallanca2010}
{Pallanca}, C., {Dalessandro}, E., {Ferraro}, F.~R., {et~al.} 2010, \apj, 725,
  1165

\bibitem[{{Pallanca} {et~al.}(2014){Pallanca}, {Ransom}, {Ferraro},
  {Dalessandro}, {Lanzoni}, {Hessels}, {Stairs}, \& {Freire}}]{Pallanca2014}
{Pallanca}, C., {Ransom}, S.~M., {Ferraro}, F.~R., {et~al.} 2014, \apj, 795, 29

\bibitem[{{Papitto} {et~al.}(2013){Papitto}, {Bozzo}, {Ferrigno}, \&
  {Rea}}]{Papitto2013}
{Papitto}, A., {Bozzo}, E., {Ferrigno}, C., \& {Rea}, N. 2013, The Astronomer's
  Telegram, 5534, 1

\bibitem[{{Paresce} {et~al.}(1992){Paresce}, {de Marchi}, \&
  {Ferraro}}]{Paresce1992}
{Paresce}, F., {de Marchi}, G., \& {Ferraro}, F.~R. 1992, \nat, 360, 46

\bibitem[{{Penny}(1976)}]{Penny1976}
{Penny}, A.~J. 1976, {PhD thesis, Sussex University}

\bibitem[{{Pietrukowicz} {et~al.}(2008){Pietrukowicz}, {Kaluzny},
  {Schwarzenberg-Czerny}, {Thompson}, {Pych}, {Krzeminski}, \&
  {Mazur}}]{Pietrukowicz2008}
{Pietrukowicz}, P., {Kaluzny}, J., {Schwarzenberg-Czerny}, A., {et~al.} 2008,
  \mnras, 388, 1111

\bibitem[{{Piotto} {et~al.}(2015){Piotto}, {Milone}, {Bedin}, {Anderson},
  {King}, {Marino}, {Nardiello}, {Aparicio}, {Barbuy}, {Bellini}, {Brown},
  {Cassisi}, {Cool}, {Cunial}, {Dalessandro}, {D'Antona}, {Ferraro}, {Hidalgo},
  {Lanzoni}, {Monelli}, {Ortolani}, {Renzini}, {Salaris}, {Sarajedini}, {van
  der Marel}, {Vesperini}, \& {Zoccali}}]{Piotto2015}
{Piotto}, G., {Milone}, A.~P., {Bedin}, L.~R., {et~al.} 2015, \aj, 149, 91

\bibitem[{{Pooley} \& {Hut}(2006)}]{Pooley2006}
{Pooley}, D. \& {Hut}, P. 2006, \apjl, 646, L143

\bibitem[{{Pooley} {et~al.}(2002){Pooley}, {Lewin}, {Homer}, {Verbunt},
  {Anderson}, {Gaensler}, {Margon}, {Miller}, {Fox}, {Kaspi}, \& {van der
  Klis}}]{Pooley2002}
{Pooley}, D., {Lewin}, W. H.~G., {Homer}, L., {et~al.} 2002, \apj, 569, 405

\bibitem[{{Reynolds} {et~al.}(2007){Reynolds}, {Callanan}, {Fruchter},
  {Torres}, {Beer}, \& {Gibbons}}]{Reynolds2007}
{Reynolds}, M.~T., {Callanan}, P.~J., {Fruchter}, A.~S., {et~al.} 2007, \mnras,
  379, 1117

\bibitem[{{Roberts}(2013)}]{Roberts2013}
{Roberts}, M. S.~E. 2013, in Neutron Stars and Pulsars: Challenges and
  Opportunities after 80 years, ed. J.~{van Leeuwen}, Vol. 291, 127--132

\bibitem[{{Rozyczka} {et~al.}(2016){Rozyczka}, {Thompson}, {Narloch}, {Pych},
  \& {Schwarzenberg-Czerny}}]{Rozyczka2016}
{Rozyczka}, M., {Thompson}, I.~B., {Narloch}, W., {Pych}, W., \&
  {Schwarzenberg-Czerny}, A. 2016, \actaa, 66, 307

\bibitem[{{Rucinski}(2000)}]{Rucinski2000}
{Rucinski}, S.~M. 2000, \aj, 120, 319

\bibitem[{{Saito} {et~al.}(1997){Saito}, {Kawai}, {Kamae}, {Shibata}, {Dotani},
  \& {Kulkarni}}]{Saito1997}
{Saito}, Y., {Kawai}, N., {Kamae}, T., {et~al.} 1997, \apjl, 477, L37

\bibitem[{{Scargle}(1982)}]{Scargle1982}
{Scargle}, J.~D. 1982, \apj, 263, 835

\bibitem[{{Shara} {et~al.}(2005){Shara}, {Hinkley}, {Zurek}, {Knigge}, \&
  {Dieball}}]{Shara2005}
{Shara}, M.~M., {Hinkley}, S., {Zurek}, D.~R., {Knigge}, C., \& {Dieball}, A.
  2005, \aj, 130, 1829

\bibitem[{{Sirianni} {et~al.}(2005){Sirianni}, {Jee}, {Ben{\'\i}tez},
  {Blakeslee}, {Martel}, {Meurer}, {Clampin}, {De Marchi}, {Ford}, {Gilliland},
  {Hartig}, {Illingworth}, {Mack}, \& {McCann}}]{Sirianni2005}
{Sirianni}, M., {Jee}, M.~J., {Ben{\'\i}tez}, N., {et~al.} 2005, \pasp, 117,
  1049

\bibitem[{{Smedley} {et~al.}(2015){Smedley}, {Tout}, {Ferrario}, \&
  {Wickramasinghe}}]{Smedley2015}
{Smedley}, S.~L., {Tout}, C.~A., {Ferrario}, L., \& {Wickramasinghe}, D.~T.
  2015, \mnras, 446, 2540

\bibitem[{{Stetson}(1987)}]{Stetson1987}
{Stetson}, P.~B. 1987, \pasp, 99, 191

\bibitem[{{Stetson}(1994)}]{Stetson1994}
{Stetson}, P.~B. 1994, \pasp, 106, 250

\bibitem[{{Stetson}(1996)}]{Stetson1996}
{Stetson}, P.~B. 1996, \pasp, 108, 851

\bibitem[{{Tauris} \& {Savonije}(1999)}]{TaurisSavoije1999}
{Tauris}, T.~M. \& {Savonije}, G.~J. 1999, \aap, 350, 928

\bibitem[{{Turk} \& {Lorimer}(2013)}]{Turk2013}
{Turk}, P.~J. \& {Lorimer}, D.~R. 2013, \mnras, 436, 3720

\bibitem[{{van Dokkum}(2001)}]{VanDokkum2001}
{van Dokkum}, P.~G. 2001, \pasp, 113, 1420

\bibitem[{{Verbunt} {et~al.}(1996){Verbunt}, {Kuiper}, {Belloni}, {Johnston},
  {de Bruyn}, {Hermsen}, \& {van der Klis}}]{Verbunt1996}
{Verbunt}, F., {Kuiper}, L., {Belloni}, T., {et~al.} 1996, \aap, 311, L9

\bibitem[{{Verbunt} {et~al.}(2008){Verbunt}, {Pooley}, \&
  {Bassa}}]{Verbunt2008}
{Verbunt}, F., {Pooley}, D., \& {Bassa}, C. 2008, in Dynamical Evolution of
  Dense Stellar Systems, ed. E.~{Vesperini}, M.~{Giersz}, \& A.~{Sills}, Vol.
  246, 301--310

\bibitem[{{Wadiasingh} {et~al.}(2017){Wadiasingh}, {Harding}, {Venter},
  {B{\"o}ttcher}, \& {Baring}}]{Wadiasingh2017}
{Wadiasingh}, Z., {Harding}, A.~K., {Venter}, C., {B{\"o}ttcher}, M., \&
  {Baring}, M.~G. 2017, \apj, 839, 80

\bibitem[{{Witham} {et~al.}(2006){Witham}, {Knigge}, {G{\"a}nsicke},
  {Aungwerojwit}, {Corradi}, {Drew}, {Greimel}, {Groot}, {Morales-Rueda},
  {Rodriguez-Flores}, {Rodriguez-Gil}, \& {Steeghs}}]{Witham2006}
{Witham}, A.~R., {Knigge}, C., {G{\"a}nsicke}, B.~T., {et~al.} 2006, \mnras,
  369, 581

\bibitem[{{Zavlin} {et~al.}(2002){Zavlin}, {Pavlov}, {Sanwal}, {Manchester},
  {Tr{\"u}mper}, {Halpern}, \& {Becker}}]{Zavlin2002}
{Zavlin}, V.~E., {Pavlov}, G.~G., {Sanwal}, D., {et~al.} 2002, \apj, 569, 894

\bibitem[{{Zhang} {et~al.}(2023){Zhang}, {Freire}, {Ridolfi}, {Pan}, {Zhao},
  {Heinke}, {Chen}, {Cadelano}, {Pallanca}, {Hou}, {Fu}, {Dai},
  {G{\"u}gercino{\u{g}}lu}, {Guo}, {Hessels}, {Hu}, {Li}, {Ni}, {Pan},
  {Ransom}, {Ruan}, {Stairs}, {Tsai}, {Wang}, {Wang}, {Wang}, {Wu}, {Yuan},
  {Zhang}, {Zhu}, {Zhang}, \& {Li}}]{Zhang2023}
{Zhang}, L., {Freire}, P. C.~C., {Ridolfi}, A., {et~al.} 2023, \apjs, 269, 56

\bibitem[{{Zhang} {et~al.}(2022){Zhang}, {Ridolfi}, {Blumer}, {Freire},
  {Manchester}, {McLaughlin}, {Kremer}, {Cameron}, {Zhang}, {Behrend},
  {Burgay}, {Buchner}, {Champion}, {Chen}, {Dai}, {Feng}, {Fu}, {Guo}, {Hobbs},
  {Keane}, {Kramer}, {Levin}, {Li}, {Ni}, {Pan}, {Padmanabh}, {Possenti},
  {Ransom}, {Tsai}, {Venkatraman Krishnan}, {Wang}, {Zhang}, {Zhi}, {Zhang}, \&
  {Li}}]{Zhang2022}
{Zhang}, L., {Ridolfi}, A., {Blumer}, H., {et~al.} 2022, \apjl, 934, L21

\bibitem[{{Zhao} \& {Heinke}(2022)}]{ZhaoHeinke2022}
{Zhao}, J. \& {Heinke}, C.~O. 2022, \mnras, 511, 5964

\end{thebibliography}
\begin{appendix}
\onecolumn
\begin{landscape}
\section{Summary table}\label{summaryt}
\begin{table}[h!]
\caption{\label{summarytable}Summary of the high confidence optical counterparts identified in this work.}
\centering
\resizebox{1.35\textwidth}{!}{%
\begin{tabular}{lcccccccccccccc}
\hline\hline
Name & HUGS ID & R.A. & Dec. & $d$ & $m_{\rm F225W}$ & $m_{\rm F275W}$ & $m_{\rm F390W}$ & $m_{\rm F435W}$ & $m_{\rm F555W}$ & $m_{\rm F814W}$ & Membership & Criteria & Notes \\
& & [$^\circ$] & [$^\circ$] & [$\arcsec$] & [mag]& [mag]& [mag]& [mag]& [mag]& [mag] & probability & & \\
\hline
$\mathrm{X1  ^{c}}$ & R0077953  & 15.8119957 & $-$70.8488898 & 0.06 & 20.32  & 19.27  & 18.86  &  18.64  & 18.34 & 17.61 & 1.00 &   Unique candidate  &  qLMXB $^*$ (?) \\
$\mathrm{X2  ^{c}}$ & - & 15.8188075  & $-$70.8502508 & 0.08 &  22.89 & 22.20  & - & - & - & - &  -  &  CMD  & CV  \\
$\mathrm{X4  ^{c}}$ & R0113456 & 15.7570819 & $-$70.8381431 & 0.09 & 22.70  & 22.23  & 22.5  &  -   &  22.57  & 21.54 &  0.00 &   CMD  & Galaxy $^*$ \\
$\mathrm{X5  ^{c}}$ & R0088605 & 15.8181305 & $-$70.8467920 & 0.03 & 25.54   & 25.05  & 24.41 &   -   & 23.76  & 22.99 &  0.31 &  CMD + Variability   & BW MSP or CV \\
$\mathrm{X6  ^{c}}$ & - & 15.8111520 & $-$70.8482723 & 0.13 &23.77  & 23.44  & 23.27  &  22.03   &  23.02  & 21.80 &  0.09 & CMD + H$\alpha$  & CV \\
$\mathrm{X7  ^{c}}$ & - &15.8139031 & $-$70.8485905 & 0.06 & 22.26  & 21.17  & 20.66    &  20.31   &  20.05   & 19.17 &  -  & Variability  & AB or RB MSP \\
$\mathrm{X8  ^{c}}$ & R0004525 & 15.8058206 & $-$70.8503698 &  0.11 & 21.36  & 19.79  & 18.70 &  18.35    & 17.93   & 16.99 & 0.50  & H$\alpha$    & AB  \\
$\mathrm{X9  ^{c}}$ & R0005396 & 15.8031461 & $-$70.8481876 &  0.10 & 19.65  & 18.59  & 17.96 &  17.78    & 17.27 & 16.19 & 0.87  &  CMD   & RG + blue object $^*$ \\
$\mathrm{X10 ^{c}}$ & R0006360 & 15.8422514 & $-$70.8457921 &  0.18 & 19.82  & 18.66  & 17.87 &   -   & 16.82 & 15.67 &  0.32 & CMD  + RV var  &  semi-detached binary system \\
$\mathrm{X12 ^{c}}$ & R0005116 & 15.8187085 & $-$70.8489012 &  0.11 & 21.34  & 19.73  & 18.59  & 18.25     & 17.62 & 16.51 & 0.91   &  CMD + RV var & RSS $^*$ \\
$\mathrm{X13 ^{c}}$ & R0068161 & 15.8175106 & $-$70.8513261 &  0.46 & - & 23.97  & 22.35    &  21.92    & 21.44    & 20.41 &  0.52  &   H$\alpha$  & AB \\
$\mathrm{X15 ^{c}}$ & R0089306 & 15.8056443 & $-$70.8462507 &  0.05 & -  & - & 19.16    &  18.87    & 18.21    & 17.08 & 0.46  &  CMD  &  SSG $^*$ or MSP\\
$\mathrm{X16A^{c}}$\tablefootnote{This corresponds to the candidate counterpart identified as X16 in \citet{Kumawat2024}} & R0078991 & 15.8038772 & $-$70.8494455 &  0.15 & 22.59  & 21.08  & 19.7    &  19.34    & 18.78    & 17.68 &  0.68  &  CMD  & SSG $^*$ or MSP\\
$\mathrm{X16B^{c}}$ & R0072542 & 15.8040431 & $-$70.8495953 &  0.53 & 18.62  & 17.98  & 17.84 &  17.61  & 17.53 & 17.08 & 0.34  &  CMD + Variability + RV var& WUMa\\
$\mathrm{X17 ^{c}}$ & - & 15.8088114 & $-$70.8448532 & 0.13 & 22.56 & 22.18  & - & - & - & - & - & CMD  & CV in outburst \\ 
$\mathrm{X18A^{c}}$ & - & 15.8025839 & $-$70.8511724 & 0.24 & 23.45  & 23.25   & 24.80    &   -   & 22.64     & 21.48 &  -  &  CMD   & CV or canonical MSP  \\
$\mathrm{X18B^{c}}$ & - & 15.8026443 & $-$70.8510299 & 0.45 & 24.19   & 23.74  & 22.96    &  23.01    & 21.35     & 20.45  &  -  &  CMD   & CV or canonical MSP \\
$\mathrm{X19 ^{c}}$ & - & 15.8127189 & $-$70.8580322 & 0.67 & 25.37  & 24.36 & 25.08    &  -    & 25.05    & - & -  &  CMD   & CV or canonical MSP \\
$\mathrm{X20 ^{c}}$ & R0002866 & 15.8033216 & $-$70.8549707 & 0.11 & 22.13   & 20.50  & 19.25    &  -    & 18.32    & 17.22 &  0.89  &  CMD + H$\alpha$   & SSG $^*$ or MSP\\
$\mathrm{X21A^{c}}$ & R0004837 & 15.7943940 &  $-$70.8496118 & 0.22 & 21.24   & 19.52   & 18.16    &  -    & 17.15   & 16.06 &  0.48  & CMD + RV var& RSS $^*$ \\
$\mathrm{X21B^{c}}$\tablefootnote{This is not the same counterpart as the X21B identified by \citet{Kumawat2024}} & - & 15.7944000   &  $-$70.8497595 & 0.48 & -   & -  & 24.84    &      & 24.07     & 21.87  &  0.00 & CMD  &  BW MSP or CV  \\
$\mathrm{X23 ^{c}}$\tablefootnote{Our candidate counterpart for the X23 X-ray source does not match any of the three candidates, X23A, X23B, and X23C, suggested by \citet{Kumawat2024} for the same source. Indeed, their X23A, X23B, and X23C do not fulfill any of the criteria established in this paper for identifying high-confidence counterparts. On the other side, our $\mathrm{X23 ^{c}}$ shows both significant photometric variability and H$\alpha$ excess, in addition to its peculiar CMD position.} & R0005147 & 15.8085802 & $-$70.8488293 & 0.18 & 17.97   & 17.27   & 17.47     &  16.77    & 17.08    & 16.92  & 0.92   &  CMD + Variability + H$\alpha$   & close eclipsing binary BSS \\
$\mathrm{X24A^{c}}$ & R0029822 &  15.8526754 & $-$70.8636441 & 0.38 & 22.05   & 21.75  & 22.61     &   -   & 21.80    & 20.72  & 0.01  &  CMD   & Galaxy $^*$ \\
$\mathrm{X24B^{c}}$ & R0001102 & 15.8526677 & $-$70.8637183 & 0.39 & 21.01  & 19.75  & 19.07     &   -   & 18.42 & 17.54  & 0.86   & CMD    & SSG or MSP \\
$\mathrm{X25A^{c}}$ & - & 15.8072321 & $-$70.8517652 & 0.60 & 21.54  & 21.45  & 22.42     &   22.20   & 21.52     & 20.44  & 0.02   &   CMD  & CV \\
$\mathrm{X25B^{c}}$ & R0067240 & 15.8081627 & $-$70.8517508 & 0.50 & 21.48  & 20.52  & 20.16  &  19.92    & 19.70    & 18.99 &  0.40  &  CMD + H$\alpha$   & AB\\
$\mathrm{X28 ^{c}}$ & R0067261 & 15.8112529 & $-$70.8511236 & 0.67 & 23.55 & 23.31  & -  &  -    & -  & 22.35 &  0.00  & CMD  & CV or canonical MSP \\
$\mathrm{X33 ^{c}}$ & - & 15.8085784 & $-$70.8477017 & 0.77 & 21.53 & 20.38 & - & - & - & - & - & Variability & RB MSP or AB \\
\hline
\end{tabular}%
}
\tablefoot{The table provides, from left to right: the name of the optical counterpart, its ID in the HUGS catalogue ('-' if the source is absent from the catalogue), right ascension, declination, distance from the X-ray source position,  $m_{\rm F225W}$, $m_{\rm F275W}$, $m_{\rm F390W}$, $m_{\rm F435W}$, $m_{\rm F555W}$, and $m_{\rm F814W}$ magnitudes obtained from our photometric analysis ('-' if not detected), and membership probability ('-' if unavailable). We mark the name of the optical counterparts with a "c" as a superscript. For X-ray sources with multiple candidate counterparts, we distinguish among different candidates by appending a capital letter to their names, starting from "A". The "Criteria" column outlines the selection criteria that we used for including the counterparts in the list, where "CMD" refers to a peculiar position in the CMD, "RV var" marks the presence of variability in the RV curves and "H$\alpha$" indicates the presence of H$\alpha$ excess. The last column provides the possible interpretation of the system, with the '*' superscript indicating that the interpretation was proposed by \cite{Kumawat2024}.}
\end{table}
\end{landscape}
\section{Finding Charts}\label{Appendix} 
\begin{figure}[h!]
\includegraphics[width=0.9\hsize]{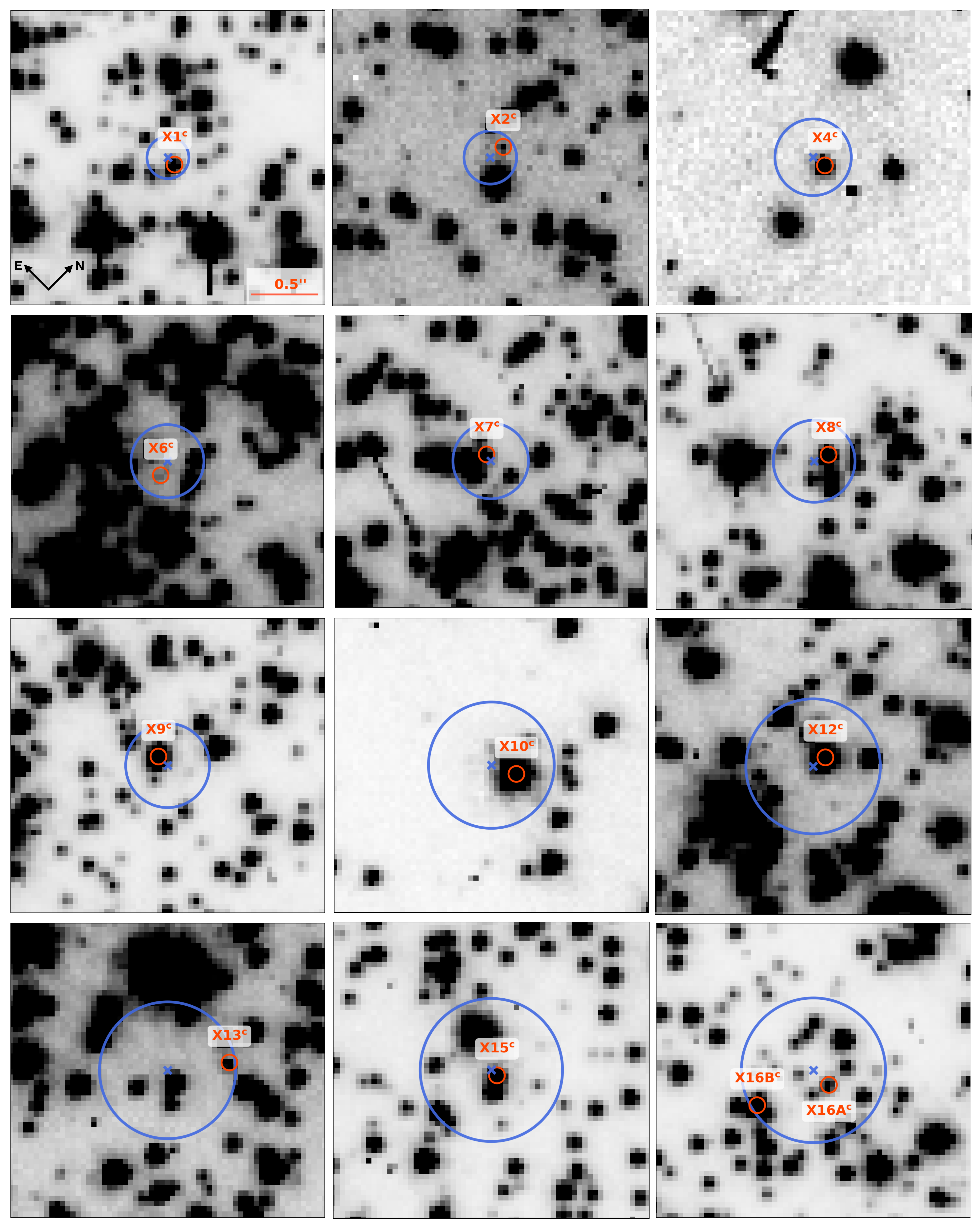}
\centering
\caption{Finding charts of our high-confidence counterparts. Each square covers an area of $2.4\arcsec$ on each side. In each panel, the X-ray source position is indicated by a blue cross, along with a blue circle whose radius represents the uncertainty in the X-ray position, $\mathrm{unc_{X}}$. The red circle marks the location of the optical counterpart. All the finding charts are oriented in the same way as indicated by the arrows in the first finding chart for $\mathrm{X1^{c}}$.}
\label{appendix1}
\end{figure}

In Fig.~\ref{appendix1}, we present the finding charts for our 28 high-confidence counterparts. All images are captured using the F390W filter of the WFC3, except for the finding charts of $\mathrm{X2^{c}}$, $\mathrm{X18A-B^{c}}$, $\mathrm{X28^{c}}$, and $\mathrm{X33^{c}}$, whose images were taken with the F275W filter of the WFC3. The finding charts for $\mathrm{X5^{c}}$ and $\mathrm{X17^{c}}$ are shown in the main text of the paper, in Fig.~\ref{finding} and Fig.~\ref{figx17}, respectively.
\begin{figure}
\includegraphics[width=0.9\hsize]{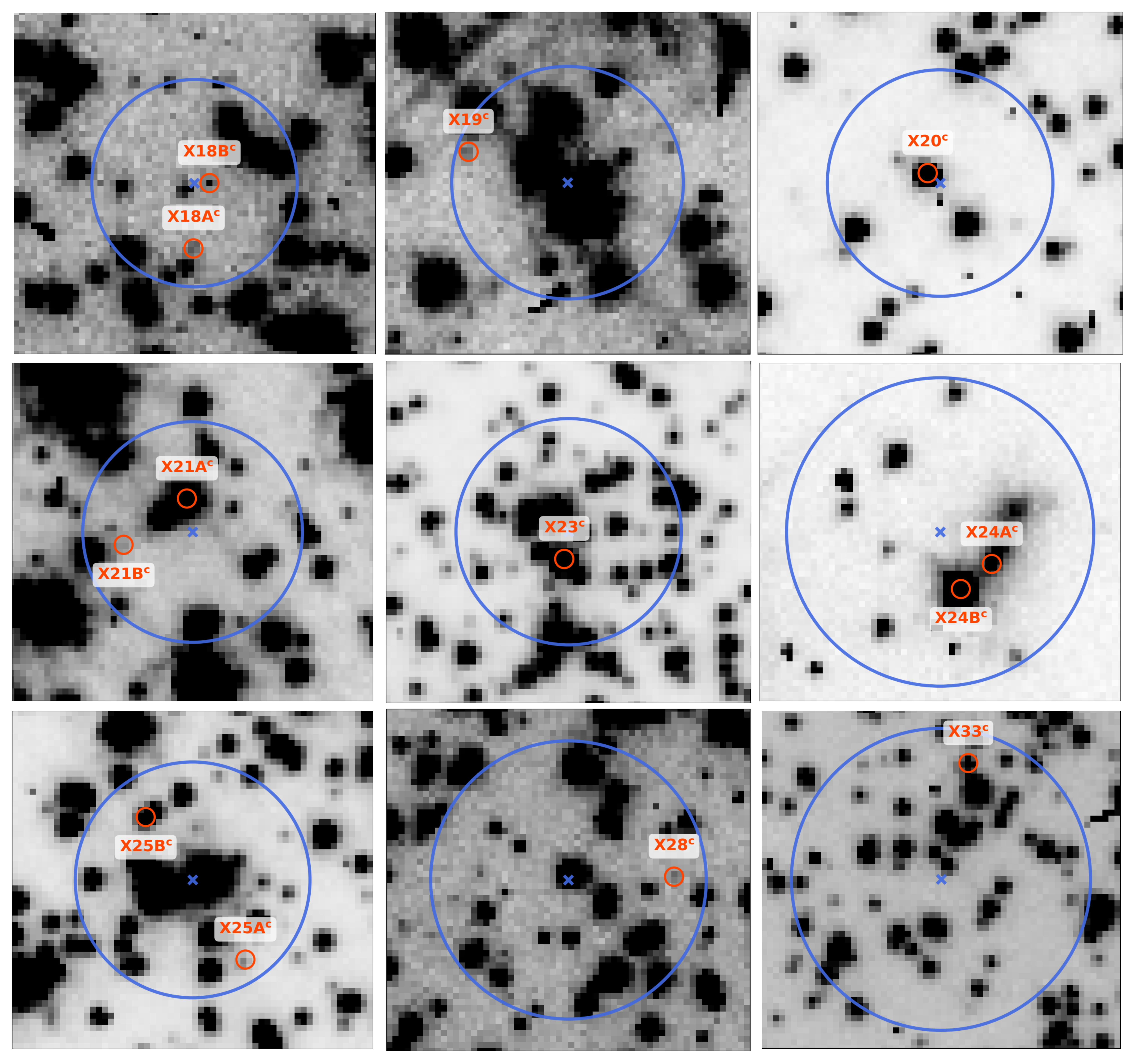}
\centering
\caption{Same as Fig.~\ref{appendix1}}
\label{appendix2}
\end{figure}
\end{appendix}
\end{document}